\pdfoutput=1
\documentclass[11pt,a4paper]{scrartcl}
\PassOptionsToPackage{obeyspaces}{url}

\usepackage[mathlines]{lineno}
\usepackage[shortlabels]{enumitem}
\usepackage{pgffor}
\usepackage{amssymb,amsmath}
\usepackage{fancyvrb}
\usepackage[normalem]{ulem}
\usepackage{array}
\usepackage{ccicons}
\usepackage{listings}
\usepackage{seqsplit}
\usepackage{multirow}
\usepackage[frozencache]{minted}
\usepackage{hepnames}
\usepackage{relsize}

\usepackage{CLICdp}

\usepackage{CLICdp_definitions}

\providecommand{\tightlist}{%
  \setlength{\itemsep}{0pt}\setlength{\parskip}{0pt}}

\ifx\paragraph\undefined\else
\let\oldparagraph\paragraph
\renewcommand{\paragraph}[1]{\oldparagraph{#1}\mbox{}}
\fi
\ifx\subparagraph\undefined\else
\let\oldsubparagraph\subparagraph
\renewcommand{\subparagraph}[1]{\oldsubparagraph{#1}\mbox{}}
\fi

\providecommand{\subtitle}[1]{}

\DeclareMathAlphabet\mathbfcal{OMS}{cmsy}{b}{n}
\newcommand{\corry}{Corryvreckan\xspace}
\newcommand{\corrybold}{\textbf{Corryvreckan}\xspace}

\newcommand{\apsq}{Allpix\textsuperscript{2}\xspace}

\newcommand{\comment}[1]{} 

\newcommand{\passthrough}[1]{\lstset{mathescape=false}#1\lstset{mathescape=true}}

\DeclareUrlCommand\parameter{\bfseries\urlstyle{tt}}
\newcommand{\command}[1]{\parameter{#1}}

\newcommand{\CPP}{C\nolinebreak[4]\hspace{-.05em}\raisebox{.2ex}{\relsize{-1}{\textbf{++}}}\xspace}

\DeclareUrlCommand\dir{\urlstyle{tt}}
\newcommand{\file}[1]{\dir{#1}}

\DeclareUrlCommand\dir{\urlstyle{tt}}
\newcommand{\module}[1]{\dir{[#1]}}

\lstdefinelanguage{Ini}
{
    basicstyle=\ttfamily\small,
    columns=fullflexible,
    morecomment=[s][\color{blue}\bfseries]{[}{]},
    morecomment=[l]{\#},
    morecomment=[l]{;},
    commentstyle=\color{gray}\ttfamily,
    alsoletter={=},
    morekeywords={=},
    otherkeywords={},
    keywordstyle={\color{green}\bfseries}
}

\newsavebox{\warningbox}
\newenvironment{warning}
  {\newcommand\colboxcolor{pink}%
   \begin{lrbox}{\warningbox}%
   \begin{minipage}{\dimexpr\linewidth-2em\relax}}
  {\end{minipage}\end{lrbox}%
   \begin{center}
     \setlength\fboxsep{0pt}
     \colorbox{\colboxcolor}{\setlength\fboxsep{1em}\fbox{\usebox{\warningbox}}}
  \end{center}}

\newcommand{\modulehead}[4]{
    \renewcommand{\arraystretch}{1.1}
    \rowcolors*{1}{gray!5}{gray!5}
    \begin{tabular}{|p{\dimexpr 0.2\linewidth-2\tabcolsep}p{\dimexpr 0.8\linewidth-2\tabcolsep}|}
      \hline
      \textbf{Maintainer} & {#1} \\
      \textbf{Module Type} & \emph{#2} \\
      \textbf{Detector Type} & \emph{#3} \\
      \textbf{Status} & #4 \\
      \hline
    \end{tabular}
}

\newcommand{\moduleheadGlobal}[2]{
    \renewcommand{\arraystretch}{1.1}
    \rowcolors*{1}{gray!5}{gray!5}
    \begin{tabular}{|p{\dimexpr 0.2\linewidth-2\tabcolsep}p{\dimexpr 0.8\linewidth-2\tabcolsep}|}
      \hline
      \textbf{Maintainer} & {#1} \\
      \textbf{Module Type} & \emph{GLOBAL} \\
      \textbf{Status} & #2 \\
      \hline
    \end{tabular}
}

\newcommand{\includemodulesmd}{\def\temp{{modules/AlignmentDUTResidual.tex,modules/AlignmentMillepede.tex,modules/AlignmentTrackChi2.tex,modules/AnalysisDUT.tex,modules/AnalysisEfficiency.tex,modules/AnalysisTelescope.tex,modules/AnalysisTimingATLASpix.tex,modules/Clustering4D.tex,modules/ClusteringSpatial.tex,modules/Correlations.tex,modules/DUTAssociation.tex,modules/EtaCalculation.tex,modules/EtaCorrection.tex,modules/EventLoaderATLASpix.tex,modules/EventLoaderCLICpix.tex,modules/EventLoaderCLICpix2.tex,modules/EventLoaderEUDAQ.tex,modules/EventLoaderEUDAQ2.tex,modules/EventLoaderMuPixTelescope.tex,modules/EventLoaderTimepix1.tex,modules/EventLoaderTimepix3.tex,modules/FileReader.tex,modules/FileWriter.tex,modules/ImproveReferenceTimestamp.tex,modules/MaskCreator.tex,modules/MaskCreatorTimepix3.tex,modules/Metronome.tex,modules/OnlineMonitor.tex,modules/Prealignment.tex,modules/TextWriter.tex,modules/Tracking4D.tex,modules/TrackingSpatial.tex,modules/TreeWriterDUT.tex}}\ifx\temp\empty
  \textit{Module documentation not added because Markdown to \LaTeX~conversion was not possible. Pandoc is required for the conversion.}
\else
  \hypertarget{alignmentdutresidual}{%
\subsection{AlignmentDUTResidual}\label{alignmentdutresidual}}

\modulehead{Daniel Hynds (\href{mailto:daniel.hynds@cern.ch}{\nolinkurl{daniel.hynds@cern.ch}}),\newline Simon Spannagel (\href{mailto:simon.spannagel@cern.ch}{\nolinkurl{simon.spannagel@cern.ch}})}{DUT}{all}{Functional}

\paragraph{Description}

This module performs translational and rotational DUT alignment. The
alignment is performed with respect to the reference plane set in the
configuration file.
This module uses tracks for alignment. The module moves the detector it
is instantiated for and minimises the unbiased residuals calculated from
the track intercepts with the plane.

\paragraph{Parameters}

\begin{itemize}
\tightlist
\item
  \parameter{iterations}: Number of times the chosen
  alignment method is to be iterated. Default value is
  \parameter{3}.
\item
  \parameter{align_position}: Boolean to select whether
  to align the X and Y displacements of the detector. Note that
  the Z displacement is never changed. The default value is
  \parameter{true}.
\item
  \parameter{align_orientation}: Boolean to select
  whether to align the three rotations of the detector under
  consideration. The default value is
  \parameter{true}.
\item
  \parameter{prune_tracks}: Boolean to set if tracks with
  a number of associated clusters \textgreater{}
  \parameter{max_associated} \parameter{_clusters} or with a \parameter{track_chi2ndof} \textgreater{} \parameter{max_track_chi2ndof}
  should be excluded from use in the alignment. The number of discarded
  tracks is output on terminal. Default is
  \parameter{false}.
\item
  \parameter{max_associated_clusters}: Maximum number of
  associated clusters per track allowed when
  \parameter{prune_tracks = true} for the track to be
  used in the alignment. Default value is \parameter{1}.
\item
  \parameter{max_track_chi2ndof}: Maximum \parameter{track_chi2ndof}
  value allowed when \parameter{prune_tracks = true} for
  the track to be used in the alignment. Default value is
  \parameter{10.0}.
\end{itemize}

\paragraph{Plots produced}

For the DUT, the following plots are produced:
\begin{itemize}
\tightlist
\item Histograms of the track residuals in X/Y
\item Various profile plots of the residuals
\end{itemize}

\paragraph{Usage}

\begin{minted}[frame=single,framesep=3pt,breaklines=true,tabsize=2,linenos]{ini}
[Corryvreckan]
# The global track limit can be used to reduce the run time:
number_of_tracks = 200000

[AlignmentDUTResidual]
log_level = INFO
\end{minted}

\hypertarget{alignmentmillepede}{%
\subsection{AlignmentMillepede}\label{alignmentmillepede}}

\moduleheadGlobal{Simon Spannagel (\href{mailto:simon.spannagel@cern.ch}{\nolinkurl{simon.spannagel@cern.ch}})}{Work in progress}

\paragraph{Description}

This implementation of the \texttt{[AlignmentMillepede]} module has been taken from the
\href{https://gitlab.cern.ch/lhcb/Kepler}{Kepler framework} used for
test beam data analysis within the LHCb collaboration. It has been
written by Christoph Hombach and has seen contributions from Tim Evans
and Heinrich Schindler. This version is adapted to the
Corryvreckan framework.

The Millepede algorithm allows a simultaneous fit of both the tracks and
the alignment constants.
The modules stops if the convergence, i.e.~the absolute sum of all
corrections over the total number of parameters, is smaller than the
configured value.

\paragraph{Parameters}

\begin{itemize}
\tightlist
\item
  \parameter{exclude_dut} : Exclude the DUT from the
  alignment procedure. Default value is \parameter{false}.
\item
  \parameter{iterations}: Number of times the chosen
  alignment method is to be iterated. Default value is
  \parameter{3}.
\item
  \parameter{dofs}: Degrees of freedom to be aligned. This
  parameter should be given as vector of six boolean values for the
  parameters ``Translation X'', ``Translation Y'', ``Translation Z'',
  ``Rotation X'', ``Rotation Y'' and ``Rotation Z''. The default setting
  is an alignment of all parameters except for ``Translation Z'', i.e.
  \parameter{dofs = true, true, false, true, true, true}.
\item
  \parameter{residual_cut}: Residual cut to reject a
  track as an outlier. Default value is
  \parameter{0.05mm};
\item
  \parameter{residual_cut_init}: Initial residual cut
  for outlier rejection in the first iteration. This value is applied
  for the first iteration and replaced by
  \parameter{residual_cut} thereafter. Default value is
  \parameter{0.6mm}.
\item
  \parameter{number_of_stddev}: Cut to reject track
  candidates based on their $\chi^2$/ndof value. Default value is
  \parameter{0}, i.e.~the feature is disabled.
\item
  \parameter{sigmas}: Uncertainties for each of the
  alignment parameters described above, in their respective units. Defaults to
  \parameter{50um, 50um, 50um,} \parameter{0.005rad, 0.005rad, 0.005rad}.
\item
  \parameter{convergence}: Convergence value at which the
  module stops iterating. It is defined as the sum of all residuals divided by the number of free parameters. Default value is \parameter{10e-5}.
\end{itemize}

\paragraph{Plots produced}

No plots are produced.

\paragraph{Usage}

\begin{minted}[frame=single,framesep=3pt,breaklines=true,tabsize=2,linenos]{ini}
[Millepede]
iterations = 10
dofs = true, true, false, true, true, true
exclude_dut = false
\end{minted}

\hypertarget{alignmenttrackchi2}{%
\subsection{AlignmentTrackChi2}\label{alignmenttrackchi2}}

\moduleheadGlobal{Daniel Hynds (\href{mailto:daniel.hynds@cern.ch}{\nolinkurl{daniel.hynds@cern.ch}}),\newline
Simon Spannagel (\href{mailto:simon.spannagel@cern.ch}{\nolinkurl{simon.spannagel@cern.ch}})}{Functional}

\paragraph{Description}

This module performs translational and rotational telescope plane
alignment. The alignment is performed with respect to the reference
plane set in the configuration file.

This module uses tracks on the clipboard to align the telescope planes.
For each telescope detector except the reference plane, this method
moves the detector, refits all of the tracks, and minimises the $\chi^2$
of these new tracks. This method automatically iterates through all
devices contributing to the track.

\paragraph{Parameters}

\begin{itemize}
\tightlist
\item
  \parameter{iterations}: Number of times the chosen
  alignment method is to be iterated. Default value is
  \parameter{3}.
\item
  \parameter{align_position}: Boolean to select whether
  to align the X and Y displacements of the detector or not. Note that
  the Z displacement is never aligned. The default value is
  \parameter{true}.
\item
  \parameter{align_orientation}: Boolean to select
  whether to align the three rotations of the detector under
  consideration. The default value is
  \parameter{true}.
\item
  \parameter{prune_tracks}: Boolean to set if tracks with
  a number of associated clusters \textgreater{}
  \parameter{max_associated_clusters} or with a track
  $\chi^2$/ndof \textgreater{} \parameter{max_track_chi2ndof}
  should be excluded from use in the alignment. The number of discarded
  tracks is output on terminal. Default is
  \parameter{false}.
\item
  \parameter{max_associated_clusters}: Maximum number of
  associated clusters per track allowed when
  \parameter{prune_tracks = true} for the track to be
  used in the alignment. Default value is \parameter{1}.
\item
  \parameter{max_track_chi2ndof}: Maximum track $\chi^2$/ndof
  value allowed when \parameter{prune_tracks = true} for
  the track to be used in the alignment. Default value is
  \parameter{10.0}.
\end{itemize}

\paragraph{Plots produced}

For each detector, the following plots are produced:
\begin{itemize}
\tightlist
\item Graphs of the translational shifts along the X/Y-axis vs.~the iteration number
\item Graphs of the rotational shifts along the X/Y/Z-axis vs.~the iteration number
\end{itemize}

\paragraph{Usage}

\begin{minted}[frame=single,framesep=3pt,breaklines=true,tabsize=2,linenos]{ini}
[Corryvreckan]
# The global track limit can be used to restrict the alignment:
number_of_tracks = 200000

[AlignmentTrackChi2]
log_level = INFO
\end{minted}

\hypertarget{analysisdut}{%
\subsection{AnalysisDUT}\label{analysisdut}}

\modulehead{Simon Spannagel (\href{mailto:simon.spannagel@cern.ch}{\nolinkurl{simon.spannagel@cern.ch}})}{DUT}{all}{Functional}

\paragraph{Description}

Generic analysis module for all prototypes.

\paragraph{Parameters}

\begin{itemize}
\tightlist
\item
  \parameter{time_cut_frameedge}: Parameter to discard
  telescope tracks at the frame edges (start and end of the current
  CLICpix2 frame). Defaults to \parameter{20ns}.
\item
  \parameter{chi2ndof_cut}: Acceptance criterion for the maximum telescope track $\chi^2$/ndof, defaults to a value of \parameter{3}.
\item
  \parameter{use_closest_cluster}: If \parameter{true} the cluster with the smallest distance to
  the track is used if a track has more than one associated cluster. If \parameter{false}, loop over all associated clusters. Defaults to \parameter{true}.
\end{itemize}

\paragraph{Plots produced}

For the DUT, the following plots are produced:
\begin{itemize}
\item 2D histograms:
    \begin{itemize}
    \tightlist
    \item Maps of the position, size, and charge/raw value of associated clusters, of all pixels of associated clusters and those within a region-of-interest
    \item Maps of the in-pixel efficiencies in local/global coordinates
    \item Maps of matched/non-matched track positions
    \end{itemize}
\item 1D histograms:
    \begin{itemize}
    \tightlist
    \item Histograms of the cluster size of associated clusters in X/Y
    \item Histogram of the charge/raw values of associated clusters
    \item Varios histograms for track residuals for different cluster sizes
    \end{itemize}
\end{itemize}

\paragraph{Usage}

\begin{minted}[frame=single,framesep=3pt,breaklines=true,tabsize=2,linenos]{ini}
[AnalysisDUT]
timeCutFrameEdge = 50ns
chi2ndof_cut = 5.
use_closest_cluster = false
\end{minted}

\hypertarget{analysisefficiency}{%
\subsection{AnalysisEfficiency}\label{analysisefficiency}}

\modulehead{Simon Spannagel (\href{mailto:simon.spannagel@cern.ch}{\nolinkurl{simon.spannagel@cern.ch}}), \newline Jens
Kroeger (jens.kroeger@cern.ch)}{DUT}{all}{Functional}

\paragraph{Description}

This module measures the efficiency of the device-under-test by
comparing its cluster positions with the interpolated track position at
the DUT. It also comprises a range of histograms to investigate where
inefficiencies might come from.

\paragraph{Parameters}

\begin{itemize}
\tightlist
\item
  \parameter{time_cut_frameedge}: Parameter to discard
  telescope tracks at the frame edges (start and end of the current
  event window). Defaults to \parameter{20ns}.
\item
  \parameter{chi2ndof_cut}: Acceptance criterion for
  telescope tracks, defaults to a value of \parameter{3}.
\item
  \parameter{inpixel_bin_size}: Parameter to set the bin
  size of the in-pixel 2D efficiency histogram. This should be given in
  units of distance and the same value is used in both axes. Defaults to
  \parameter{1.0um}.
\end{itemize}

\paragraph{Plots produced}

For the DUT, the following plots are produced:
\begin{itemize}
\tightlist
\item
  2D histograms:

  \begin{itemize}
  \tightlist
  \item
    2D maps of in-pixel and chip and efficiency in local and global coordinates
  \item
    2D maps of the position difference of a track (with/without associated
    cluster) to the previous track
  \end{itemize}
\item
  1D histograms:

  \begin{itemize}
  \tightlist
  \item
    Histogram of all single-pixel efficiencies
  \item
    Histogram of time/position difference of the matched/non-matched track (with/without
    associated cluster) to the previous track
  \end{itemize}

\item
  Other:

  \begin{itemize}
  \tightlist
  \item
    Value of total efficiency as \parameter{TEfficiency}
    including (asymmetric) error bars
  \item
    Value of total efficency as \parameter{TName} so it
    can be read off easily by eye from the root file
  \end{itemize}
\end{itemize}

\paragraph{Usage}

\begin{minted}[frame=single,framesep=3pt,breaklines=true,tabsize=2,linenos]{ini}
[AnalysisEfficiency]
chi2ndof_cut = 5
\end{minted}

\hypertarget{analysistelescope}{%
\subsection{AnalysisTelescope}\label{analysistelescope}}

\moduleheadGlobal{Simon Spannagel (\href{mailto:simon.spannagel@cern.ch}{\nolinkurl{simon.spannagel@cern.ch}})}{Functional}

\paragraph{Description}

This module produces reference histograms for the telescope performance
used for tracking. It produces local and global biased residuals for
each of the telescope planes, and if Monte Carlo information is
available, calculates the residuals between track and Monte Carlo
particle.

Furthermore, the telescope resolution at the position of the DUT
detector is plotted of Monte Carlo information is available. The Monte
Carlo particle position is compared with the track interception with the
DUT.

\paragraph{Parameters}

\begin{itemize}
\tightlist
\item
  \parameter{chi2ndof_cut}: track $\chi^2$ over number of degrees
  of freedom for the track to be taken into account. Tracks with a
  larger value are discarded. Default value is
  \parameter{3}.
\end{itemize}

\paragraph{Plots produced}

For the DUT, the following plots are produced:
\begin{itemize}
\tightlist
\item
  Histogram of the resolution at position of DUT in X/Y
\end{itemize}

For each detector participating in tracking, the following plots are
produced:

\begin{itemize}
\tightlist
\item
  Histograms of the biased local/global track residual for X/Y
\item
  Local/global residuals with track and Monte Carlo particle for X/Y
\end{itemize}

\paragraph{Usage}

\begin{minted}[frame=single,framesep=3pt,breaklines=true,tabsize=2,linenos]{ini}
[NAME]
chi2ndof_cut = 3
\end{minted}

\hypertarget{analysistimingatlaspix}{%
\subsection{AnalysisTimingATLASpix}\label{analysistimingatlaspix}}

\modulehead{Jens Kroeger (\href{mailto:jens.kroeger@cern.ch}{\nolinkurl{jens.kroeger@cern.ch}})}{DUT}{ATLASpix}{Functional}

\paragraph{Description}

This module can be used for an in-depth timing
analysis of the ATLASpix, including row and timewalk corrections.
The ATLASpix shows a row-dependent shift of the time residuals which can be corrected for.
Timewalk describes a shift of the time residual depending on the signal strength which can be deduced from the time-over-threshold in the ATLASpix. 
Also this can be corrected for for an improved time resolution.

Before being able to apply the row and timewalk correction, correction
files need to be provided. These can be generated by setting
\parameter{calc_corrections=true}. The calculation of the
timewalk correction is based on a row-corrected histogram such that this
procedure needs to be performed in 2 steps.

The row correction should be generally valid whereas the timewalk correction is threshold dependent and thus needs to be repeated each time the threshold is changed.

\begin{enumerate}
\def\labelenumi{\arabic{enumi}.}
\tightlist
\item
  Calculate the row correction.
\item
  Use row correction file and calculate timewalk correction. After this,
  both corrections can be applied consecutively.
\end{enumerate}

\paragraph{Parameters}

\begin{itemize}
\tightlist
\item
  \parameter{time_cut_rel}: Number of standard deviations the \parameter{time_resolution} of the detector plane will be multiplied by. This value then defines the range of the track time correlation histograms and should be set according to the value chosen for the \module{DUTAssociation}. By default, a relative time cut is applied. Absolute and relative time cuts are mutually exclusive. Defaults to \parameter{3.0}.
\item
  \parameter{time_cut_abs}: Specifies an absolute value for the range of the track time correlation histograms and should be set according to the value chosen for the \module{DUTAssociation}. Absolute and relative time cuts are mutually exclusive. No default value.
\item
  \parameter{chi2ndof_cut}: Acceptance criterion for
  telescope tracks, defaults to a value of \parameter{3}.
\item
  \parameter{time_cut_frameedge}: Parameter to discard
  telescope tracks at the frame edges (start and end of the current
  frame). Defaults to \parameter{20ns}.
\item
  \parameter{cluster_charge_cut}: Parameter to discard
  clusters with a charge larger than the cut. Defaults to
  \parameter{100000e} (inifitely large).
\item
  \parameter{cluster_size_cut}: Parameter to discard
  clusters with a size too large, only for debugging purposes, default
  is 100 (inifitely large).
\item
  \parameter{high_tot_cut}: Cut dividing `low' and
  `high' ToT events (based on seed pixel ToT). Defaults to
  \parameter{40}.
\item
  \parameter{high_charge_cut}: Cut dividing `low' and
  `high' charge events (based on cluster charge). Defaults to
  \parameter{40e}.
\item
  \parameter{left_tail_cut}: Cut to divide into left
  tail and main peak of time correlation histogram. Only used to
  investigate characteristics of left tail. Defaults to
  \parameter{-10ns}.
\item
  \parameter{calc_corrections}: If
  \parameter{true}, TGraphErrors for row and timewalk
  corrections are produced. Defaults to false.
\item
  \parameter{correction_file_row}: Path to file which
  contains TGraphErrors for row correction. If this parameter is set,
  also \parameter{correction_graph_row} needs to be set.
  No default.
\item
  \parameter{correction_file_timewalk}: Path to file
  which contains TGraphErrors for timewalk correction. If this parameter
  is set, also \parameter{correction_graph_timewalk}
  needs to be set. No default.
\item
  \parameter{correction_graph_row}: Name of the
  TGraphErrors including its path in the root file used for row
  correction. E.g.
  \dir{AnalysisTimingATLASpix/apx_0/gTimeCorrelationVsRow}. No default.
\item
  \parameter{correction_graph_timewalk}: Name of the
  TGraphErrors including its path in the root file used for row
  correction. E.g.
  \dir{AnalysisTimingATLASpix/apx_0/gTimeCorrelationVsTot}. No default.
\end{itemize}

\paragraph{Plots produced}

For the DUT, the following plots are produced:
\begin{itemize}
\tightlist
\item
  1D histograms:

  \begin{itemize}
  \tightlist
  \item
    Various histograms of track time correlations
  \item
    Histograms of the pixel ToT
  \item
    Histograms of the pixel timestamp
  \item
    Histograms of the cluster size
  \end{itemize}
\item
  2D histograms:

  \begin{itemize}
  \tightlist
  \item
    Histograms of the track time correlation vs.~time/column/row/seed pixel ToT for various cluster sizes, before and after row and timewalk corrections
  \item
    Cluster size vs.~cluster ToT (only associated clusters)
  \item
    Various 2D maps of clusters and all pixels from clusters
  \item
    2D maps of in-pixel distribution of tracks
  \item
    Map of associated clusters
  \item
    2D maps of pixel ToT vs.~time
  \end{itemize}
\item
  Other:

  \begin{itemize}
  \tightlist
  \item
    Graphs of the peak of time correlation vs.~row/ToT before and after row/timewalk corrections
  \end{itemize}
\end{itemize}

\paragraph{Usage}

\begin{minted}[frame=single,framesep=3pt,breaklines=true,tabsize=2,linenos]{ini}
[AnalysisTimingATLASpix]
calc_corrections = false
correction_file_row = "correction_files/row_correction_file.root"
correction_graph_row = "AnalysisTimingATLASpix/apx0/gRTimeCorrelationVsRow"
correction_file_timewalk = "correction_files/timewalk_correction_file.root"
correction_graph_timewalk = "AnalysisTimingATLASpix/apx0/gTimCorrelationVsTot"
\end{minted}

\hypertarget{clustering4d}{%
\subsection{Clustering4D}\label{clustering4d}}

\modulehead{Daniel Hynds (\href{mailto:daniel.hynds@cern.ch}{\nolinkurl{daniel.hynds@cern.ch}})}{DETECTOR}{all}{Functional}

\paragraph{Description}

This module performs clustering for detectors with valid individual hit
timestamps. The clustering method is either an arithmetic mean or a a
charge-weighted centre-of-gravity calculation, using a positional cut
and a timing cut on proximity. If the pixel information is binary
(i.e.~no valid charge-equivalent information is available), the
arithmetic mean is calculated for the position. Also, if one pixel of a
cluster has charge zero, the arithmetic mean is calculated even if
charge-weighting is selected because it is assumed that the zero-reading
is false and does not to represent a low charge but an unknown value.
Thus, the arithmetic mean is safer.

Split clusters can be recovered using a larger search radius for
neighbouring pixels.

\paragraph{Parameters}

\begin{itemize}
\tightlist
\item
   \parameter{time_cut_rel}: Number of standard deviations the \parameter{time_resolution} of the detector plane will be multiplied by. This value is then used as the maximum time difference allowed between pixels for association to a cluster. By default, a relative time cut is applied. Absolute and relative time cuts are mutually exclusive. Defaults to \parameter{3.0}.
\item 
  \parameter{time_cut_abs}: Specifies an absolute value for the maximum time difference allowed between pixels for association to a cluster. Absolute and relative time cuts are mutually exclusive. No default value.
\item
  \parameter{neighbour_radius_col}: Search radius for
  neighbouring pixels in column direction, defaults to
  \parameter{1} (do not allow split clusters)
\item
  \parameter{neighbour_radius_row}: Search radius for
  neighbouring pixels in row direction, defaults to
  \parameter{1} (do not allow split clusters)
\item
  \parameter{charge_weighting}: If true, calculate a
  charge-weighted mean for the cluster centre. If false, calculate the
  simple arithmetic mean per column/row. Defaults to \parameter{true}.
\end{itemize}

\paragraph{Plots produced}

For each detector the following plots are produced:

\begin{itemize}
\tightlist
\item
  Histograms of the cluster size, charge, seed charge, width in X and Y, timestamps, and multiplicity
\item
  2D map of the cluster positions in global coordinates
\end{itemize}

\paragraph{Usage}

\begin{minted}[frame=single,framesep=3pt,breaklines=true,tabsize=2,linenos]{ini}
[Clustering4D]
time_cut_abs = 200ns
\end{minted}

\hypertarget{clusteringspatial}{%
\subsection{ClusteringSpatial}\label{clusteringspatial}}

\modulehead{Daniel Hynds (\href{mailto:daniel.hynds@cern.ch}{\nolinkurl{daniel.hynds@cern.ch}})}{DETECTOR}{all}{Functional}

\paragraph{Description}

This module clusters the input data of a detector without individual hit
timestamps. The clustering method only uses positional information:
either charge-weighted centre-of-gravity or arithmetic mean calculation
using touching neighbours method, and no timing information. If the
pixel information is binary (i.e.~no valid charge-equivalent information
is available), the arithmetic mean is calculated for the position. Also,
if one pixel of a cluster has charge zero, the arithmetic mean is
calculated even if charge-weighting is selected because it is assumed
that the zero-reading is false and does not to represent a low charge
but an unknown value. These clusters are stored on the clipboard for
each device.

\paragraph{Parameters}

\begin{itemize}
\tightlist
\item
  \parameter{use_trigger_timestamp}: If true, set
  trigger timestamp of Corryvreckan event as cluster timestamp. If
  false, set pixel timestamp. Default value is
  \parameter{false}.
\item
  \parameter{charge_weighting}: If true, calculate a
  charge-weighted mean for the cluster centre. If false, calculate the
  simple arithmetic mean. Defaults to \parameter{true}.
\end{itemize}

\paragraph{Plots produced}

For each detector the following plots are produced:

\begin{itemize}
\tightlist
\item
  Histograms of the cluster size, charge, seed charge, width in X and Y, timestamps, and multiplicity
\item
  2D map of the cluster positions in global and local coordinates
\end{itemize}

\paragraph{Usage}

\begin{minted}[frame=single,framesep=3pt,breaklines=true,tabsize=2,linenos]{ini}
[ClusteringSpatial]
use_trigger_timestamp = true
\end{minted}

\hypertarget{correlations}{%
\subsection{Correlations}\label{correlations}}

\modulehead{Simon Spannagel (\href{mailto:simon.spannagel@cern.ch}{\nolinkurl{simon.spannagel@cern.ch}}), \newline
Daniel Hynds (\href{mailto:daniel.hynds@cern.ch}{\nolinkurl{daniel.hynds@cern.ch}})}{DETECTOR}{all}{Functional}

\paragraph{Description}

This module collects pixel and
cluster objects from the clipboard and creates
correlation and timing plots with respect to the reference detector.
No plots are produced for \parameter{aux} devices.

\paragraph{Parameters}

\begin{itemize}
\tightlist
\item
  \parameter{do_time_cut}: Boolean to switch on/off the cut on cluster times for correlations. Defaults to \parameter{false}.
\item
  \parameter{time_cut_rel}: Number of standard deviations the \parameter{time_resolution} of the detector plane will be multiplied by. This value is then used as the maximum time difference for cluster correlation if \parameter{do_time_cut = true}. A relative time cut is applied by default when \parameter{do_time_cut = true}. Absolute and relative time cuts are mutually exclusive. Defaults to \parameter{3.0}.
\item
  \parameter{time_cut_abs}: Specifies an absolute value for the maximum time difference allowed for cluster correlation if \parameter{do_time_cut = true}. Absolute and relative time cuts are mutually exclusive. No default value.
  \parameter{do_time_cut} is set to
  \parameter{true}, defaults to
  \parameter{100ns}.
\end{itemize}

\paragraph{Plots produced}

For each device the following plots are produced:
\begin{itemize}
\item 2D histograms:
    \begin{itemize}
    \tightlist
    \item
      Hitmaps on pixel and cluster-level
    \item
      Various 2D histograms of the spatial correlations in local/global coordinates
    \item
      2D histogram of the timing correlation over time
    \end{itemize}

\item 1D histograms:
    \begin{itemize}
    \tightlist
    \item
      Various histograms of the spatial correlations in X/Y in local/global coordinates
    \item
      Various histograms of the time correlation
    \end{itemize}
\end{itemize}

\paragraph{Usage}

\begin{minted}[frame=single,framesep=3pt,breaklines=true,tabsize=2,linenos]{ini}
[Correlations]
do_time_cut = true
time_cut_rel = 5.0
\end{minted}

\hypertarget{dutassociation}{%
\subsection{DUTAssociation}\label{dutassociation}}

\modulehead{Simon Spannagel (\href{mailto:simon.spannagel@cern.ch}{\nolinkurl{simon.spannagel@cern.ch}})}{DUT}{all}{Functional}

\paragraph{Description}

Module to establish an association between clusters on a DUT plane and a
reference track. The association allows for cuts in position and time.
For the spatial cut, two options are implemented which can be chosen
using \parameter{use_cluster_centre}.

By default, the distance of the closest pixel of a cluster to the track
intercept is compared to the spatial cut in
local coordinates. If larger than the cut, the cluster is not associated
to the track. This option can be chosen, e.g.~for an efficiency
analysis, when the cluster centre might be pulled away from the track
intercept by a delta electron in the silicon. The other option is to
compare the distance between the cluster centre and the track intercept
to the spatial cut (also in local
coordinates).

The module assigns the DUT clusters found to be within the specified criteria to the respective track.
This information can later on be exploited in analysis modules.

\paragraph{Parameters}

\begin{itemize}
\tightlist
\item
  \parameter{spatial_cut_rel}: Factor by which the \parameter{spatial_resolution} in X and Y of each detector plane will be multiplied. These calculated value are defining an ellipse which is then used as the maximum distance in the XY plane allowed between clusters and a track for association to the track. By default, a relative spatial cut is applied. Absolute and relative spatial cuts are mutually exclusive. Defaults to \parameter{3.0}.
\item
  \parameter{spatial_cut_abs}: Specifies a set of absolute value (X and Y) which defines an ellipse for the maximum spatial distance in the XY plane between clusters and a track for association to the track. Absolute and relative spatial cuts are mutually exclusive. No default value.
\item
  \parameter{time_cut_rel}: Number of standard deviations the \parameter{time_resolution} of the detector plane will be multiplied by. This value is then used as the maximum time difference allowed between a DUT cluster and track for association. By default, a relative time cut is applied. Absolute and relative time cuts are mutually exclusive. Defaults to \parameter{3.0}.
\item
  \parameter{time_cut_abs}: Specifies an absolute value for the maximum time difference allowed between DUT cluster and track for association. Absolute and relative time cuts are mutually exclusive. No default value.
\item
  \parameter{use_cluster_centre}: If set true, the
  cluster centre will be compared to the track position for the spatial
  cut. If false, the nearest pixel in the cluster will be used. Defaults
  to \parameter{false}.
\end{itemize}

\paragraph{Plots produced}

For the DUT, the following plots are produced:
\begin{itemize}
\tightlist
\item
  Histograms of the distance in X/Y from the cluster to the pixel closest to the track for various cluster sizes
\item
  Histogram of the number of associated clusters per track
\item
  Histogram of the number of clusters discarded by a given cut
\end{itemize}

\paragraph{Usage}

\begin{minted}[frame=single,framesep=3pt,breaklines=true,tabsize=2,linenos]{ini}
[DUTAssociation]
spatial_cut_abs = 100um, 50um
time_cut_abs = 200ns
use_cluster_centre = false
\end{minted}

\hypertarget{etacalculation}{%
\subsection{EtaCalculation}\label{etacalculation}}

\modulehead{Daniel Hynds (\href{mailto:daniel.hynds@cern.ch}{\nolinkurl{daniel.hynds@cern.ch}}), \newline Simon Spannagel
(\href{mailto:simon.spannagel@cern.ch}{\nolinkurl{simon.spannagel@cern.ch}})}{DETECTOR}{all}{Work in progress}

\paragraph{Description}

This module performs a fit to obtain corrections for non-linear charge sharing, also know as the $\eta$-distribution. The distributions are calculated for two-pixel
clusters of any detector in the analysis by comparing the in-pixel track
position and the calculated cluster centre position. Histograms for all
available detectors are filled for both X and Y coordinate. At the end
of the run, fits to the recorded profiles are performed using the
provided formulas. A printout of the resulting fit parameters is
provided in the format read by the \module{EtaCorrection} module for convenience.

In order to measure the correct $\eta$-distribution, no additional
$\eta$-correction should be applied before this calculation, i.e.~by
using the \texttt{[EtaCorrection]} module.

\paragraph{Parameters}

\begin{itemize}
\tightlist
\item
  \parameter{chi2ndof_cut}: Track quality cut on its $\chi^2$
  over numbers of degrees of freedom. Default value is
  \parameter{100}.
\item
  \parameter{eta_formula_x} /
  \parameter{eta_formula_y}: Formula for the
  recorded $\eta$-distributions, defaults to a polynomial of fifth
  order, i.e.
  $$[0] + [1]*x + [2]*x^2 + [3]*x^3 + [4]*x^4 + [5]*x^5$$
\end{itemize}

\paragraph{Plots produced}

For each detector the following plots are produced:

\begin{itemize}
\tightlist
\item
  2D histogram of the calculated $\eta$-distribution, for x and y
\item
  Profile plot of the calculated $\eta$-distribution, for x and y
\end{itemize}

\paragraph{Usage}

\begin{minted}[frame=single,framesep=3pt,breaklines=true,tabsize=2,linenos]{ini}
[EtaCalculation]
chi2ndof_cut = 100
eta_formula_x = [0] + [1]*x + [2]*x^2 + [3]*x^3 + [4]*x^4 + [5]*x^5
\end{minted}

\hypertarget{etacorrection}{%
\subsection{EtaCorrection}\label{etacorrection}}

\modulehead{Daniel Hynds (\href{mailto:daniel.hynds@cern.ch}{\nolinkurl{daniel.hynds@cern.ch}}), \newline Simon Spannagel
(\href{mailto:simon.spannagel@cern.ch}{\nolinkurl{simon.spannagel@cern.ch}})}{DETECTOR}{all}{Work in progress}

\paragraph{Description}

This module applies the previously determined $\eta$-corrections to cluster positions of any detector in order to correct for non-linear charge sharing. Corrections can be applied to any
cluster read from the clipboard. The correction function as well as the
parameters for each of the detectors can be given separately for X and Y
via the configuration file.

This module does not calculate the $\eta$ distribution.

\paragraph{Parameters}

\begin{itemize}
\tightlist
\item
  \parameter{eta_formula_x} /
  \parameter{eta_formula_y}: The formula for the
  $\eta$ correction to be applied for the X an Y coordinate,
  respectively. It defaults to a polynomial of fifth ordern, i.e.
  $$[0] + [1]*x + [2]*x^2 + [3]*x^3 + [4]*x^4 + [5]*x^5$$
\item
  \parameter{eta_constants_x_<detector>} /
  \parameter{eta_constants_y_<detector>}: Vector of
  correction factors, representing the parameters of the above
  correction function, in X and Y coordinates, respectively. Defaults to
  an empty vector, i.e.~by default no correction is applied. The
  \parameter{<detector>} part of the variable has to be
  replaced with the respective unique name of the detector given in the
  setup file.
\end{itemize}

\paragraph{Plots produced}

No plots are produced. The result of the
$\eta$-correction is best observed in residual distributions of the
respective detector.

\paragraph{Usage}

\begin{minted}[frame=single,framesep=3pt,breaklines=true,tabsize=2,linenos]{ini}
[EtaCorrection]
eta_formula_x = [0] + [1]*x + [2]*x^2 + [3]*x^3 + [4]*x^4 + [5]*x^5
eta_constants_x_dut = 0.025 0.038  6.71 -323.58  5950.3 -34437.5
\end{minted}

\hypertarget{eventloaderatlaspix}{%
\subsection{EventLoaderATLASpix}\label{eventloaderatlaspix}}

\modulehead{Simon Spannagel (\href{mailto:simon.spannagel@cern.ch}{\nolinkurl{simon.spannagel@cern.ch}}), \newline Tomas Vanat (\href{mailto:tomas.vanat@cern.ch}{\nolinkurl{tomas.vanat@cern.ch}}), \newline Jens Kroeger (\href{mailto:jens.kroeger@cern.ch}{\nolinkurl{jens.kroeger@cern.ch}})}{DETECTOR}{ATLASpix}{Functional}

\paragraph{Description}

This module reads in data for the ATLASpix device from an input file
created by the Caribou readout system. It supports the binary output format,~i.e. \parameter{output  = 'binary'} in the configuration file for Caribou.

It requires either another event loader of another detector
type before, which defines the event start and end times (variables
\parameter{eventStart} and
\parameter{eventEnd} on the clipboard) or an instance of
the \texttt{Metronome} module which provides this information.

The module opens and reads one data file named
\file{data.bin} in the specified input directory and for each hit with a timestamp between \parameter{eventStart} and \parameter{eventEnd} it stores the detectorID, row, column, timestamp, and ToT on the clipboard.
Since a calibration is not yet implemented, the pixel charge is set to the pixel ToT.

\paragraph{Parameters}

\begin{itemize}
\tightlist
\item
  \parameter{input_directory}: Path to the directory
  containing the \file{data.bin} file. This path
  should lead to the directory above the ALTASpix directory, as this
  string is added to the input directory in the module.
\item
  \parameter{clock_cycle}: Period of the clock used to
  count the trigger timestamps in, defaults to
  \parameter{6.25ns}.
\item
  \parameter{clkdivend2}: Value of clkdivend2 register in
  ATLASPix specifying the speed of TS2 counter. Default is
  \parameter{0}.
\item
  \parameter{high_tot_cut}: ``high ToT'' histograms are
  filled if pixel ToT is larger than this cut. Default is
  \parameter{40}.
\item
  \parameter{buffer_depth}: Depth of buffer in which
  pixel hits are timesorted before being added to an event. If set to
  \parameter{1}, no timesorting is done.
  Default is \parameter{1000}.
\end{itemize}

\paragraph{Plots produced}

For the DUT, the following plots are produced:

\begin{itemize}
\item
    2D histograms:
    \begin{itemize}
    \tightlist
    \item
      Regular and ToT-weighted hit maps
    \item
      2D map of the difference of the event start and the pixel timestamp over time
    \item
      2D map of the difference of the trigger time and the pixel timestamp over time
    \end{itemize}

\item 1D histograms:
    \begin{itemize}
    \tightlist
    \item
      Various histograms of the pixel information: ToT, charge, ToA, TS1, TS2, TS1-bits, TS2-bits
    \item
      Histograms of the pixel multiplicity, triggers per event, and pixels over time
    \item
      Various timing histograms
    \end{itemize}
\end{itemize}

\paragraph{Usage}

\begin{minted}[frame=single,framesep=3pt,breaklines=true,tabsize=2,linenos]{ini}
[ATLASpixEventLoader]
input_directory = /user/data/directory
clock_cycle = 8ns
clkdivend2 = 7
buffer_depth = 100
\end{minted}

\hypertarget{eventloaderclicpix}{%
\subsection{EventLoaderCLICpix}\label{eventloaderclicpix}}

\modulehead{Daniel Hynds (\href{mailto:daniel.hynds@cern.ch}{\nolinkurl{daniel.hynds@cern.ch}})}{DETECTOR}{CLICpix}{Functional}

\paragraph{Description}

This module reads in data for a CLICpix device from an input file.

The module opens and reads one data file in the specified input
directory with the ending \file{.dat}. For each hit
it stores the detectorID, row, column, and ToT. The shutter rise and
fall time information are used to define the current event on the clipboard.

\paragraph{Parameters}

\begin{itemize}
\tightlist
\item
  \parameter{input_directory}: Path to the directory
  containing the \file{.dat} file.
\end{itemize}

\paragraph{Plots produced}

For each detector, the following plots are produced:
\begin{itemize}
\tightlist
\item
  2D hit map
\item
  Histogram of the pixel ToT, multiplicity, and shutter length
\end{itemize}

\paragraph{Usage}

\begin{minted}[frame=single,framesep=3pt,breaklines=true,tabsize=2,linenos]{ini}
[CLICpixEventLoader]
input_directory = /user/data/directory/CLICpix/
\end{minted}

\hypertarget{eventloaderclicpix2}{%
\subsection{EventLoaderCLICpix2}\label{eventloaderclicpix2}}

\modulehead{Daniel Hynds (\href{mailto:daniel.hynds@cern.ch}{\nolinkurl{daniel.hynds@cern.ch}}), \newline Simon Spannagel
(\href{mailto:simon.spannagel@cern.ch}{\nolinkurl{simon.spannagel@cern.ch}}), \newline Morag Williams
(\href{mailto:morag.williams@cern.ch}{\nolinkurl{morag.williams@cern.ch}})}{DETECTOR}{CLICpix2}{Functional}

\paragraph{Description}

This module reads in data for a CLICpix2 device from an input file. This module always attempts to define the current event on the clipboard using the begin and end of the shutter read from data.
Thus, this module can only be used as the first event loader module in the reconstruction chain.

The module opens and reads one data file in the specified input
directory. The input directory is searched for a data file with the file
extension \file{.raw} and a pixel matrix
configuration file required for decoding with the file extension
\file{.cfg} and a name starting with
\parameter{matrix}. The data is decoded using the CLICpix2
data decoder shipped with the Peary DAQ framework. For each pixel hit,
the detectorID, the pixel's column and row address as well as ToT and
ToA values are stored, depending on their availability from data. If no ToA information is available, the pixel timestamp is set to the center of the frame.
The shutter rise and fall time information are used to set the current time
and event length as described above.

\paragraph{Parameters}

\begin{itemize}
\tightlist
\item
  \parameter{input_directory}: Path to the directory
  containing the \dir{.csv} file. This path should
  lead to the directory above the CLICpix directory, as this string is
  added onto the input directory in the module.
\item
  \parameter{discard_zero_tot}: Discard all pixel hits
  with a ToT value of zero. Defaults to \parameter{false}.
\end{itemize}

\paragraph{Plots produced}

For each detector, the following plots are produced:
\begin{itemize}
\item 2D histograms:
    \begin{itemize}
    \tightlist
    \item
      Hit map
    \item
      Maps of masked pixels
    \item
      Maps of pixel ToT vs.~ToA
    \end{itemize}

\item 1D histograms:
    \begin{itemize}
    \tightlist
    \item
      Histograms of the pixel ToT, ToA, and particle count (if values are available)
    \item
      Histogram of the pixel multiplicity
    \end{itemize}
\end{itemize}

\paragraph{Usage}

\begin{minted}[frame=single,framesep=3pt,breaklines=true,tabsize=2,linenos]{ini}
[CLICpix2EventLoader]
input_directory = /user/data/directory
\end{minted}

\hypertarget{eventloadereudaq}{%
\subsection{EventLoaderEUDAQ}\label{eventloadereudaq}}

\moduleheadGlobal{Simon Spannagel
(\href{mailto:simon.spannagel@cern.ch}{\nolinkurl{simon.spannagel@cern.ch}})}{Functional}

\paragraph{Description}

This module allows data recorded by EUDAQ and stored in the EUDAQ-native
raw format to be read into Corryvreckan. The EUDAQ decoder plugins are
used to transform the data into the
\parameter{StandardPlane} event type before storing the
individual pixel objects on the Corryvreckan
clipboard.

The detector IDs are taken from the plane name and IDs, two possible
naming options for Corryvreckan are available: When setting
\parameter{long_detector_id = true}, the name of the
sensor plane and the ID are used in the form
\parameter{<name>_<ID>}, while only the ID is used
otherwise as \parameter{plane<ID>}. Only detectors listed in the Corryvreckan geometry are decoded and stored, data from other detectors available in the same EUDAQ event are
ignored.

\paragraph{Requirements}

This module requires an installation of
\href{https://github.com/eudaq/eudaq}{EUDAQ 1.x}. The installation path
should be set as environment variable via

\begin{minted}[frame=single,framesep=3pt,breaklines=true,tabsize=2,linenos]{ini}
export EUDAQPATH=/path/to/eudaq
\end{minted}

for CMake to find the library link against and headers to include.

\paragraph{Parameters}

\begin{itemize}
\tightlist
\item
  \parameter{file_name}: File name of the EUDAQ raw data
  file. This parameter is mandatory.
\item
  \parameter{long_detector_id}: Boolean switch to
  configure using the long or short detector ID in Corryvreckan,
  defaults to \parameter{true}.
\end{itemize}

\paragraph{Plots produced}

No plots are produced.

\paragraph{Usage}

\begin{minted}[frame=single,framesep=3pt,breaklines=true,tabsize=2,linenos]{ini}
[EUDAQEventLoader]
file_name = "rawdata/eudaq/run020808.raw"
long_detector_id = true
\end{minted}

\hypertarget{eventloadereudaq2}{%
\subsection{EventLoaderEUDAQ2}\label{eventloadereudaq2}}

\modulehead{Jens Kroeger
(\href{mailto:jens.kroeger@cern.ch}{\nolinkurl{jens.kroeger@cern.ch}}), \newline Simon Spannagel (\href{mailto:simon.spannagel@cern.ch}{\nolinkurl{simon.spannagel@cern.ch}})}{DETECTOR}{all}{Functional}

\paragraph{Description}

This module allows data recorded by EUDAQ2 and stored in a EUDAQ2 binary
file as raw detector data to be read into \corry. For each
detector type, the corresponding converter module in EUDAQ2 is used to
transform the data into the \parameter{StandardPlane}
event type before storing the individual pixel
objects on the \corry clipboard.

The detectors need to be named according to the following scheme:
\parameter{<detector_type>_<plane_number>} where
\parameter{detector_type} is the type specified in the
detectors file and \parameter{<plane_number>} is an
iterative number over the planes of the same type.

If the data of different detectors is stored in separate files, the
parameters \parameter{name} or
\parameter{type} can be used as shown in the usage example
below. It should be noted that the order of the detectors is crucial.
The first detector that appears in the configuration defines the event
window to which the hits of all other detectors are compared. In the
example below this is the CLICpix2.

If the data of multiple detectors is stored in the same file as
sub-events, it must be ensured that the event defining the time frame is
processed first. This is achieved by instantiating two event loaders in
the desired order and providing them with the same input data file. The
individual (sub-) events are compared against the detector type.

For each event, the algorithm checks for an event on the clipboard. If
none is available, the current event defines the event on the clipboard.
Otherwise, it is checked whether or not the current event lies within
the clipboard event. If yes, the corresponding pixels are added to the
clipboard for this event. If earlier, the next event is read until a
matching event is found. If later, the pointer to this event is kept and
it continues with the next detector.

If no detector is capable of defining events, the
\texttt{[Metronome]} module needs to be used.

Tags stored in the EUDAQ2 event header are read, a conversion to a
double value is attempted and, if successful, a profile with the value
over the number of events in the respective run is automatically
allocated and filled. This feature can e.g.~be used to log temperatures
of the devices during data taking, simply storing the temperature as
event tags.

\paragraph{Requirements}

This module requires an installation of
\href{https://eudaq.github.io/}{EUDAQ2}. The installation path needs to
be set to

\begin{minted}[frame=single,framesep=3pt,breaklines=true,tabsize=2,linenos]{ini}
export EUDAQ2PATH=/path/to/eudaq2
\end{minted}

when running CMake to find the library to link against and headers to
include.

It is recommended to only build the necessary libraries of EUDAQ2 to
avoid linking against unnecessary third-party libraries such as Qt5.
This can be achieved e.g.~by using the following CMake configuration for
EUDAQ2:

\begin{minted}[frame=single,framesep=3pt,breaklines=true,tabsize=2,linenos]{ini}
cmake -DEUDAQ_BUILD_EXECUTABLE=OFF -DEUDAQ_BUILD_GUI=OFF ..
\end{minted}

It should be noted that individual EUDAQ2 modules should be enabled for
the build, depending on the data to be processed with \corry. For instance, to
allow decoding of Caribou data, the respective EUDAQ2 module has to
be built using

\begin{minted}[frame=single,framesep=3pt,breaklines=true,tabsize=2,linenos]{ini}
cmake -DUSER_CARIBOU_BUILD=ON ..
\end{minted}

\hypertarget{contract-between-eudaq-decoder-and-eventloader}{%
\paragraph{Contract between EUDAQ Decoder and
Event Loader}\label{contract-between-eudaq-decoder-and-eventloader}}

The decoder guarantees to
\begin{itemize}
\item return \parameter{true} only
when there is a fully decoded event available and
\parameter{false} in all other cases.
\item not return any event before a possible T0 signal in the data. This is a signal that indicates the clock reset at the beginning of the run. It can be a particular data word or the observation of the pixel timestamp jumping back to zero, depending on data format of each the detector.
\item return the smallest
possible granularity of data in time either as event or as sub-events
within one event.
\item always return valid event time stamps. If the device
does not have timestamps, it should return zero for the beginning of the
event and have a valid trigger number set.
\item provide the detector type
via the \parameter{GetDetectorType()} function in the
decoded StandardEvent.
\end{itemize}

\hypertarget{configuring-eudaq2-event-converters}{%
\paragraph{Configuring EUDAQ2 Event
Converters}\label{configuring-eudaq2-event-converters}}

Some data formats depend on external configuration parameters for
interpretation. The \texttt{[EventLoaderEUDAQ2]} takes all key-value pairs
available in the configuration and forwards them to the appropriate
StandardEvent converter on the EUDAQ side. It should be kept in mind
that the resulting configuration strings are parsed by EUDAQ2, not by
\corry, and that therefore the functionality is reduced. For
example, it does not interpret \parameter{true} or
\parameter{false} alphabetic value of a Boolean variable
but will return false in both cases. Thus \parameter{key = 0} or \parameter{key = 1}
have to be used in these cases. Also, more complex constructs such as
arrays or matrices read by the \corry configuration are simply
interpreted as strings.

\paragraph{Parameters}

\begin{itemize}
\tightlist
\item
  \parameter{file_name}: File name of the EUDAQ2 raw data
  file. This parameter is mandatory.
\item
  \parameter{inclusive}: Boolean parameter to select
  whether new data should be compared to the existing \corry event
  in inclusive or exclusive mode. The inclusive interpretation will
  allow new data to be added to the event as soon as there is some
  overlap between the data time frame and the existing event, i.e.~as
  soon as the end of the time frame is later than the event start or as
  soon as the time frame start is before the event end. In the exclusive
  mode, the frame will only be added to the existing event if its start
  and end are both within the defined event.
\item
  \parameter{skip_time}: Time that is skipped at the
  start of a run. All hits with earlier timestamps are discarded. Default is \parameter{0ms}.
\item
  \parameter{get_time_residuals}: Boolean to choose if
  time residual plots are created. Default value is
  \parameter{false}.
\item
  \parameter{get_tag_vectors}: Boolean to enable
  creation of EUDAQ2 event tag histograms. Default value is
  \parameter{false}.
\item
  \parameter{ignore_bore}: Boolean to ignore
  the Begin-of-Run event (BORE) from EUDAQ2. Default value is
  \parameter{true}.
\item
  \parameter{adjust_event_times}: Matrix that allows the
  user to shift the event start/end of all different types of EUDAQ
  events before comparison to any other \corry data. The first
  entry of each row specifies the data type, the second is the offset
  which is added to the event start and the third entry is the offset
  added to the event end. A usage example is shown below, double
  brackets are required if only one entry is provided.
\item
  \parameter{buffer_depth}: Depth of buffer in which
  EUDAQ2 \parameter{StandardEvents} are timesorted. This
  algorithm only works for \parameter{StandardEvents} with
  well-defined timestamps. Setting it to \parameter{0}
  disables timesorting. Default is \parameter{0}.
\item
  \parameter{shift_triggers}: Shift the trigger ID up or
  down when assigning it to the \corry event. This allows to
  correct trigger ID offsets between different devices such as the TLU
  and MIMOSA26.
\end{itemize}

\paragraph{Plots produced}

For all detectors, the following plots are produced:
\begin{itemize}
\item 2D histograms:
    \begin{itemize}
    \tightlist
    \item
      Hit map
    \item
      2D map of the pixel time minus event begin residual over time
    \item
      2D map of the difference of the event start and the pixel timestamp over time
    \item
      2D map of the difference of the trigger time and the pixel timestamp over time
    \item
      profiles of the event tag data
    \end{itemize}

\item 1D histograms:
    \begin{itemize}
    \tightlist
    \item
      Histograms of the pixel hit times, raw values, multiplicities, and pixels per event
    \item
      Histograms of the eudaq/clipboard event start/end, and durations
    \item
      Histogram of the pixel time minus event begin residual
    \end{itemize}
\end{itemize}

\paragraph{Usage}

\begin{minted}[frame=single,framesep=3pt,breaklines=true,tabsize=2,linenos]{ini}
[EventLoaderEUDAQ2]
name = "CLICpix2_0"
file_name = /path/to/data/examplerun_clicpix2.raw

[EventLoaderEUDAQ2]
type = "MIMOSA26"
file_name = /path/to/data/examplerun_telescope.raw
adjust_event_times = [["TluRawDataEvent", -115us, +230us]]
buffer_depth = 1000
\end{minted}

\hypertarget{eventloadermupixtelescope}{%
\subsection{EventLoaderMuPixTelescope}\label{eventloadermupixtelescope}}

\moduleheadGlobal{Lennart Huth (\href{mailto:lennart.huth@desy.de}{\nolinkurl{lennart.huth@desy.de}})}{Work in progress}

\paragraph{Description}

This module reads in and converts data taken with the MuPix telescope.
It requires one input file which contains the data of all planes in the telescope.
The data contains time-ordered readout frames.
For frame, the module loops over all hits and stores the row, column, timestamp, and ToT are stored on the clipboard.

It cannot be combined with other event loaders and does not require a \module{Metronome}.

\paragraph{Parameters}

\begin{itemize}
\tightlist
\item
  \parameter{input_directory}: Defines the input file. No
  default.
\item
  \parameter{Run}: not in use. Defaults to
  \parameter{-1}.
\item
  \parameter{is_sorted}: Defines if data recorded is on
  FPGA timestamp sorted. Defaults to \parameter{false}.
\item
  \parameter{ts2_is_gray}: Defines if the timestamp is
  gray encoded or not. Defaults to \parameter{false}.
\end{itemize}

\paragraph{Plots produced}

For all detectors, the following plots are produced:
\begin{itemize}
\tightlist
\item
  2D histogram of pixel hit positions
\item
  1D histogram of the pixel timestamps
\end{itemize}

\paragraph{Usage}

\begin{minted}[frame=single,framesep=3pt,breaklines=true,tabsize=2,linenos]{ini}
[EventLoaderMuPixTelescope]
input_directory = "/path/to/file"
Run = -1
is_sorted = false
ts2_is_gray = false
\end{minted}

\hypertarget{eventloadertimepix1}{%
\subsection{EventLoaderTimepix1}\label{eventloadertimepix1}}

\moduleheadGlobal{Daniel Hynds (\href{mailto:daniel.hynds@cern.ch}{\nolinkurl{daniel.hynds@cern.ch}})}{Functional}

\paragraph{Description}

This module loads raw data from Timepix1 devices~\cite{timepix} and adds it to the
clipboard as pixel objects. The input file
must have extension \file{.txt}, and these files are
sorted into time order using the file titles.

\paragraph{Parameters}

\begin{itemize}
\tightlist
\item
  \parameter{input_directory}: Path of the directory
  above the data files.
\end{itemize}

\paragraph{Plots produced}

No plots are produced.

\paragraph{Usage}

\begin{minted}[frame=single,framesep=3pt,breaklines=true,tabsize=2,linenos]{ini}
[Timepix1EventLoader]
input_directory = "path/to/directory"
\end{minted}

\hypertarget{eventloadertimepix3}{%
\subsection{EventLoaderTimepix3}\label{eventloadertimepix3}}

\modulehead{Daniel Hynds (\href{mailto:daniel.hynds@cern.ch}{\nolinkurl{daniel.hynds@cern.ch}}), \newline Simon Spannagel (\href{mailto:simon.spannagel@cern.ch}{\nolinkurl{simon.spannagel@cern.ch}})}{DETECTOR}{Timepix3}{Functional}

\paragraph{Description}

This module loads raw data from a Timepix3~\cite{timepix3} device and adds it to the
clipboard. The input files must have the extension
\parameter{.dat} and are sorted into time order via the
data file serial numbers. This code also identifies
\parameter{trimdac} files and applies this mask to the
pixels.

The data can be split into events using an event length in time, or
using a maximum number of hits on a detector plane.
SpidrSignal and
Pixel objects are loaded to the clipboard for
each detector.

The hit timestamps are derived from the 40 MHz ToA counter and the fast
on-pixel oscillator, which is measuring the precise hit arrival phase
within to the global 40 MHz clock. In Timepix3, the phase of the 40 MHz
clock can be shifted from one double column to the next by 22.5 degree
by the clock generator in order to minimize the instant digital power
supply due to the pixel matrix clock tree. This mode is used in the
CLICdp telescope, and thus, the column-to-column phase shift is taken
into account when calculating the hit arrival times. See also the
Timepix3 chip manual version 1.9, section 3.2.1 and/or
\autocite{timepix3-talk}, slides 25 and 48.

This module requires either another event loader of another detector
type before which defines the event start and end times (Event object on
the clipboard) or an instance of the \texttt{[Metronome]} module which provides
this information.
The frame-based readout mode of the Timepix3 is not supported.

The calibration is performed as described in~\cite{Pitters_2019, cds-timepix3-calibration} and requires a Timepix3 plane to be set as \parameter{role = DUT}.

\paragraph{Parameters}

\begin{itemize}
\tightlist
\item
  \parameter{input_directory}: Path to the directory
  above the data directory for each device. The device name is added to
  the path during the module.
\item
  \parameter{trigger_latency}:
\item
  \parameter{calibration_path}: Path to the calibration
  directory. If this parameter is set, a calibration will be applied to
  the Timepix3 plane set as \parameter{role = DUT}. The assumed folder structure is
  \dir{[calibration_path]/[detector name]/cal_thr_[threshold]_ik_[ikrum dac]/}. In this directory the two files \file{[detector name]_cal_[tot/toa].txt} need to exist.\\
  For the ToT calibration, the file format needs to be\\
  \parameter{col | row | row | a (ADC/mV) | b (ADC) | c (ADC*mV) | t (mV) | chi2/ndf}.\\
  For the ToA calibration, it needs to be\\
  \parameter{column | row | c (ns*mV) | t (mV) | d (ns) | chi2/ndf}.
\item
  \parameter{threshold}: String defining the
  \parameter{[threshold]} DAC value for loading the
  appropriate calibration file.
\end{itemize}

\paragraph{Plots produced}

For all detectors, the following plots are produced:
\begin{itemize}
\tightlist
\item
  2D map of pixel positions
\item
  Histogram with pixel ToT before and after calibration
\item
  Map for each calibration parameter if calibration is used
\end{itemize}

\paragraph{Usage}

\begin{minted}[frame=single,framesep=3pt,breaklines=true,tabsize=2,linenos]{ini}
[Timepix3EventLoader]
input_directory = "path/to/directory"
calibration_path = "path/to/calibration"
threshold = 1148
number_of_pixelhits = 0
\end{minted}

\hypertarget{filereader}{%
\subsection{FileReader}\label{filereader}}

\moduleheadGlobal{Simon Spannagel (\href{mailto:simon.spannagel@cern.ch}{\nolinkurl{simon.spannagel@cern.ch}})}{Functional}

\paragraph{Description}

Converts all object data stored in the ROOT data file produced by the
\texttt{[FileWriter]} module back into the clipboard (see the description of
\texttt{[FileWriter]} for more information about the format). Reads all trees
defined in the data file that contain \corry objects. Places all
objects read from the tree onto the clipboard storage.

If the requested number of events for the run is less than the number of
events the data file contains, all additional events in the file are
skipped. If more events than available are requested, a warning is
displayed and the other events of the run are skipped.

Currently it is not yet possible to exclude objects from being read. In
case not all objects should be converted to messages, these objects need
to be removed from the file before the anlysis is started.

\paragraph{Parameters}

\begin{itemize}
\tightlist
\item
  \parameter{file_name} : Location of the ROOT file
  containing the trees with the object data.
\item
  \parameter{include} : Array of object names (without
  \parameter{corryvreckan::} prefix) to be read from the
  ROOT trees, all other object names are ignored (cannot be used
  simulateneously with the \parameter{exclude} parameter).
\item
  \parameter{exclude}: Array of object names (without
  \parameter{corryvreckan::} prefix) not to be read from
  the ROOT trees (cannot be used simultaneously with the \parameter{include}
  parameter).
\end{itemize}

\paragraph{Plots produced}

No plots are produced.

\paragraph{Usage}

This module should be placed at the beginning of the main configuration.
An example to read only Cluster and Pixel objects from the file
\file{data.root} is:

\begin{minted}[frame=single,framesep=3pt,breaklines=true,tabsize=2,linenos]{ini}
[FileReader]
file_name = "data.root"
include = "Cluster", "Pixel"
\end{minted}

\hypertarget{filewriter}{%
\subsection{FileWriter}\label{filewriter}}

\moduleheadGlobal{Simon Spannagel (\href{mailto:simon.spannagel@cern.ch}{\nolinkurl{simon.spannagel@cern.ch}})}{Functional}

\paragraph{Description}

Reads all objects from the clipboard into a vector of base class object
pointers. The first time a new type of object is received, a new tree is
created bearing the class name of this object. For every detector, a new
branch is created within this tree. A leaf is automatically created for
every member of the object. The vector of objects is then written to the
file for every event it is dispatched, saving an empty vector if an
event does not include the specific object.

\paragraph{Parameters}

\begin{itemize}
\tightlist
\item
  \parameter{file_name} : Name of the data file to
  create, relative to the output directory of the framework. The file
  extension \parameter{.root} will be appended if not
  present.
\item
  \parameter{include} : Array of object names (without
  \parameter{corryvreckan::} prefix) to write to the ROOT
  trees, all other object names are ignored (cannot be used together
  simultaneously with the \parameter{exclude} parameter).
\item
  \parameter{exclude}: Array of object names (without
  \parameter{corryvreckan::} prefix) that are not written
  to the ROOT trees (cannot be used together simultaneously with the
  \parameter{include} parameter).
\end{itemize}

\paragraph{Plots produced}

No plots are produced.

\paragraph{Usage}

To create the default file (with the name \file{data.root}) containing
trees for all objects except for Cluster, the following configuration
can be placed at the end of the main configuration:

\begin{minted}[frame=single,framesep=3pt,breaklines=true,tabsize=2,linenos]{ini}
[FileWriter]
file_name = "data.root"
exclude = "Cluster"
\end{minted}

\hypertarget{improvereferencetimestamp}{%
\subsection{ImproveReferenceTimestamp}\label{improvereferencetimestamp}}

\moduleheadGlobal{Florian Pitters (\href{mailto:florian.pitters@cern.ch}{\nolinkurl{florian.pitters@cern.ch}})}{Work in progress}

\paragraph{Description}

Replaces the existing reference timestamp (earliest hit on reference
plane) by either the trigger timestamp (method 0) or the average track
timestamp (method 1). For method 0 to work, a trigger timestamp has to
be saved as SPIDR signal during data taking.

\paragraph{Parameters}

\begin{itemize}
\tightlist
\item
  \parameter{improvement_method}: Determines which method
  to use. Trigger timestamp is 0, average track timestamp is 1. Default
  value is \parameter{1}.
\item
  \parameter{signal_source}: Determines which detector
  plane carries the trigger signals. Only relevant for method 0. Default
  value is \parameter{"W0013_G02"}.
\item
  \parameter{trigger_latency}: Adds a latency to the
  trigger timestamp to shift time histograms. Default
  value is \parameter{0}.
\end{itemize}

\paragraph{Plots produced}

No plots are produced.

\paragraph{Usage}

\begin{minted}[frame=single,framesep=3pt,breaklines=true,tabsize=2,linenos]{ini}
[ImproveReferenceTimestamp]
improvement_method = 0
signal_source = "W0013_G02"
trigger_latency = 20ns
\end{minted}

\hypertarget{maskcreator}{%
\subsection{MaskCreator}\label{maskcreator}}

\modulehead{Simon Spannagel (\href{mailto:simon.spannagel@cern.ch}{\nolinkurl{simon.spannagel@cern.ch}})}{DETECTOR}{all}{Functional}

\paragraph{Description}

This module reads in pixel objects for each
device from the clipboard, and masks pixels considered noisy.

Currently, two methods are available. The
\parameter{localdensity} noise estimation method is taken
from the Proteus framework~\cite{proteus-repo}. It uses a local estimate
of the expected hit rate to find pixels that are a certain number of
standard deviations away from this estimate. The second method,
\parameter{frequency}, is a simple cut on a global pixel
firing frequency which masks pixels with a hit rate larger than
\parameter{frequency_cut} times the mean global hit rate.

The module appends the pixels to be masked to the mask files provided in the geometry file for each device.
If no mask file is specified there, a new file \file{mask_<detector_name>.txt} is created in the globally configured output directory. 
Already existing masked pixels are maintained. 
No masks are applied in this module as this is done by the respective event loader modules when reading input data.

\paragraph{Parameters}

\begin{itemize}
\tightlist
\item
  \parameter{method}: Select the method to evaluate noisy
  pixels. Can be either \parameter{localdensity} or
  \parameter{frequency}, where the latter is chosen by
  default.
\item
  \parameter{frequency_cut}: Frequency threshold to
  declare a pixel as noisy, defaults to 50. This means, if a pixel
  exhibits 50 times more hits than the average pixel on the sensor, it
  is considered noisy and is masked. Only used in
  \parameter{frequency} mode.
\item
  \parameter{bins_occupancy}: Number of bins for
  occupancy distribution histograms, defaults to 128.
\item
  \parameter{density_bandwidth}: Bandwidth for local
  density estimator, defaults to \parameter{2} and is only
  used in \parameter{localdensity} mode.
\item
  \parameter{sigma_above_avg_max}: Cut for noisy
  pixels, number of standard deviations above average, defaults to \parameter{5}.
  Only used in \parameter{localdensity} mode.
\item
  \parameter{rate_max}: Maximum rate, defaults to
  \parameter{1}. Only used in
  \parameter{localdensity} mode.
\end{itemize}

\paragraph{Plots produced}

For each detector the following plots are produced:
\begin{itemize}
\item 2D map of masked pixels
\item 2D histogram of occupancy
\end{itemize}

\paragraph{Usage}

\begin{minted}[frame=single,framesep=3pt,breaklines=true,tabsize=2,linenos]{ini}
[MaskCreator]
frequency_cut = 10
\end{minted}

\hypertarget{maskcreatortimepix3}{%
\subsection{MaskCreatorTimepix3}\label{maskcreatortimepix3}}

\modulehead{Daniel Hynds
(\href{mailto:daniel.hynds@cern.ch}{\nolinkurl{daniel.hynds@cern.ch}})}{DETECTOR}{Timepix3}{Functional}

\paragraph{Description}

This module reads in pixel objects for each
device from the clipboard, masks all pixels with a hit rate larger than
10 times the mean hit rate, and updates the trimdac chip configuration file for the device
with these masked pixels.
If this file does not exist yet, a new file named \file{<detectorID>_trimdac_masked.txt} will be created.

\paragraph{Parameters}

No parameters are used from the configuration file.

\paragraph{Plots produced}

No plots are produced.

\paragraph{Usage}

\begin{minted}[frame=single,framesep=3pt,breaklines=true,tabsize=2,linenos]{ini}
[Timepix3MaskCreator]
\end{minted}

\hypertarget{metronome}{%
\subsection{Metronome}\label{metronome}}

\moduleheadGlobal{Simon Spannagel (\href{mailto:simon.spannagel@cern.ch}{\nolinkurl{simon.spannagel@cern.ch}})}{Functional}

\paragraph{Description}

The \texttt{[Metronome]} module can be used to slice data
without strict event structure in arbitrarily long time frames, which
serve as event definition for subsequent modules. A new Event object is
created and stored on the clipboard, with the begin and end of the frame
calculated from the \parameter{event_length} parameter.

The module also provides the option to add trigger IDs to the generated
event by specifying the number of triggers to be set per event via the
\parameter{triggers} parameter. Trigger IDs are consecutive numbers,
starting at zero. With a setting of
\parameter{triggers = 2}, the first event would receive
trigger IDs 0 and 1, the subsequent event 2 and 3 and so forth.
A more detailed description is provided in Section~\ref{sec:metronome}.

\paragraph{Parameters}

\begin{itemize}
\tightlist
\item
  \parameter{event_length}: Length of the event to be
  defined in physical units (not clock cycles of a specific device).
  Default value is \parameter{10us}.
\item
  \parameter{skip_time}: Time to skip at the begin of the
  run before processing the first event. Defaults to
  \parameter{0us}.
\item
  \parameter{triggers}: Number of triggers to generate and
  add to each event. All trigger timestamps are set to the center of the
  configured metronome time frame. Defaults to 0, i.e.~no triggers
  are added.
\end{itemize}

\paragraph{Plots produced}

No plots are produced.

\paragraph{Usage}

\begin{minted}[frame=single,framesep=3pt,breaklines=true,tabsize=2,linenos]{ini}
[Metronome]
event_length = 500ns
\end{minted}

\hypertarget{onlinemonitor}{%
\subsection{OnlineMonitor}\label{onlinemonitor}}

\moduleheadGlobal{Daniel Hynds (\href{mailto:daniel.hynds@cern.ch}{\nolinkurl{daniel.hynds@cern.ch}}), \newline Simon Spannagel (\href{mailto:simon.spannagel@cern.ch}{\nolinkurl{simon.spannagel@cern.ch}})}{Functional}

\paragraph{Description}

This module opens a GUI to monitor the progress of the reconstruction.
Since Linux allows concurrent (reading) file access, this can already be
started while a run is recorded to disk and thus serves as online
monitoring tool during data taking. A set of canvases is available to
display a variety of information ranging from hitmaps and basic
correlation plots to more advances results such as tracking quality and
track angles. The plots on each of the canvases contain real time data,
automatically updated every \parameter{update} events.

The displayed plots and their source can be configured via the framework
configuration file. Here, each canvas is configured via a matrix
containing the path of the plot and its plotting options in each row,
e.g.

\begin{minted}[frame=single,framesep=3pt,breaklines=true,tabsize=2,linenos]{ini}
DUTPlots =  [["EventLoaderEUDAQ2/%DUT%/hitmap", "colz"],
             ["EventLoaderEUDAQ2/%DUT%/hPixelRawValues", "log"]]
\end{minted}

The allowed plotting options comprise all drawing options offered by
ROOT. In addition, the \parameter{log} keyword is
supported, which switches the Y axis of the respective plot to a
logarithmic scale.

Several keywords can be used in the plot path, which are parsed and
interpreted by the OnlineMonitor module:

\begin{itemize}
\tightlist
\item
  \parameter{
  plot is added for each of the available detectors by replacing the
  keyword with the respective detector name.
\item
  \parameter{
  value of the \parameter{DUT} configuration key of the
  framework.
\item
  \parameter{
  the value of the \parameter{reference} configuration key
  of the framework.
\end{itemize}

The \texttt{corryvreckan} namespace is not required to be added to the plot
path.

\paragraph{Parameters}

\begin{itemize}
\item
  \parameter{update}: Number of events after which to
  update, defaults to \parameter{500}.
\item
  \parameter{canvas_title}: Title of the GUI window to be
  shown, defaults to
  \texttt{Corryvreckan Testbeam Monitor}. This
  parameter can be used to e.g.~display the current run number in the
  window title.
\item
  \parameter{ignore_aux}: With this boolean variable set,
  detectors with \parameter{auxiliary} roles are ignored
  and none of their histograms are added to the UI. Defaults to
  \parameter{true}.
\item
  \parameter{overview}: List of plots to be placed on the
  ``Overview'' canvas of the online monitor. The list of plots created
  in the default configuration is listed below.
\item
  \parameter{dut_plots}: List of plots to be placed on
  the ``DUTPlots'' canvas of the online monitor. By default, this canvas
  contains plots collected from the
  \texttt{EventLoaderEUDAQ2} as well as the
  \texttt{AnalysisDUT} modules for the each configured
  DUT. This canvas should be customized for the respective DUT.
\item
  \parameter{hitmaps}: List of plots to be placed on the
  ``HitMaps'' canvas of the online monitor. By default, this canvas
  displays \parameter{Correlations/
  for all detectors.
\item
  \parameter{tracking}: List of plots to be placed on the
  ``Tracking'' canvas of the online monitor. The list of plots created
  in the default configuration is listed below.
\item
  \parameter{residuals}: List of plots to be placed on the
  ``Residuals'' canvas of the online monitor. By default, this canvas
  displays \parameter{Tracking4D/
  for all detectors.
\item
  \parameter{correlation_x}: List of plots to be placed
  on the ``CorrelationX'' canvas of the online monitor. By default, this
  canvas displays
  \parameter{Correlations/
  all detectors.
\item
  \parameter{correlation_y}: List of plots to be placed
  on the ``CorrelationY'' canvas of the online monitor. By default, this
  canvas displays
  \parameter{Correlations/
  all detectors.
\item
  \parameter{correlation_x2d}: List of plots to be placed
  on the ``CorrelationX2D'' canvas of the online monitor. By default,
  this canvas displays
  \parameter{Correlations/
  for all detectors.
\item
  \parameter{correlation_y2d}: List of plots to be placed
  on the ``CorrelationY2D'' canvas of the online monitor. By default,
  this canvas displays
  \parameter{Correlations/
  for all detectors.
\item
  \parameter{charge_distributions}: List of plots to be
  placed on the ``ChargeDistributions'' canvas of the online monitor. By
  default, this canvas displays
  \parameter{Clustering4D/
  all detectors.
\item
  \parameter{event_times}: List of plots to be placed on
  the ``EventTimes'' canvas of the online monitor. By default, this
  canvas displays
  \parameter{Correlations/
  all detectors.
\end{itemize}

\paragraph{Plots produced}

The following plots are displayed by default (can be configured):

\begin{itemize}
\item Overview canvas:
    \begin{itemize}
    \tightlist
    \item
    Histogram of the cluster ToT of reference plane
    \item
    2D hit map of reference plane
    \item
    Histogram of the residual in X of reference plane
    \end{itemize}
\item Tracking canvas:
    \begin{itemize}
    \tightlist
    \item
    Histogram of the track $\chi^2$
    \item
    Histogram of the track angle in X
    \end{itemize}
\end{itemize}

\paragraph{Usage}

\begin{minted}[frame=single,framesep=3pt,breaklines=true,tabsize=2,linenos]{ini}
[OnlineMonitor]
update = 200
dut_plots = [["EventLoaderEUDAQ2/%DUT%/hitmap", "colz"],
             ["EventLoaderEUDAQ2/%DUT%/hPixelTimes"],
             ["EventLoaderEUDAQ2/%DUT%/hPixelRawValues"],
             ["EventLoaderEUDAQ2/%DUT%/pixelMultiplicity", "log"],
             ["AnalysisDUT/clusterChargeAssociated"],
             ["AnalysisDUT/associatedTracksVersusTime"]]
\end{minted}

\hypertarget{prealignment}{%
\subsection{Prealignment}\label{prealignment}}

\modulehead{Morag Williams (\href{mailto:morag.williams@cern.ch}{\nolinkurl{morag.williams@cern.ch}})}{DETECTOR}{all}{Functional}

\paragraph{Description}

This module performs a translational telescope plane prealignment. 
The rotational alignment is not changed.

This initial alignment along the X and Y axes is designed to be
performed before the \texttt{[Alignment]} module, which
carries out translational and rotational alignment of the planes. To not
include the DUT in this translational alignment, it will need to be
masked in the configuration file.

The required translational shifts in X and Y are calculated for each
detector as the mean of the 1D correlation histogram along the axis.
As described in Chapter~\ref{ch:howtoalign}, the spatial correlations in X and Y should not be forced to be centered around zero for the final alignment as they correspond to the \emph{physical displacement} of the detector plane in X and Y with respect to the reference plane.
However, for the prealignment this is a an acceptable estimation which works without any tracking.

\paragraph{Parameters}

\begin{itemize}
\tightlist
\item
  \parameter{damping_factor}: A factor to change the
  percentage of the calcuated shift applied to each detector. Default
  value is \parameter{1}.
\item
  \parameter{max_correlation_rms}: The maximum RMS of
  the 1D correlation histograms allowed for the shifts to be applied.
  This factor should be tuned for each run, and aims to combat the effect of
  flat distributions. Default value is \parameter{6mm}.
\item
  \parameter{time_cut_rel}: Number of standard deviations the \parameter{time_resolution} of the detector plane will be multiplied by. This value is then used as the maximum time difference between a cluster on the current detector and a cluster on the reference plane to be considered in the prealignment. Absolute and relative time cuts are mutually exclusive. Defaults to \parameter{3.0}.
\item
  \parameter{time_cut_abs}: Specifies an absolute value for the maximum time difference between a cluster on the current detector and a cluster on the reference plane to be considered in the prealignment. Absolute and relative time cuts are mutually exclusive. No default value.
\end{itemize}

\paragraph{Plots produced}

For each detector the following plots are produced:

\begin{itemize}
\tightlist
\item
  1D histograms of the correlations in X/Y (comparing to the reference plane)
\item
  2D histograms of the correlation plot for X/Y in local/global coordinates (comparing to the reference
  plane)
\end{itemize}

\paragraph{Usage}

\begin{minted}[frame=single,framesep=3pt,breaklines=true,tabsize=2,linenos]{ini}
[Prealignment]
log_level = INFO
max_correlation_rms = 6.0
damping_factor = 1.0
\end{minted}

\hypertarget{textwriter}{%
\subsection{TextWriter}\label{textwriter}}

\moduleheadGlobal{Paul Schuetze (\href{paul.schuetze@desy.de}{\nolinkurl{paul.schuetze@desy.de}})}{Functional}

\paragraph{Description}

This module writes objects to an ASCII text file. For this, the data
content of the selected objects available on the clipboard are printed
to the specified file.

Events are separated by an event header

\begin{verbatim}
=== <event number> ===
\end{verbatim}

and individual object types by the type marker:

\begin{verbatim}
--- <detector name> ---
\end{verbatim}

With \parameter{include} and
\parameter{exclude} certain object types can be selected
to be printed.

\paragraph{Parameters}

\begin{itemize}
\tightlist
\item
  \parameter{file_name} : Name of the data file to
  create, relative to the output directory of the framework. The file
  extension \parameter{.txt} will be appended if not
  present.
\item
  \parameter{include} : Array of object names to write to
  the ASCII text file, all other object names are ignored (cannot be
  used together simultaneously with the \parameter{exclude} parameter).
\item
  \parameter{exclude}: Array of object names that are not
  written to the ASCII text file (cannot be used together simultaneously
  with the \parameter{include} parameter).
\end{itemize}

\paragraph{Plots produced}

No plots are produced.

\paragraph{Usage}

\begin{minted}[frame=single,framesep=3pt,breaklines=true,tabsize=2,linenos]{ini}
[TextWriter]
file_name = "exampleFileName"
include = "Pixel"
\end{minted}

\hypertarget{tracking4d}{%
\subsection{Tracking4D}\label{tracking4d}}

\moduleheadGlobal{Simon Spannagel (\href{mailto:simon.spannagel@cern.ch}{\nolinkurl{simon.spannagel@cern.ch}}), \newline Daniel Hynds (\href{mailto:daniel.hynds@cern.ch}{\nolinkurl{daniel.hynds@cern.ch}})}{Functional}

\paragraph{Description}

This module performs a basic tracking method.

Clusters from the first plane in Z (named the seed plane) are related to
clusters close in time on the other detector planes using straight line
tracks. The DUT plane can be excluded from the track finding.

\paragraph{Parameters}

\begin{itemize}
\tightlist
\item
 \parameter{time_cut_rel}: Factor by which the \parameter{time_resolution} of each detector plane will be multiplied, either the \parameter{time_resolution} of the first plane in Z or the current telescope plane, whichever is largest. This calculated value is then used as the maximum time difference allowed between clusters and a track for association to the track. This allows the time cuts between different planes to be detector appropriate. By default, a relative time cut is applied. Absolute and relative time cuts are mutually exclusive. Defaults to \parameter{3.0}.
\item 
  \parameter{time_cut_abs}: Specifies an absolute value for the maximum time difference allowed between clusters and a track for association to the track. Absolute and relative time cuts are mutually exclusive. No default value.
\item 
  \parameter{spatial_cut_rel}: Factor by which the \parameter{spatial_resolution} in X and Y of each detector plane will be multiplied. These calculated value are defining an ellipse which is then used as the maximum distance in the XY plane allowed between clusters and a track for association to the track. This allows the spatial cuts between different planes to be detector appropriate. By default, a relative spatial cut is applied. Absolute and relative spatial cuts are mutually exclusive. Defaults to \parameter{3.0}.
\item
  \parameter{spatial_cut_abs}: Specifies a set of absolute value (X and Y) which defines an ellipse for the maximum spatial distance in the XY plane between clusters and a track for association to the track. Absolute and relative spatial cuts are mutually exclusive. No default value.
\item
  \parameter{min_hits_on_track}: Minimum number of
  associated clusters needed to create a track, equivalent to the
  minimum number of planes required for each track. Default value is
  \parameter{6}.
\item
  \parameter{exclude_dut}: Boolean to choose if the DUT
  plane is included in the track finding. Default value is
  \parameter{true}.
\item
  \parameter{require_detectors}: Names of detectors which
  are required to have a cluster on the track. If a track does not have
  a cluster from all detectors listed here, it is rejected. If empty, no
  detector is required. Default is empty.
\item
  \parameter{timestamp_from}: Defines the detector which
  provides the track timestamp. This detector also needs to be set as
  \parameter{required_detector}. If empty, the average
  timestamp of all clusters on the track will be used. Empty by default.
\end{itemize}

\paragraph{Plots produced}

The following plots are produced only once:
\begin{itemize}
\tightlist
\item
  Histograms of the track $\chi^2$ and track $\chi^2$/degrees of freedom
\item
  Histogram of the clusters per track, and tracks per event
\item
  Histograms of the track angle with respect to the X/Y-axis
\end{itemize}

For each detector, the following plots are produced:
\begin{itemize}
\tightlist
\item
  Histograms of the track residual in X/Y for various cluster sizes
\end{itemize}

\paragraph{Usage}

\begin{minted}[frame=single,framesep=3pt,breaklines=true,tabsize=2,linenos]{ini}
[Tracking4D]
min_hits_on_track = 4
spatial_cut_abs = 300um
timing_cut_abs = 200ns
exclude_dut = true
require_detectors = "ExampleDetector_0", "ExampleDetector_1"
\end{minted}

\hypertarget{trackingspatial}{%
\subsection{TrackingSpatial}\label{trackingspatial}}

\moduleheadGlobal{Daniel Hynds (\href{mailto:daniel.hynds@cern.ch}{\nolinkurl{daniel.hynds@cern.ch}})}{Functional}

\paragraph{Description}

This module performs track finding using only positional information (no
timing information). It is based on a linear extrapolation along the
Z-axis, followed by a nearest neighbour search.

\paragraph{Parameters}

\begin{itemize}
\tightlist
\item 
  \parameter{spatial_cut_rel}: Factor by which the \parameter{spatial_resolution} in X and Y of each detector plane will be multiplied. These calculated value are defining an ellipse which is then used as the maximum distance in the XY plane allowed between clusters and a track for association to the track. This allows the spatial cuts between different planes to be detector appropriate. By default, a relative spatial cut is applied. Absolute and relative spatial cuts are mutually exclusive. Defaults to \parameter{3.0}.
\item 
  \parameter{spatial_cut_abs}: Specifies a set of absolute value (X and Y) which defines an ellipse for the maximum spatial distance in the XY plane between clusters and a track for association to the track. Absolute and relative spatial cuts are mutually exclusive. No default value.
\item
  \parameter{min_hits_on_track}: The minimum number of
  planes with clusters associated to a track for it to be stored.
  Default value is \parameter{6}.
\item
  \parameter{exclude_dut}: Boolean to set if the DUT
  should be included in the track fitting. Default value is
  \parameter{true}.
\end{itemize}

\paragraph{Plots produced}

The following plots are produced only once:
\begin{itemize}
\tightlist
\item
  Histograms of the track $\chi^2$ and track $\chi^2$/degrees of freedom
\item
  Histogram of the clusters per track, and tracks per event
\item
  Histograms of the track angle with respect to the X/Y-axis
\end{itemize}

For each detector, the following plots are produced:
\begin{itemize}
\tightlist
\item
  Histograms of the residual in X/Y
\end{itemize}

\paragraph{Usage}

\begin{minted}[frame=single,framesep=3pt,breaklines=true,tabsize=2,linenos]{ini}
[TrackingSpatial]
spatial_cut_rel = 5.0
min_hits_on_track = 5
exclude_dut = true
\end{minted}

\hypertarget{treewriterdut}{%
\subsection{TreeWriterDUT}\label{treewriterdut}}

\modulehead{Morag Williams
(\href{mailto:morag.williams@cern.ch}{\nolinkurl{morag.williams@cern.ch}})}{DUT}{all}{Functional}

\paragraph{Description}

This module writes out data from a Timepix3 DUT for timing analysis. The
output ROOT TTree contains data in branches. This is intended for
analysis of the timing capabilities of Timepix3 devices of different
thicknesses.

For each track associated DUT cluster object
the following information is written out:

\begin{itemize}
\tightlist
\item
  Event ID
\item
  Size in X
\item
  Size in Y
\item
  Number of pixels in the cluster
\end{itemize}

For each pixel object in an associated
cluster the follwing information is written
out:

\begin{itemize}
\tightlist
\item
  X position
\item
  Y position
\item
  ToT
\item
  ToA
\end{itemize}

For each track with associated DUT
clusters the following information is written
out:

\begin{itemize}
\tightlist
\item
  Intercept with the DUT (3D position vector)
\end{itemize}

\paragraph{Parameters}

\begin{itemize}
\tightlist
\item
  \parameter{file_name}: Name of the data file to create,
  relative to the output directory of the framework. The file extension
  \file{.root} will be appended if not present.
  Default value is \file{outputTuples.root}.
\item
  \parameter{tree_name}: Name of the tree inside the
  output ROOT file. Default value is \parameter{tree}.
\end{itemize}

\paragraph{Plots produced}

No plots are produced.

\paragraph{Usage}

\begin{minted}[frame=single,framesep=3pt,breaklines=true,tabsize=2,linenos]{ini}
[TreeWriterDUT]
file_name = "myOutputFile.root"
tree_name = "myTree"
\end{minted}

\fi}

\newcommand{\inputmd}[1]{\def\temp{md//jobsub.tex}\ifx\temp\empty
  \textit{This section is missing Markdown to \LaTeX~conversion was not possible. Pandoc is required for the conversion.}
\else
  \input{md/#1}
\fi}

\addbibresource{usermanual/references.bib}

\newcommand{\version}{\lstinline|v0.9.14+4^g789e31c5|}
\newcommand{\project}{Corryvreckan}

\newcommand{\addreferencesline}{\addcontentsline{toc}{chapter}{References}}

\newcommand{\addlicense}{
\begin{table}[H]
\centering
\renewcommand{\arraystretch}{1.5}
\begin{tabular}{>{\centering\arraybackslash}m{.10\textwidth}>{\raggedright\arraybackslash}m{.90\textwidth}}
 \Large{\ccLogo \ccAttribution} & \footnotesize{This manual is licensed under the Creative Commons Attribution 4.0 International License.\newline To view a copy of this license, visit \url{http://creativecommons.org/licenses/by/4.0/}.} \\
\end{tabular}
\end{table}
}

\setlist[description]{style=nextline}

\clicdpnote{2019}{006}  

\date{\today}

\addauthor{J.~Kr\"oger}{\institute{1}\institute{2}}
\addauthor{S.~Spannagel\thanks{simon.spannagel@cern.ch}}{\institute{1}}
\addauthor{M.~Williams}{\institute{1}\institute{3}}

\addinstitute{1}{CERN, Switzerland}
\addinstitute{2}{University of Heidelberg, Germany}
\addinstitute{3}{University of Glasgow, UK}

\abstract{
Test beam data reconstruction is a task that requires a large amount of flexibility due to the heterogeneous data acquisition environments found in these experiments.
Often, detectors with different readout schemes such as triggered, frame-based or data driven approaches are combined in a single setup.
In order to correlate these data and to reconstruct particle tracks, a versatile event building algorithm and analysis framework is required.
Corryvreckan is a flexible, fast and lightweight test beam data reconstruction framework based on a modular concept of the reconstruction chain.
It is written in modern \CPP, requires a minimum of external dependencies and is designed to be fully configurable by the user without the need to alter code.
This document presents the user manual of the software as of release version~1.0.
}

\titlecomment{This work was carried out in the framework of the CLICdp Collaboration} 

\title{User Manual for the \corrybold Test Beam Data Reconstruction Framework, Version 1.0} 

\graphicspath{{figures/}}

\begin{document}
\titlepage

\clearpage
\tableofcontents

\clearpage

\section{Introduction}
\label{ch:introduction}

\corry is a flexible, fast and lightweight test beam data reconstruction framework based on a modular concept of the reconstruction chain.
It is designed to fulfil the requirements for offline event building in complex data-taking environments combining detectors with very different readout architectures.
\corry reduces external dependencies to a minimum by implementing its own flexible but simple data format to store intermediate reconstruction steps as well as final results.

The modularity of the reconstruction chain allows users to add their own functionality (such as event loaders to support different data formats or analysis modules to investigate specific features of detectors), without having to deal with centrally provided functionality, such as coordinate transformations, input and output, parsing of user input, and configuration of the analysis.
In addition, tools for batch submission of runs to a cluster scheduler such as \command{HTCondor} are provided to ease the (re-)analysis of complete test beam campaigns within a few minutes.

This project strongly profits from the developments undertaken for the \apsq project~\cite{apsq,apsq-website}: \emph{A Generic Pixel Detector Simulation Framework}.
Both frameworks employ very similar philosophies for configuration and modularity, and users of one framework will find it easy to get started with the other companion.
Some parts of the code base are shared explicitly, such as the configuration class or the module instantiation logic.
In addition, the \module{FileReader} and \module{FileWriter} modules have profited heavily from their corresponding framework components in \apsq.
The relevant sections of the \apsq manual~\cite{apsq-manual,clicdp-apsq-manual} have been adapted for this document.

It is also possible to combine the usage of both software frameworks: data produced by \apsq can be read in and analysed with \textit{Corryvreckan}.
This allows, for instance, to perform data/Monte Carlo comparisons when simulating a beam telescope configuration and analysing it with the same parameters as the read test-beam data.

\subsection{Scope of this Manual}
This document is meant to be the primary user guide for \corry.
It contains both an extensive description of the user interface and configuration possibilities, and a detailed introduction to the code base for potential developers.
This manual is designed to:
\begin{itemize}
\item Guide new users through the installation;
\item Introduce new users to the toolkit for the purpose of running their own test beam data reconstruction and analysis;
\item Explain the structure of the core framework and the components it provides to the reconstruction modules;
\item Provide detailed information about all modules and how to use and configure them;
\item Describe the required steps for implementing new reconstruction modules and algorithms.
\end{itemize}

More detailed information on the code itself can be found in the Doxygen reference manual~\cite{corry-doxygen} available online.
No programming experience is required from novice users, but knowledge of (modern) \CPP will be useful in the later chapters and may contribute to the overall understanding of the mechanisms.

\subsection{Getting Started}
An installation guideline is provided in Chapter~\ref{ch:installation}.
To get started with the analysis, some working examples can be found in the \dir{testing/} directory of the repository.
In addition, tutorials are available on \url{https://cern.ch/corryvreckan}.

\subsection{Support and Reporting Issues}
As for most of the software used within the high-energy particle physics community, only limited support on a best-effort basis can be offered for this software.
The authors are, however, happy to receive feedback on potential improvements or problems.
Reports on issues, questions concerning the software and documentation, and suggestions for improvements are very much appreciated.
These should preferably be brought up on the issues tracker of the project which can be found in the repository~\cite{corry-issue-tracker}.

\subsection{Contributing Code}
\label{sub:contributing}
\corry is a community project that benefits from active participation in the development and code contributions from users.
Users and prospective developers are encouraged to discuss their needs via the issue tracker of the repository~\cite{corry-issue-tracker} to receive ideas and guidance on how to implement a specific feature.
Getting in touch with other developers early in the development cycle avoids spending time on features that already exist or are currently under development by other users.

The repository contains a few tools to facilitate contributions and to ensure code quality, as detailed in Chapter~\ref{ch:testing}.

\section{Installation}
\label{ch:installation}

This section aims to provide details and instructions on how to build and install \corry.
An overview of possible build configurations is given.
After installing and loading the required dependencies, there are various options to customize the installation of \corry.
This chapter contains details on the standard installation process and information about custom build configurations.

\subsection{Supported Operating Systems}
\label{sec:os}
\corry is designed to run without issues on either a recent Linux distribution or Mac OS\,X.
Furthermore, the continuous integration of the project ensures correct building and functioning of the software framework on CentOS\,7 (with GCC and LLVM), SLC\,6 (with GCC and LLVM) and Mac OS Mojave (OS X 10.14, with AppleClang).

\subsection{CMVFS}
\label{sec:cvmfs_install}
The software is automatically deployed to CERN's VM file system (CVMFS)~\cite{cvmfs} for every new tag.
In addition, the \parameter{master} branch is built and deployed every night.
New versions are published to the folder \dir{/cvmfs/clicdp.cern.ch/software/corryvreckan/} where a new folder is created for every new tag, while updates via the \parameter{master} branch are always stored in the \dir{latest} folder.

The deployed version currently comprises of all modules that are active by default and do not require additional dependencies.
A \file{setup.sh} is placed in the root folder of the respective release, which allows all runtime dependencies necessary for executing this version to be set up.
Versions for both SLC\,6 and CentOS\,7 are provided.

\subsection{Docker}
\label{sec:docker}
Docker images are provided for the framework allowing anyone to run analyses without needing to install \corry on their system.
The only required program is the Docker executable as all other dependencies are provided within the Docker images.
In order to exchange configuration files and output data between the host system and the Docker container, a folder from the host system should be mounted to the container's data path \dir{/data}, which also acts as the Docker \parameter{WORKDIR} location.

The following command creates a container from the latest Docker image in the project registry and starts an interactive shell session with the \command{corry} executable already in the \texttt{\$PATH}.
Here, the current host system path is mounted to the \dir{/data} directory of the container.

\begin{verbatim}
$ docker run --interactive --tty                               \
             --volume "$(pwd)":/data                           \
             --name=corryvreckan                               \
             gitlab-registry.cern.ch/corryvreckan/corryvreckan \
             bash
\end{verbatim}

Alternatively it is also possible to directly start the reconstruction process instead of an interactive shell, e.g. using the following command:
\begin{verbatim}
$ docker run --tty --rm                                        \
             --volume "$(pwd)":/data                           \
             --name=corryvreckan                               \
             gitlab-registry.cern.ch/corryvreckan/corryvreckan \
             "corry -c my_analysis.conf"
\end{verbatim}
where an analysis described in the configuration \file{my_analysis.conf} is directly executed and the container terminated and deleted after completing the data processing.
This closely resembles the behavior of running \corry natively on the host system.
Of course, any additional command line arguments known to the \command{corry} executable described in Section~\ref{sec:executable} can be appended.

For tagged versions, the tag name should be appended to the image name, e.g.\ \parameter{gitlab-registry.cern.ch/corryvreckan/corryvreckan:v1.0}, and a full list of available Docker containers is provided via the project container registry~\cite{corry-container-registry}.

\subsection{Binaries}

Binary release tarballs are deployed to EOS to serve as downloads from the web to the directory \dir{/eos/project/c/corryvreckan/www/releases}.
New tarballs are produced for every tag as well as for nightly builds of the \parameter{master} branch, which are deployed with the name \file{corryvreckan-latest-<system-tag>-opt.tar.gz}.

\subsection{Compilation from Source}

The following paragraphs describe how to compile the \corry framework and its individual analysis and reconstruction modules from the source code.

\subsubsection{Prerequisites}
\label{sec:prerequisites}
The core framework is compiled separately from the individual modules, therefore \corry has only one required dependency: ROOT 6 (versions below 6 are not supported)~\cite{root}.
Please refer to~\cite{rootinstallation} for instructions on how to install ROOT.
ROOT has several components and to run \corry the GenVector package is required, a package that is included in the default build.

\subsubsection{Downloading the source code}
The latest version of \corry can be downloaded from the CERN Gitlab repository~\cite{corry-repo}.
For production environments, it is recommended to only download and use tagged software versions as many of the available git branches are considered development versions and might exhibit unexpected behavior.

For developers, it is recommended to always use the latest available version from the git \texttt{master} branch.
The software repository can be cloned as follows:

\begin{verbatim}
$ git clone https://gitlab.cern.ch/corryvreckan/corryvreckan.git
$ cd corryvreckan
\end{verbatim}

\subsubsection{Configuration via CMake}
\label{sec:cmake_config}
\corry uses the CMake build system to configure, build, and install the core framework as well as all modules.
An out-of-source build is recommended: this means CMake should not be directly executed in the source folder.
Instead, a \texttt{build} folder should be created from which CMake should be run.
For a standard build without any additional flags this entails executing:

\begin{verbatim}
$ mkdir build
$ cd build
$ cmake ..
\end{verbatim}

CMake can be run with several extra arguments to change the type of installation.
These options can be set with -D\textit{option}.
The following options are noteworthy:
\begin{itemize}
\item \parameter{CMAKE_INSTALL_PREFIX}: The directory to use as a prefix for installing the binaries, libraries, and data.
Defaults to the source directory (where the folders \dir{bin/} and \dir{lib/} are added).
\item \parameter{CMAKE_BUILD_TYPE}: The type of build to install, which defaults to \parameter{RelWithDebInfo} (compiles with optimizations and debug symbols).
Other possible options are \texttt{Debug} (for compiling with no optimizations, but with debug symbols and extended tracing using the Clang Address Sanitizer library) and \texttt{Release} (for compiling with full optimizations and no debug symbols).
\item \textbf{\texttt{BUILD\_\textit{ModuleName}}}: If the specific module \parameter{ModuleName} should be installed or not.
Defaults to \texttt{ON} for most modules, however some modules with additional dependencies such as EUDAQ or EUDAQ2~\cite{eudaq,eudaq2} are disabled by default.
This set of parameters allows to configure the build for minimal requirements as detailed in Section~\ref{sec:prerequisites}.
\item \parameter{BUILD_ALL_MODULES}: Build all included modules, defaulting to \texttt{OFF}.
This overwrites any selection using the parameters described above.
\end{itemize}

An example of a custom debug build, including the \module{EventLoaderEUDAQ2} module and with installation to a custom directory, is shown below:
\begin{verbatim}
$ mkdir build
$ cd build
$ cmake -DCMAKE_INSTALL_PREFIX=../install/ \
        -DCMAKE_BUILD_TYPE=DEBUG \
        -DBUILD_EventLoaderEUDAQ2=ON ..
\end{verbatim}
It should be noted that the \module{EventLoaderEUDAQ2} module requires additional dependencies and is therefore not built by default.

\subsubsection{Compilation and installation}
Compiling the framework is now a single command in the build folder created earlier, where \parameter{<number_of_cores>} is replaced with the number of cores to use for compilation:
\begin{verbatim}
$ make -j<number_of_cores>
\end{verbatim}
The compiled (non-installed) version of the executable can be found at \file{src/exec/corry} in the \dir{build} folder.
Running \corry directly without installing can be useful for developers.
It is not recommended for normal users, because the correct library and model paths are only fully configured during installation.

To install the library to the selected installation location (defaulting to the source directory of the repository) requires the following command:
\begin{verbatim}
$ make install
\end{verbatim}

The binary is now available as \file{bin/corry} in the installation directory.

\section{The \corrybold Framework}
\label{ch:framework}

\corry  is based on a collection of modules read from a configuration file and an event loop which sequentially processes data from all detectors assigned to one event.
In each loop iteration, all modules are executed in the linear order they appear in the configuration file and have access to all data created by previous modules.
At the end of the loop iteration, the data cache is cleared and the framework continues with processing the next iteration.
This chapter provides basic information about the different components of the framework, its executable as well as the available configuration parameters that are processed on a global level.

\subsection{The \texttt{corry} Executable}
\label{sec:executable}
The \corry executable \command{corry} functions as the interface between the user and the framework.
It is primarily used to provide the main configuration file, but also allows options from the main configuration file to be added or overwritten.
This is useful for quick testing as well as for batch processing of many runs to be reconstructed.

The executable handles the following arguments:
\begin{itemize}
\item \texttt{-c <file>}: Specifies the configuration file to be used for the reconstruction, relative to the current directory.
This is the only \textbf{required} argument and the program will fail to start if this argument is not given.
\item \texttt{-l <file>}: Specify an additional location, such as a file to forward log output into. This is used as an additional destination alongside the standard output and the location specified in the framework parameters described in Section~\ref{sec:framework_parameters}.
\item \texttt{-v <level>}: Sets the global log verbosity level, overwriting the value specified in the configuration file described in Section~\ref{sec:framework_parameters}.
Possible values are \texttt{FATAL}, \texttt{STATUS}, \texttt{ERROR}, \texttt{WARNING}, \texttt{INFO} and \texttt{DEBUG}, where all options are case-insensitive.
The module specific logging level introduced in Section~\ref{sec:logging_verbosity} is not overwritten.
\item \texttt{-o <option>}: Passes extra framework or module options that adds to and/or overwrites those in the main configuration file.
This argument may be specified multiple times to add multiple options.
Options are specified as key-value pairs in the same syntax as used in the configuration files described in Chapter~\ref{ch:configuration_files}, but the key is extended to include a reference to a configuration section or instantiation in shorthand notation.
There are three types of keys that can be specified:
\begin{itemize}
\item Keys to set \textbf{framework parameters}: These have to be provided in exactly the same way as they would be in the main configuration file (a section does not need to be specified). An example to overwrite the standard output directory would be \texttt{corry -c <file> -o output\_directory="run123456"}.
\item Keys for \textbf{module configurations}: These are specified by adding a dot (\texttt{.}) between the module and the key as it would be given in the configuration file (thus \textit{module}.\textit{key}). An example to overwrite the number of hits required for accepting a track candidate would be \command{corry -c <file> -o Tracking4D.min_hits_on_track=5}.
\item Keys to specify values for a particular \textbf{module instantiation}: The identifier of the instantiation and the name of the key are split by a dot (\texttt{.}), in the same way as for keys for module configurations (thus \textit{identifier}.\textit{key}). The unique identifier for a module can contain one or more colons (\texttt{:}) to distinguish between various instantiations of the same module. The exact name of an identifier depends on the name of the detector. An example to change the neighbor pixel search radius of the clustering algorithm for a particular instantiation of the detector named \textit{my\_dut} could be \command{corry -c <file> -o Clustering4D:my_dut.neighbour_radius_row=2}.
\end{itemize}
Note that only the single argument directly following the \texttt{-o} is interpreted as the option. If there is whitespace in the key-value pair this should be properly enclosed in quotation marks to ensure that the argument is parsed correctly.
\item \texttt{-g <option>}: Passes extra detector options that are added to and/or overwritten in the detector configuration file.
This argument can be specified multiple times to add multiple options.
The options are parsed in the same way as described above for module options, but only one type of key can be specified to overwrite an option for a single detector.
These are specified by adding a dot (\texttt{.}) between the detector and the key as it would be given in the detector configuration file (thus \textit{detector}.\textit{key}). An example to change the orientation of a particular detector named \texttt{detector1} would be \texttt{corry -c <file> -g detector1.orientation=0deg,0deg,45deg}.
\end{itemize}

No direct interaction with the framework is possible during the reconstruction. Signals can be sent using keyboard shortcuts to terminate the run, either gracefully or with force. The executable understands the following signals:
\begin{itemize}
\item SIGINT (\texttt{CTRL+C}): Request a graceful shutdown of the reconstruction. This means the event currently being processed is finished, while all other events requested in the configuration file are ignored. After finishing the event, the finalization stage is executed for every module to ensure all modules finish properly.
\item SIGTERM: Same as SIGINT, request a graceful shutdown of the reconstruction. This signal is emitted e.g.\ by the \command{kill} command or by cluster computing schedulers to ask for a termination of the job.
\item SIGQUIT (\texttt{CTRL+\textbackslash}): Forcefully terminates the framework. It is not recommended to use this signal as it will normally lead to the loss of all generated data. This signal should only be used when graceful termination is not possible.
\end{itemize}

\subsection{The Clipboard}

The clipboard is the framework infrastructure for temporarily storing information during the event processing.
Every module can access the clipboard to both read and write information.
Collections or individual elements on the clipboard are accessed via their data type, and an optional key can be used in addition to identify them with a certain detector by its name.

The clipboard consists of three parts: the event, a temporary data storage, and a persistent storage space.

\subsubsection{The Event}

The event is the central element storing meta-information about the data currently processed.
This includes the time frame (or data slice) within which all data are located as well as trigger numbers associated with these data.
New data to be added are always compared to the event on the clipboard to determine whether they should be discarded, included, or buffered for later use.
A detailed description of the event building process is provided in Chapter~\ref{ch:events}.

\subsubsection{Temporary Data Storage}
The temporary data storage is only available during the processing of each individual event.
It is automatically cleared at the end of the event processing and has to be populated with new data in the new event to be processed.

The temporary storage acts as the main data structure to communicate information between different modules and can hold multiple collections of \corry objects such as pixel hits, clusters, or tracks.
In order to be able to flexibly store different data types on the clipboard, the access methods for the temporary data storage are implemented as templates, and vectors of any data type deriving from \parameter{corry::Object} can be stored and retrieved.

\subsubsection{Persistent Storage}
The persistent storage is not cleared at the end of processing each event and can be used to store information used in multiple events.
Currently this storage only allows for the caching of double-precision floating point numbers.

\subsection{Global Framework Parameters}
\label{sec:framework_parameters}
The \corry framework provides a set of global parameters that control and alter its behavior. These parameters are inherited by all modules.
The currently available global parameters are:

\begin{itemize}
\item \parameter{detectors_file}: Location of the file describing the detector configuration described in Section~\ref{sec:detector_config}.
The only \textit{required} global parameter as the framework will fail to start if it is not specified.
This parameter can take multiple paths; all provided files will be combined into one geometry description of the setup.
\item \parameter{detectors_file_updated}: Location of the file that the (potentially) updated detector configuration should be written into. If this file does not already exist, it will be created. If the same file is given as for \parameter{detectors_file}, the file is overwritten. If no file is specified using this parameter then the updated geometry is not written to file.
\item \parameter{histogram_file}: Location of the file where the ROOT output histograms of all modules will be written to. The file extension \texttt{.root} will be appended if not present. Directories within the ROOT file will be created automatically for all modules.
\item \parameter{number_of_events}: Determines the total number of events the framework will process, where negative numbers allow for the processing of all data available.
After reaching the specified number of events, the reconstruction is stopped.
Defaults to $-1$.
\item \parameter{number_of_tracks}: Determines the total number of tracks the framework should reconstruct, where negative numbers indicate that there is no limit on the number of reconstructed tracks.
After reaching the specified number of events, the reconstruction is stopped.
Defaults to $-1$.
\item \parameter{run_time}: Determines the wall-clock time of data acquisition the framework should reconstruct up until. Negative numbers indicate that there is no limit on the time slice to reconstruct.
Defaults to $-1$.
\item \parameter{log_level}: Specifies the lowest log level that should be reported.
Possible values are \texttt{FATAL}, \texttt{STATUS}, \texttt{ERROR}, \texttt{WARNING}, \texttt{INFO}, and \texttt{DEBUG}, where all options are case-insensitive.
Defaults to the \texttt{INFO} level.
More details and information about the log levels, including how to change them for a particular module, can be found in Section~\ref{sec:logging_verbosity}.
Can be overwritten by the \texttt{-v} parameter on the command line (see Section~\ref{sec:executable}).
\item \parameter{log_format}: Determines the log message format to be displayed.
Possible options are \texttt{SHORT}, \texttt{DEFAULT}, and \texttt{LONG}, where all options are case-insensitive.
More information can be found in Section~\ref{sec:logging_verbosity}.
\item \parameter{log_file}: File where the log output will be written, in addition to the standard printing (usually the terminal).
Another (additional) location to write to can be specified on the command line using the \texttt{-l} parameter (see Section~\ref{sec:executable}).
\item \parameter{library_directories}: Additional directories to search in for module libraries, before searching the default paths.
\item \parameter{output_directory}: Directory to write all output files into.
Subdirectories are created automatically for all module instantiations.
This directory will also contain the \parameter{histogram_file} specified via the parameter described above.
Defaults to the current working directory with the subdirectory \dir{output/} attached.
\item \parameter{purge_output_directory}: Decides whether the content of an already existing output directory is deleted before a new run starts. Defaults to \texttt{false}, i.e. files are kept but will be overwritten by new files created by the framework.
\item \parameter{deny_overwrite}: Forces the framework to abort the run and throw an exception when attempting to overwrite an existing file. Defaults to \texttt{false}, i.e. files are overwritten when requested. This setting is inherited by all modules, but can be overwritten in the configuration section of each of the modules.
\end{itemize}

\subsection{Modules and the Module Manager}
\label{sec:module_manager}
\corry is a modular framework and one of the core ideas is to partition functionality into independent modules that can be inserted or removed as required.
These modules are located in the subdirectory \dir{src/modules/} of the repository, with the name of the directory as the unique name of the module.
The suggested naming scheme is CamelCase, thus an example module name would be \module{OnlineMonitor}.
Any \emph{specifying} part of a module name should precede the \emph{functional} part of the name, e.g.\ \module{EventLoaderCLICpix2} rather than \module{CLICpix2EventLoader}.
There are three different kinds of modules that can be defined:
\begin{itemize}
    \item \textbf{Global}: Modules for which a single instance runs, irrespective of the number of detectors.
    \item \textbf{Detector}: Modules that are concerned with only a single detector at a time.
    These are then replicated for all required detectors.
    \item \textbf{DUT}: Similar to the Detector modules, these modules run for a single detector.
    However, they are only replicated if the respective detector is marked as device-under-test (DUT).
\end{itemize}
The type of module determines the constructor used, the internal unique name, and the supported configuration parameters.
For more details about the instantiation logic for the different types of modules, see Section~\ref{sec:module_instantiation}.

Furthermore, detector modules can be restricted to certain types of detectors.
This allows the framework to automatically determine for which of the given detectors the respective module should be instantiated.
For example, the \module{EventLoaderCLICpix2} module will only be instantiated for detectors with the correct type \parameter{CLICpix2} without any additional configuration effort required by the user.
This procedure is described in more detail in Section~\ref{sec:module_files}.

\subsubsection{Module Status Codes}

The \command{run()} function of each module returns a status code to the module manager to indicate the status of the module.
These codes can be used to e.g. request the end of the current run or to signal problems in data processing.
The following status codes are currently supported:

\begin{description}
    \item{\command{Success}}: Indicates that the module successfully finished processing all data. The framework continues with the next module.
    \item{\command{NoData}}: Indicates that the respective module did not find any data to process. The framework continues with the next module.
    \item{\command{DeadTime}}: Indicates that the detector handled by the respective module is currently in data acquisition dead time. The framework skips all remaining modules for this event and continues with the subsequent event. By employing this method, measurements such as efficiency are not affected by known inefficiencies during detector dead times.
    \item{\command{EndRun}}: Allows the module to request a premature end of the run. This can e.g.\ be used by alignment modules to stop the run after they have accumulated enough tracks for the alignment procedure. The framework executes all remaining modules for the current event and then enters the finalization stage.
    \item{\command{Failure}}: Indicates that there was a severe problem when processing data in the respective module. The framework skips all remaining modules for this event and enters the finalization stage.
\end{description}

\subsubsection{Execution Order}

Modules are executed in the order in which they appear in the configuration file; the sequence is repeated for every time frame or event.
If one module in the chain raises e.g.\ a \command{DeadTime} status, the processing of the next event is begun without executing the remaining modules for the current event.

The execution order is of special importance in the event building process, since the first module will always have to define the event while subsequent event loader modules will have to adhere to the time frame defined by the first module.
A complete description of the event building process is provided in Chapter~\ref{ch:events}.

This behavior should also be taken into account when choosing the order of modules in the configuration file, since e.g.\ data from detectors catered by subsequent event loaders is not processed and hit maps are not updated if an earlier module requested to skip the rest of the module chain.

\subsubsection{Module instantiation}
\label{sec:module_instantiation}
Modules are dynamically loaded and instantiated by the Module Manager.
They are constructed, initialized, executed, and finalized in the linear order in which they are defined in the configuration file. For this reason, the configuration file should follow the order of the real process.
For each section in the main configuration file (see Chapter~\ref{ch:configuration_files} for more details), a corresponding library is searched for which contains the module (the exception being the global framework section).
Module libraries are always named following the scheme \textbf{libCorryvreckanModule\texttt{ModuleName}}, reflecting the \texttt{ModuleName} configured via CMake.
The module search order is as follows:
\begin{enumerate}
\item Modules already loaded from an earlier section header
\item All directories in the global configuration parameter \parameter{library_directories} in the provided order, if this parameter exists.
\item The internal library paths of the executable, which should automatically point to the libraries that are built and installed together with the executable.
These library paths are stored in \dir{RPATH} on Linux, see the next point for more information.
\item The other standard locations to search for libraries depending on the operating system.
Details about the procedure Linux follows can be found in~\cite{linuxld}.
\end{enumerate}

If the loading of the module library is successful, the module is checked to determine if it is a global, detector, or DUT module.
A single module may be called multiple times in the configuration with overlapping requirements (such as a module that runs on all detectors of a given type, followed by the same module but with different parameters for one specific detector, also of this type). Hence the Module Manager must establish which instantiations to keep and which to discard.
The instantiation logic determines a unique name and priority for every instantiation. Internally these are handled as numerical values, where a lower number indicates a higher priority.
The name and priority for the instantiation are determined differently for the two types of modules:
\begin{itemize}
\item \textbf{Global}: Name of the module, the priority is always \emph{high}.
\item \textbf{Detector/DUT}: Combination of the name of the module and the name of detector this module is executed for.
If the name of the detector is specified directly by the \parameter{name} parameter, the priority is \emph{high}.
If the detector is only matched by the \parameter{type} parameter, the priority is \emph{medium}.
If the \parameter{name} and \parameter{type} are both unspecified and the module is instantiated for all detectors, the priority is \emph{low}.
\end{itemize}
In the end, only a single instance for every unique name is allowed.
If there are multiple instantiations with the same unique name, the instantiation with the highest priority is kept.
If multiple instantiations with the same unique name and the same priority exist, an exception is raised.

\subsection{Logging and Verbosity Levels}
\label{sec:logging_verbosity}
\corry is designed to identify mistakes and implementation errors as early as possible and to provide the user with clear indications of the location and cause the problem.
The amount of feedback can be controlled using different log levels, which are inclusive, i.e.\ lower levels also include messages from all higher levels.
The global log level can be set using the global parameter \parameter{log_level}.
The log level can be overridden for a specific module by adding the \parameter{log_level} parameter to the respective configuration section.
The following log levels are supported:
\begin{itemize}
\item \textbf{FATAL}: Indicates a fatal error that will lead to the direct termination of the application.
Typically only emitted in the main executable after catching exceptions, as they are the preferred way of fatal error handling.
An example of a fatal error is an invalid value for an existing configuration parameter.
\item \textbf{STATUS}: Important information about the status of the reconstruction.
Is only used for messages that have to be logged in every run, such as initial information on the module loading, opened data files, and the current progress of the run.
\item \textbf{ERROR}: Severe error that should not occur during a normal, well-configured reconstruction.
Frequently leads to a fatal error and can be used to provide extra information that may help in finding the problem. For example, it is used to indicate the reason a dynamic library cannot be loaded.
\item \textbf{WARNING}: Indicate conditions that should not occur normally and possibly lead to unexpected results.
The framework will however continue without problems after a warning.
A warning is, for example, issued to indicate that a calibration file for a certain detector cannot be found and that the reconstruction is therefore performed with uncalibrated data.
\item \textbf{INFO}: Information messages about the reconstruction process.
Contains summaries of the reconstruction details for every event and for the overall run.
This should typically produce at maximum one line of output per event and module.
\item \textbf{DEBUG}: In-depth details about the progress of the reconstruction, such as information on every cluster formed or on track fitting results.
Produces large volumes of output per event and should therefore only be used for debugging the reconstruction process.
\item \textbf{TRACE}: Messages to trace what the framework or a module is currently doing.
Unlike the \textbf{DEBUG} level, it does not contain any direct information about the physics but rather indicates which part of the module or framework is currently running.
Mostly used for software debugging or determining performance bottlenecks in the reconstruction.
\end{itemize}

\begin{warning}
    It is not recommended to set the \parameter{log_level} higher than \textbf{WARNING} in a typical reconstruction as important messages may be missed.
    Setting too low logging levels should also be avoided since printing many log messages will significantly slow down the reconstruction.
\end{warning}

The logging system supports several formats for displaying the log messages.
The following formats are supported via the global parameter \parameter{log_format} or the individual module parameter with the same name:
\begin{itemize}
\item \textbf{SHORT}: Displays the data in a short form.
Includes only the first character of the log level followed by the configuration section header and the message.
\item \textbf{DEFAULT}: The default format.
Displays the system time, log level, section header, and the message itself.
\item \textbf{LONG}: Detailed logging format.
Displays all of the above but also indicates the source code file and line where the log message was produced.
This can help when debugging modules.
\end{itemize}

\subsection{Coordinate Systems}
\label{sec:coordinate_systems}

Local coordinate systems for each detector and a global frame of reference for the full setup are defined.
The global coordinate system is chosen as a right-handed Cartesian system, and the rotations of individual devices are performed around the geometrical center of their sensor.
Here, the beam direction defines the positive z-axis at the origin of the x-y-plane.
The origin along the z-axis is fixed by the placement of the detectors in the geometry of the setup.

Local coordinate systems for the detectors are also right-handed Cartesian systems, with the x- and y-axes defining the sensor plane.
The origin of this coordinate system is the center of the lower left pixel in the grid, i.e.\ the pixel with indices (0,0).
This simplifies calculations in the local coordinate system as all positions can either be stated in absolute numbers or in fractions of the pixel pitch.

\section{Configuration Files}
\label{ch:configuration_files}
The framework is configured with human-readable key-value based configuration files.
The configuration format consists of section headers within $[$ and $]$ brackets, and a global section without a header at the beginning.
Each of these sections contains a set of key-value pairs separated by the \texttt{=} character.
Comments are indicated using the hash symbol (\texttt{\#}).

Since configuration files are highly user-specific and do not directly belong to the \corry framework, they should not be stored in the \corry repository.
However, working examples can be found in the \dir{testing/} directory of the repository.

\corry can handle any file extensions for geometry and configuration files.
The examples, however, follow the convention of using the extension \texttt{*.conf} for both the detector and configuration files.

The framework has two required layers of configuration files:
\begin{itemize}
\item The \textbf{main} configuration: It is passed directly to the binary and contains both the global framework configuration and the list of modules to instantiate, together with their configuration.

\item The \textbf{detector} configuration: Passed to the framework to determine the geometry.
Describes the detector setup and contains the position, orientation and type of all detectors along with additional properties crucial for the reconstruction.
\end{itemize}

In the following paragraphs, the available types and the unit system are explained and an introduction to the different configuration files is given.

\subsection{Parsing types and units}
\label{sec:config_values}
The \corry framework supports the use of a variety of types for all configuration values.
The module requesting the configuration key specifies how the value type should be interpreted.
An error will be raised if either a necessary key is not specified in the configuration file, the conversion to the desired type is not possible, or if the given value is outside the domain of possible options.
Please refer to the module documentation in Chapter~\ref{ch:modules} for the list of module parameters and their types.
The value is parsed in an intuitive manner, however a few special rules do apply:
\begin{itemize}
\item If the value is a \textbf{string}, it may be enclosed by a single pair of double quotation marks (\texttt{"}), which are stripped before passing the value to the module(s).
If the string is not enclosed by quotation marks, all whitespace before and after the value is erased.
If the value is an array of strings, the value is split at every whitespace or comma (\texttt{,}) that is not enclosed in quotation marks.
\item If the value is a \textbf{boolean}, either numerical (\texttt{0}, \texttt{1}) or textual (\texttt{false}, \texttt{true}) representations are accepted.
\item If the value is a \textbf{relative path}, that path will be made absolute by adding the absolute path of the directory that contains the configuration file where the key is defined.
\item If the value is an \textbf{arithmetic} type, it may have a suffix indicating the unit.
The list of base units is shown in Table~\ref{tab:units}.
\end{itemize}

The internal base units of the framework are not chosen for user convenience, but for maximum precision of the calculations and to avoid the necessity of conversions in the code.
Combinations of base units can be specified by using the multiplication sign \texttt{*} and the division sign \texttt{/} that are parsed in linear order (thus $\frac{V m}{s^2}$ should be specified as $V*m/s/s$).
The framework assumes the default units (as given in Table~\ref{tab:units}) if the unit is not explicitly specified.

\begin{warning}
  If no units are specified, values will always be interpreted in the base units of the framework.
  In some cases this can lead to unexpected results.
  E.g. specifying a pixel pitch as \parameter{pixel_pitch = 55,55} results in a detector with a pixel size of \SI{55 x 55}{\milli \meter}.
  Therefore, it is strongly recommended to always specify the units explicitly for all parameters that are not dimensionless in the configuration files.
\end{warning}

\begin{table}[tbp]
\caption{List of units supported by \corry}
\label{tab:units}
\centering
\begin{tabular}{lll}
  \toprule
\textbf{Quantity}                 & \textbf{Default unit}                   & \textbf{Auxiliary units} \\
 \midrule
\multirow{6}{*}{\textit{Length}}  & \multirow{6}{*}{mm (millimeter)}        & nm (nanometer)           \\
                                  &                                         & um (micrometer)          \\
                                  &                                         & cm (centimeter)          \\
                                  &                                         & dm (decimeter)           \\
                                  &                                         & m (meter)                \\
                                  &                                         & km (kilometer)           \\
 \midrule
\multirow{4}{*}{\textit{Time}}    & \multirow{4}{*}{ns (nanosecond)}        & ps (picosecond)          \\
                                  &                                         & us (microsecond)         \\
                                  &                                         & ms (millisecond)         \\
                                  &                                         & s (second)               \\
\midrule
\multirow{3}{*}{\textit{Energy}}  & \multirow{3}{*}{MeV (megaelectronvolt)} & eV (electronvolt)        \\
                                  &                                         & keV (kiloelectronvolt)   \\
                                  &                                         & GeV (gigaelectronvolt)   \\
\midrule
\textit{Temperature}              & K (kelvin)                              & ---                      \\
\midrule
\multirow{3}{*}{\textit{Charge}}  & \multirow{3}{*}{e (elementary charge)}  & ke (kiloelectrons)       \\
                                  &                                         & fC (femtocoulomb)        \\
                                  &                                         & C (coulomb)              \\
\midrule
\multirow{2}{*}{\textit{Voltage}} & \multirow{2}{*}{MV (megavolt)}          & V (volt)                 \\
                                  &                                         & kV (kilovolt)            \\
\midrule
\textit{Magnetic field strength}  & T (tesla)                               & mT (millitesla)                 \\
\midrule
\multirow{2}{*}{\textit{Angle}}   & \multirow{2}{*}{rad (radian)}           & deg (degree)             \\
                                  &                                         & mrad (milliradian)       \\
\bottomrule
\end{tabular}
\end{table}

\newpage
Examples of specifying key-values pairs of various types are given below:
\begin{minted}[frame=single,framesep=3pt,breaklines=true,tabsize=2,linenos]{ini}
# All whitespace at the front and back is removed
first_string =   string_without_quotation
# All whitespace within the quotation marks is preserved
second_string = "  string with quotation marks  "
# Keys are split on whitespace and commas
string_array = "first element" "second element","third element"
# Integers and floats can be specified in standard formats
int_value = 42
float_value = 123.456e9
# Units can be passed to arithmetic type
energy_value = 1.23MeV
time_value = 42ns
# Units are combined in linear order
acceleration_value = 1.0m/s/s
# Thus the two quantities below have the same units
random_quantity_a = 1.0deg*kV/m/s*K
random_quantity_b = 1.0deg*kV*K/m/s
# Relative paths are expanded to absolute
# Path below will be /home/user/test/ if the config file is in /home/user
output_path = "test/"
# Booleans can be represented in numerical or textual style
my_switch = true
my_other_switch = 0
\end{minted}

\subsubsection{File format}
\label{sec:config_file_format}
Throughout the framework, a simplified version of TOML~\cite{tomlgit} is used as standard format for configuration files.
The format is defined as follows:
\begin{enumerate}
\item All whitespace at the beginning or end of a line are stripped by the parser.
In the rest of this format specification, \textit{line} refers to the line with this whitespace stripped.
\item Empty lines are ignored.
\item Every non-empty line should start with either \texttt{\#}, \texttt{[} or an alphanumeric character.
Every other character should lead to an immediate parse error.
\item If the line starts with a hash character (\texttt{\#}), it is interpreted as comment and all other content on that line is ignored.
\item If the line starts with an open square bracket (\texttt{[}), it indicates a section header (also known as configuration header).
The line should contain a string with alphanumeric characters and underscores indicating the header name, followed by a closing square bracket (\texttt{]}) to end the header.
After any number of ignored whitespace characters there could be a \texttt{\#} character.
If this is the case, the rest of the line is handled as specified in point~3.
Otherwise, there should not be any other character on the line that is not whitespace.
Any line that does not comply to these specifications should lead to an immediate parse error.
Multiple section headers with the same name are allowed.
All key-value pairs in the line following this section header are part of this section until a new section header is started.
\item If the line starts with an alphanumeric character, the line should indicate a key-value pair.
The beginning of the line should contain a string of alphabetic characters, numbers, dots (\texttt{.}), colons (\texttt{:}), and/or underscores (\texttt{\_}), but it may only start with an alphanumeric character.
This string indicates the 'key'.
After an optional number of ignored whitespace, the key should be followed by an equality sign (\texttt{$=$}).
Any text between the \texttt{$=$} and the first \texttt{\#} character not enclosed within a pair of single or double quotes (\texttt{'} or \texttt{"}) is known as the non-stripped string.
Any character after the \texttt{\#} is handled as specified in point 3.
If the line does not contain any non-enclosed \texttt{\#} character, the value ends at the end of the line instead.
The 'value' of the key-value pair is the non-stripped string with all whitespace in front and at the end stripped.
The value may not be empty.
Any line that does not comply to these specifications should lead to an immediate parse error.
\item The value may consist of multiple nested dimensions that are grouped by pairs of square brackets (\texttt{[} and \texttt{]}).
The number of square brackets should be properly balanced, otherwise an error is raised.
Square brackets that should not be used for grouping should be enclosed in quotation marks.
Every dimension is split at every whitespace sequence and comma character (\texttt{,}) not enclosed in quotation marks.
Implicit square brackets are added to the beginning and end of the value, if these are not explicitly added.
A few situations require the explicit addition of outer brackets such as matrices with only one column element, i.e. with dimension 1xN.
\item The sections of the value that are interpreted as separate entities are named elements.
For a single value the element is on the zeroth dimension, for arrays on the first dimension, and for matrices on the second dimension.
Elements can be forced by using quotation marks, either single or double quotes (\texttt{'} or \texttt{"}).
The number of both types of quotation marks should be properly balanced, otherwise an error is raised.
The conversion of the elements to the actual type is performed when accessing the value.
\item All key-value pairs defined before the first section header are part of a zero-length empty section header.
\end{enumerate}

\subsubsection{Accessing parameters}
\label{sec:accessing_parameters}
Values are accessed via the configuration object.
In the following example, the key is a string called \parameter{key}, the object is named \parameter{config} and the type \parameter{TYPE} is a valid \CPP type that the value should represent.
The values can be accessed via the following methods:
\begin{minted}[frame=single,framesep=3pt,breaklines=true,tabsize=2,linenos]{c++}
// Returns true if the key exists and false otherwise
config.has("key")
// Returns the number of keys found from the provided initializer list:
config.count({"key1", "key2", "key3"});
// Returns the value in the given type, throws an exception if not existing or a conversion to TYPE is not possible
config.get<TYPE>("key")
// Returns the value in the given type or the provided default value if it does not exist
config.get<TYPE>("key", default_value)
// Returns an array of elements of the given type
config.getArray<TYPE>("key")
// Returns a matrix: an array of arrays of elements of the given type
config.getMatrix<TYPE>("key")
// Returns an absolute (canonical if it should exist) path to a file, where the second input value determines if the existence of the path is checked
config.getPath("key", true /* check if path exists */)
// Return an array of absolute paths, where the second input value determines if the existence of the paths is checked
config.getPathArray("key", false /* do not check if paths exists */)
// Returns the value as literal text including possible quotation marks
config.getText("key")
// Set the value of key to the default value if the key is not defined
config.setDefault("key", default_value)
// Set the value of the key to the default array if key is not defined
config.setDefaultArray<TYPE>("key", vector_of_default_values)
// Create an alias named new_key for the already existing old_key or throws an exception if the old_key does not exist
config.setAlias("new_key", "old_key")
\end{minted}

Conversions to the requested type are using the \parameter{from_string} and \parameter{to_string} methods provided by the framework string utility library.
These conversions largely follow standard \CPP parsing, with one important exception.
If (and only if) the value is retrieved as a C/\CPP string and the string is fully enclosed by a pair of \texttt{"} characters, these are stripped before returning the value.
Strings can thus also be provided with or without quotation marks.

\begin{warning}
    It should be noted that a conversion from string to the requested type is a comparatively heavy operation.
    For performance-critical sections of the code, one should consider fetching the configuration value once and caching it in a local variable.
\end{warning}

\subsection{Main configuration}
\label{sec:main_config}
The main configuration file consists of a set of sections specifying the modules to be used.
All modules are executed in the \emph{linear} order in which they are defined.
There are a few section names that have a special meaning in the main configuration, namely the following:
\begin{itemize}
\item The \textbf{global} (framework) header sections: These are all zero-length section headers (including the one at the beginning of the file) and all sections marked with the header \texttt{[Corryvreckan]} (case-insensitive).
These are combined and accessed together as the global configuration, which contains all parameters of the framework itself as described in Section~\ref{sec:framework_parameters}.
All key-value pairs defined in this section are also inherited by all individual configurations as long the key is not defined in the module configuration itself. This is encouraged for module parameters used in multiple modules.
\item The \textbf{ignore} header sections: All sections with name \texttt{[Ignore]} are ignored.
Key-value pairs defined in the section, as well as the section itself, are discarded by the parser.
These section headers are useful for quickly enabling and disabling individual modules by replacing their actual name by an ignore section header.
\end{itemize}

All other section headers are used to instantiate modules of the respective name.
Installed module libraries are loaded automatically at startup.
Parameters defined under the header of a module are local to that module and are not inherited by other modules.

An example for a valid albeit illustrative \corry main configuration file is:
\begin{minted}[frame=single,framesep=3pt,breaklines=true,tabsize=2,linenos]{ini}
# Key is part of the empty section and therefore the global configuration
random_string = "example1"
# The location of the detector configuration is a global parameter
detectors_file = "testbeam_setup.conf"
# The Corryvreckan section is also considered global and merged with the above
[Corryvreckan]
another_random_string = "example2"

# Stop after one thousand events:
number_of_events = 1000

# First section runs "ModuleA"
[ModuleA]
# This module takes no parameters

# Ignore this second section:
[Ignore]
my_key = "my_value"

# Third section runs "ModuleC" with configured parameters:
[ModuleC]
int_value = 2
vector_of_doubles = 23.0, 45.6, 78.9
\end{minted}

\subsection{Detector configuration}
\label{sec:detector_config}
The detector configuration file consists of a set of sections that describe the detectors in the setup.
Each section starts with a header describing the name used to identify the detector; all names are required to be unique.
Every detector should contain all of the following parameters:
\begin{itemize}
\item The \parameter{role} parameter is an array of strings indicating the function(s) of the respective detector. This can be \parameter{dut}, \parameter{reference} (\parameter{ref}), \parameter{auxiliary} (\parameter{aux}), or \parameter{none}, where the latter is the default. With the default role, the respective detector participates in tracking but is neither used as reference plane for alignment and correlations, nor treated as DUT. In a reference role, the detector is used as anchor for relative alignments and its position and orientation is used for comparison when producing correlation and residual plots. As DUT, the detector is by default excluded from tracking, and all DUT-type modules are executed for this detector. As an auxiliary device, the detector may provide additional information but does not partake in the reconstruction. This is useful to e.g. include trigger logic units (TLUs) providing only timing information.
\begin{warning}
There always has to be exactly \emph{one} reference detector in the setup. For setups with a single detector only, the role should be configured as \parameter{dut, reference} for the detector to act as both. Auxiliary devices cannot have any other role simultaneously.
\end{warning}

\item The \parameter{type} parameter is a string describing the type of detector, e.g.\ \parameter{Timepix3} or \parameter{CLICpix2}. This value might be used by some modules to distinguish between different types.
\item The \parameter{position} in the world frame.
This is the position of the geometric center of the sensor given in world coordinates as X, Y and Z as defined in Section~\ref{sec:coordinate_systems}.
\item An \parameter{orientation_mode} that determines the way that the orientation is applied.
This can be either \texttt{xyz}, \texttt{zyx}, or \texttt{zxz}, where \textbf{\texttt{xyz}} is used as default if the parameter is not specified. Three angles are expected as input, which should always be provided in the order in which they are applied.
\begin{itemize}
    \item The \texttt{xyz} option uses extrinsic Euler angles to apply a rotation around the global $X$ axis, followed by a rotation around the global $Y$ axis, and finally a rotation around the global $Z$ axis.
    \item The \texttt{zyx} option uses the \textbf{extrinsic Z-Y-X} convention for Euler angles, also known as Pitch-Roll-Yaw or 321 convention. The rotation is represented by three angles describing an initial rotation of an angle $\phi$ (yaw) about the $Z$ axis, followed by a rotation of an angle $\theta$ (pitch) about the initial $Y$ axis, followed by a third rotation of an angle $\psi$ (roll) about the initial $X$ axis.
    \item The \texttt{zxz} option uses the \textbf{extrinsic Z-X-Z} convention for Euler angles. This option is also known as the 3-1-3 or the "x-convention" and the most widely used definition of Euler angles~\cite{eulerangles}.
\end{itemize}
\begin{warning}
It is highly recommended to always explicitly state the orientation mode, rather than relying on the default configuration, to avoid unwanted behaviour.
\end{warning}

\item The \parameter{orientation} to specify the Euler angles in logical order (e.g. first $X$, then $Y$, then $Z$ for the \texttt{xyz} method), interpreted using the method above. An example for three Euler angles would be:
\begin{minted}[frame=single,framesep=3pt,breaklines=true,tabsize=2,linenos]{ini}
orientation_mode = "zyx"
orientation = 45deg 10deg 12deg
\end{minted}
which describes a rotation of \SI{45}{\degree} around the $Z$ axis, followed by a \SI{10}{\degree} rotation around the initial $Y$ axis, and finally a rotation of \SI{12}{\degree} around the initial $X$ axis.
\begin{warning}
All supported rotations are extrinsic active rotations, i.e. the vector itself is rotated, not the coordinate system. All angles in configuration files should be specified in the order they will be applied.
\end{warning}

\item The \parameter{number_of_pixels} parameter represents a two-dimensional vector with the number of pixels in the active matrix in the column and row directions respectively.
\item The \parameter{pixel_pitch} is a two-dimensional vector defining the size of a single pixel in the column and row directions respectively.
\item The intrinsic resolution of the detector has to be specified using the \parameter{spatial_resolution} parameter, a two-dimensional vector holding the position resolution for the column and row directions. This value is used to assign the uncertainty of cluster positions. This parameter defaults to the pitch$/\sqrt{12}$ of the respective detector if not specified.
\item The intrinsic time resolution of the detector should be specified using the \parameter{time_resolution} parameter with units of time. This can be used to apply detector specific time cuts in modules. This parameter is only required when using relative time cuts in the analysis.
\item The \parameter{time_offset} can be used to shift the reference time frame of an individual detector to e.g.\ account for time of flight effects between different detector planes by adding a fixed offset.
\item The \parameter{material_budget} defines the material budget of the sensor layer in fractions of the radiation length, including support. If no value is defined a default of zero is assumed. A given value has to be larger than zero.
\item Pixels to be masked in the offline analysis can be placed in a separate file specified by the \parameter{mask_file} parameter, which is explained in detail in Section~\ref{sec:masking}.
\item A region of interest in the given detector can be defined using the \parameter{roi} parameter. More details on this functionality can be found in Section~\ref{sec:roi}.
\end{itemize}

An example configuration file describing a setup with one CLICpix2~\cite{clicpix2,clicpix2-pisa} detector named \parameter{016_CP_PS} and two Timepix3~\cite{timepix3} detectors (\parameter{W0013_D04} and \parameter{W0013_J05}) is the following:

\begin{minted}[frame=single,framesep=3pt,breaklines=true,tabsize=2,linenos]{ini}
[W0013_D04]
number_of_pixels = 256, 256
orientation = 9deg, 9deg, 0deg
orientation_mode = "xyz"
pixel_pitch = 55um, 55um
position = 0um, 0um, 10mm
spatial_resolution = 4um,4um
time_resolution = 3ns
type = "Timepix3"

[016_CP_PS]
mask_file = "mask_016_CP_PS.conf"
number_of_pixels = 128,128
orientation = -0.02deg, 0.0deg, -0.015deg
orientation_mode = "xyz"
pixel_pitch = 25um, 25um
position = -0.9mm, 0.21mm, 106.0mm
spatial_resolution = 8um,8um
time_resolution = 1ms
role = "dut"
type = "CLICpix2"

[W0013_J05]
number_of_pixels = 256, 256
orientation = -9deg, 9deg, 0deg
orientation_mode = "xyz"
pixel_pitch = 55um, 55um
position = 0um, 0um, 204mm
spatial_resolution = 4um,4um
time_resolution = 3ns
role = "reference"
type = "Timepix3"
\end{minted}

\subsubsection{Masking Pixels Offline}
\label{sec:masking}

Mask files can be provided for individual detectors, which allows the user to mask specific pixels in the reconstruction.
The following syntax is used for each line within the mask file:
\begin{itemize}
    \item \command{c COL}: masking all pixels in column \parameter{COL}
    \item \command{r ROW}: masking all pixels in row \parameter{ROW}
    \item \command{p COL ROW}: masking the single pixel at address \parameter{COL, ROW}
\end{itemize}

\begin{warning}
It should be noted that the individual event loader modules have to take care of discarding masked pixels manually, the \corry framework only parses the mask file and attaches the mask information to the respective detector. The event loader modules should thus always query the detector object for masks before adding new pixels to the data collections.
\end{warning}

\subsubsection{Defining a Region of Interest}
\label{sec:roi}

The region of interest (ROI) feature of each detector allows tracks or clusters to be marked as within a certain region on the respective detector.
This information can be used in analyses to restrict the selection of tracks or clusters to certain regions of the device, e.g.\ to exclude known bad regions from the calculation of efficiencies.

The ROI is defined as a polynomial in local pixel coordinates of the device using the \parameter{roi} keyword. A rectangle could, for example, be defined by providing the four corners of the shape via the following:

\begin{minted}[frame=single,framesep=3pt,breaklines=true,tabsize=2,linenos]{ini}
roi = [1, 1], [1, 120], [60, 120], [60, 1]
\end{minted}

Internally, a winding number algorithm is used to determine whether a certain local position is within or outside the given polynomial shape.
Two functions are provided by the detector API:

\begin{minted}[frame=single,framesep=3pt,breaklines=true,tabsize=2,linenos]{c++}
// Returns "true" if the track is found to be within the ROI
bool isWithinROI(const Track* track);
// Returns "true" if the cluster, as well as all its constituent pixels, are found to be within the ROI
bool isWithinROI(const Cluster* cluster);
\end{minted}

\section{Event Building}
\label{ch:events}

\corry implements a very flexible algorithm for offline event building that allows data to be combined from devices with different readout schemes.
This is possible via the concept of detector modules, which allows data to be processed from different detectors individually as described in Section~\ref{sec:module_manager}.
Events are processed sequentially as described in Chapter~\ref{ch:framework}

The following sections provide an introduction to event building and details the procedure using a few examples.

\subsection{The Order of the Event Loaders is Key}

When building events, it is important to carefully choose the order of the event loader modules.
An event loader module is defined as a module which reads an external data source and places \corry objects on the clipboard.
The first module to be run has to define the extent of the event by defining the \parameter{Event} object on the clipboard either through trigger numbers or a time window.

Apart from modules named \module{EventLoader<...>}, also the \module{Metronome} and \module{FileReader} modules are considered event loader modules since they read external data sources and/or define the clipboard \parameter{Event} object.

\begin{warning}
Once this event is set, no changes to its start and end time are possible, and no new event definition can be set by a subsequent module.
\end{warning}

Following modules in the reconstruction chain can only access the defined event and compare its extent in time or assigned trigger IDs to the currently processed data.
A special case are triggered devices that do not provide reference timestamps, as will be discussed in Section~\ref{sec:triggered_devices}.
The event is cleared at the end of the processing chain.

However, not all event loader modules are capable of defining an event.
Detectors that run in a data-driven mode usually just provide individual measurements together with a time stamp.
In order to slice this continuous data stream into consumable frames, the \module{Metronome} module can be used as described in Section~\ref{sec:metronome}.

The order of event loader modules can be explicitly specified by using multiple instances and assigning them to individual detectors as described in Section~\ref{sec:module_instantiation}:

\begin{minted}[frame=single,framesep=3pt,breaklines=true,tabsize=2,linenos]{ini}
# First process data from the detector named "CP01_W03"
[EventLoaderEUDAQ2]
name = "CP01_W03"

# Now process all detectors of type "Timepix3"
[EventLoaderEUDAQ2]
type = "Timepix3"

# Finally, process all remaining detectors in order along the z-axis
[EventLoaderEUDAQ2]
\end{minted}

Otherwise, the order in which the detectors are placed along the z-axis in the geometry is used.

\subsection{Position of Event Data: Before, During, or After?}

After the event has been defined, subsequent modules should compare their currently processed data to the defined event.
For this, the event object can be retrieved from the clipboard storage and the event data of the current detector can be tested against it using a set of member functions.
These functions return one of the following four possible positions:
\begin{description}
        \item[\textbf{\parameter{BEFORE}:}] The data currently under consideration are dated \emph{before} the event. Therefore they should be discarded and more data should be processed.
        \item[\textbf{\parameter{DURING}:}] The data currently under consideration are \emph{within} the defined event frame and should be taken into account. More data should be processed.
        \item[\textbf{\parameter{AFTER}:}] The data are dated \emph{after} the current event has finished. The processing of data for this detector should therefore be stopped and the current data retained for consideration in the next event.
        \item[\textbf{\parameter{UNKNOWN}:}] The event does not allow the position in time of the data to be determined. The data should be skipped and the next data should be processed until one of the above conditions can be reached.
\end{description}

Depending on what information is available from detector data, different member functions are available.

\subsubsection{Data-Driven Detectors}
If the data provides a single timestamp, such as the data from a data-driven detector, the \command{getTimestampPosition()} function can be used:

\begin{minted}[frame=single,framesep=3pt,breaklines=true,tabsize=2,linenos]{c++}
// Timestamp of the data currently under scrutiny:
double my_timestamp = 1234;
// Fetching the event definition
auto event = clipboard->getEvent();
// Comparison of the timestamp to the event:
auto position = event->getTimestampPosition(my_timestamp);
\end{minted}

Since an event \emph{always} has to define its beginning and end, this function cannot return an \parameter{UNKNOWN} position.

\subsubsection{Frame-Based Detectors}
If the data to be added are from a source that defines a time frame, such as frame-based or shutter-based detectors, there are two timestamps to consider.
The appropriate function for position comparison is called \command{getFramePosition()} and takes two timestamps for beginning and end of the data frame, as well as an additional flag to choose between the interpretation modes \emph{inclusive} and \emph{exclusive} (see below).
Several modules, such as the \command{EventLoaderEUDAQ2}, employ a configuration parameter to influence the matching behavior.

\paragraph{Inclusive Selection:}
The inclusive interpretation will return \parameter{DURING} as soon as there is some overlap between the frame and the event, i.e. as soon as the end of the frame is later than the event start or the frame start is before the event end.

If the event has been defined by a reference detector, this mode can be used for the DUT to make sure the data extends well beyond the devices providing the reference tracks.
This allows for the correct measurement of e.g.\ the efficiency without being biased by DUT data that lies outside the frame of the reference detector.

\paragraph{Exclusive Selection:}
In the exclusive mode, the frame will be classified as \parameter{DURING} only if the start and end are both within the defined event.
This mode could be used if the event is defined by the DUT itself.
In this case the reference data that is added to the event should not extend beyond this boundary but should only be considered if it is fully contained within the DUT event to avoid the creation of artificial inefficiencies.

The \command{getFramePosition()} takes the start and end times as well as the matching behavior flag:

\begin{minted}[frame=single,framesep=3pt,breaklines=true,tabsize=2,linenos]{c++}
// Timestamp of the data currently under scrutiny:
double my_frame_start = 1234, my_frame_stop = 1289;
// Fetching the event definition
auto event = clipboard->getEvent();
// Comparison of the frame to the event, in this case inclusive:
auto position = event->getFramePosition(my_frame_start, my_frame_stop, true);
\end{minted}

The function returns \parameter{UNKNOWN} if the end of the given time frame is before its start and therefore ill-defined.

\subsubsection{Triggered Devices Without Timestamp}
\label{sec:triggered_devices}

Many devices return data on the basis of external trigger decisions.
If the device also provides a timestamp, the data can be directly assigned to events based on the algorithms outlined above.

The situation becomes more problematic if the respective data only have the trigger ID or number assigned but should be combined with un-triggered devices that define their events based on timestamps only.

In order to process such data, a device that relates timestamps and trigger IDs, such as a trigger logic unit, is required.
This device should be placed before the detector in question in order to assign trigger IDs to the event using the \command{addTrigger(...)} function.
Then, the triggered device without timestamps can query the event for the position of its trigger ID with respect to the event using \command{getTriggerPosition()}:

\begin{minted}[frame=single,framesep=3pt,breaklines=true,tabsize=2,linenos]{c++}
// Trigger ID of the current data
uint32_t my_trigger_id = 1234;
// Fetching the event definition
auto event = clipboard->getEvent();
// Comparison of the trigger ID to the event
auto position = event->getTriggerPosition(my_trigger_id);
\end{minted}

If the given trigger ID is smaller than the smallest trigger ID known to the event, the function places the data as \parameter{BEFORE}.
If the trigger ID is larger than the largest know ID, the function returns \parameter{AFTER}. If the trigger ID is known to the event, the respective data is dated as being \parameter{DURING} the current event.
In cases where either no trigger ID has been previously added to the event or where the given ID lies between the smallest and largest known ID but is not part of the event, \parameter{UNKNOWN} is returned.

\subsection{The Metronome}
\label{sec:metronome}

In some cases, none of the available devices require a strict event definition such as a trigger or a frame.
This is sometimes the case when all or many of the involved devices have a data-driven readout.

\begin{figure}[tbp]
        \centering
        \includegraphics[width=1.0\textwidth]{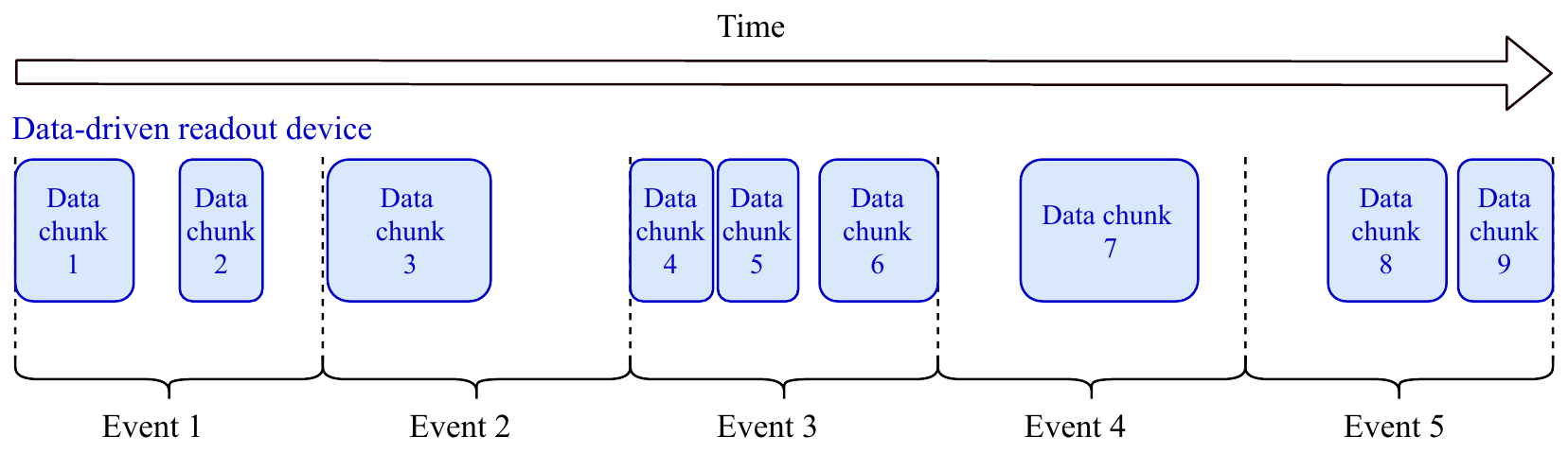}
        \caption{Event building with fixed-length time frames from the Metronome for data-driven detectors}
        \label{fig:datadriven}
\end{figure}

In this situation, the \module{Metronome} module can be used to slice the data stream into regular time frames with a defined length, as indicated in Figure~\ref{fig:datadriven}.

In addition to splitting the data stream into frames, trigger numbers can be added to each of the frames as described in the module documentation of the \module{Metronome} in Section~\ref{metronome}.
This can be used to process data exclusively recorded with triggered devices, where no timing information is available or necessary from any of the devices involved.
The module should then be set up to always add one trigger to each of its events, and all subsequent detectors will simply compare to this trigger number and add their event data.

If used, the \module{Metronome} module always has to be placed as first module in the data analysis chain because it will attempt to define the event and to add it to the clipboard.
This operation fails if an event has already been defined by a previous module.

\subsection{Example Configurations for Event Building}

In these examples, it is assumed that all data have been recorded using the \emph{EUDAQ2} framework, and that the \module{EventLoaderEUDAQ2} is used for all devices to read and decode their data.
However, this example pattern is not limited to that case and a very similar configuration could be used when the device data have been stored into device-specific native data files.

\subsubsection{Event Definition by Frame-Based DUT}
\label{sec:reco_mixedmode}
\begin{figure}[tbp]
        \centering
        \includegraphics[width=1.0\textwidth]{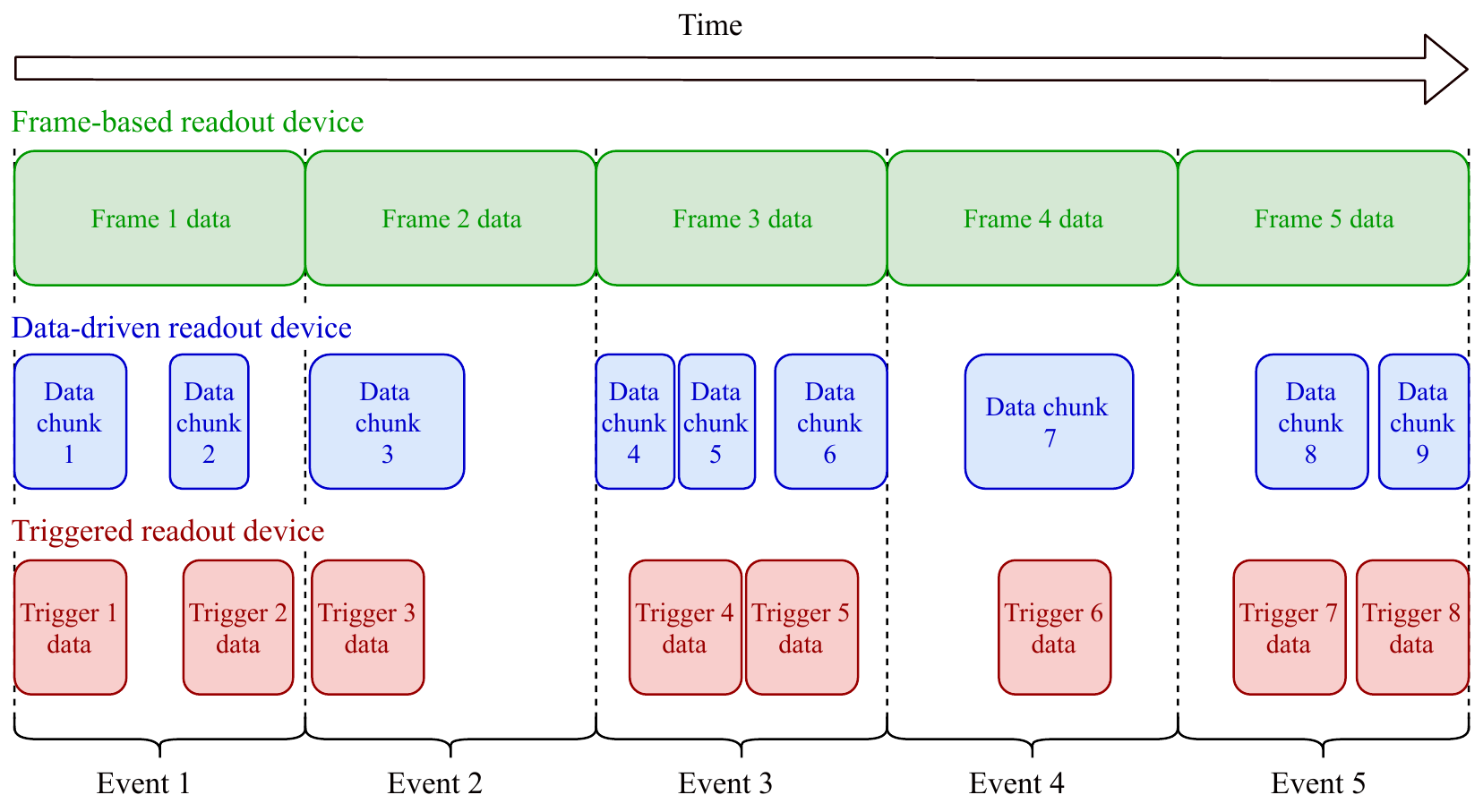}
        \caption{Event building strategy in a mixed-mode situation with devices adhering to different readout schemes. Here, the event is defined by the frame-based device and all other data are added within the event boundaries.}
        \label{fig:datadrivenandframebased}
\end{figure}

This example demonstrates the setup for event building based on a frame-based DUT.
The event building strategy for this situation is sketched in Figure~\ref{fig:datadrivenandframebased}.
The configuration contains four different devices:
\begin{description}
        \item[\parameter{CLICpix2}~\cite{clicpix2}:] This prototype acts as device under test and is a detector designed for operation at a linear collider, implementing a frame-based readout scheme. In this example, it will define the event.
        \item[\parameter{TLU}~\cite{aida-tlu}:] The trigger logic unit records scintillator coincidence signals with their respective timestamps, generates trigger signals and provides the reference clock for all other devices.
        \item[\parameter{Timepix3}~\cite{timepix3}:] This detector is operated in its data-driven mode, where the external clock from the TLU is used to timestamp each individual pixel hit. The hits are directly read out and sent to the data acquisition system in a continuous data stream.
        \item[\parameter{MIMOSA26}~\cite{mimosa26}:] These detectors are devices with a continuous rolling shutter readout, which do not record any timing information in the individual frames. The data are tagged with the incoming trigger IDs by its DAQ system.
\end{description}

In order to build proper events from these devices, the following configuration is used:

\begin{minted}[frame=single,framesep=3pt,breaklines=true,tabsize=2,linenos]{ini}
[Corryvreckan]
# ...

[EventLoaderEUDAQ2]
name = "CLICpix2_0"
file_name = /data/run001_file_caribou.raw

[EventLoaderEUDAQ2]
name = "TLU_0"
file_name = /data/run001_file_ni.raw

[EventLoaderEUDAQ2]
name = "Timepix3_0"
file_name = /data/run001_file_spidr.raw

[EventLoaderEUDAQ2]
type = "MIMOSA26"
file_name = /data/run001_file_ni.raw
\end{minted}

The first module will define the event using the frame start and end timestamps from the \parameter{CLICpix2} device.
Then, the data from the \parameter{TLU} are added by comparing the trigger timestamps to the event.
The matching trigger numbers are added to the event for later use.
Subsequently, the \parameter{Timepix3} device adds its data to the event by comparing the individual pixel timestamps to the existing event definition.
Finally, the six \parameter{MIMOSA26} detectors are added one-by-one via the automatic \parameter{type}-instantiation of detector modules described in Section~\ref{sec:module_instantiation}.
Here, the trigger numbers from the detector data are compared to the ones stored in the event as described in Section~\ref{sec:triggered_devices}.

It should be noted that in this example, the data from the \parameter{TLU} and the six \parameter{MIMOSA26} planes are read from the same file.
The event building algorithm is transparent to how the individual detector data are stored, and the very same building pattern could be used when storing the data in separate files.
In addition, it should be noted that swapping the order of the \parameter{Timepix3} and \parameter{MIMOSA26} detectors is also a valid configuration with identical results to the above scheme.

\subsubsection{Event Definition by Trigger Logic Unit}
This example demonstrates the setup for event building based on the trigger logic unit.
The configuration contains the same devices as the example in Section~\ref{sec:reco_mixedmode} but replaces the \parameter{CLICpix2} by the \parameter{ATLASpix}:
\begin{description}
        \item[\parameter{ATLASpix}~\cite{atlaspix}:] This prototype acts as device under test and is a detector initially designed for the ATLAS ITk upgrade, implementing a triggerless column-drain readout scheme.
\end{description}

In order to build proper events from these devices, the following configuration is used:

\begin{minted}[frame=single,framesep=3pt,breaklines=true,tabsize=2,linenos]{ini}
[Corryvreckan]
# ...

[EventLoaderEUDAQ2]
name = "TLU_0"
file_name = /data/run001_file_ni.raw
adjust_event_times = [["TluRawDataEvent", -115us, +230us]]

[EventLoaderEUDAQ2]
type = "MIMOSA26"
file_name = /data/run001_file_ni.raw

[EventLoaderEUDAQ2]
name = "Timepix3_0"
file_name = /data/run001_file_spidr.raw

[EventLoaderALTASpix]
name = "ATLASpix_0"
input_directory = /data/run001/
\end{minted}

The first module will define the event using the frame start and end timestamps from the \parameter{TLU} device.
As in the previous example, the \parameter{MIMOSA26} data will be added based on the trigger number.
The frame start provided by the \parameter{TLU} corresponds to the trigger timestamp and the frame end is one clockcycle later.
However, due to the rolling shutter readout scheme, the hits that are read out from the \parameter{MIMOSA26} planes when receiving a trigger may have happened in a time period before or after the trigger signal.
Consequently, the event times need to be corrected using the \parameter{adjust_event_times} parameter (described in Section~\ref{eventloadereudaq2}) such that the data from the \parameter{ATLASpix} and \parameter{Timepix3} can be added to the event in the entire time window in which the \parameter{MIMOSA26} hits may have occurred.

It should be noted that swapping the order of the \parameter{Timepix3}, \parameter{ATLASpix}, and \parameter{MIMOSA26} detectors is also a valid configuration with identical results to the above scheme.

\section{Using \corry as Online Monitor}

Reconstructing test beam data with \corry does not require many dependencies and is usually very fast due to its efficient data handling and fast reconstruction routines.
It is therefore possible to directly perform a full reconstruction including tracking and analysis of the DUT data during data taking.
On Linux machines, this is even possible on the data currently recorded since multiple read pointers are allowed per file.

The \corry framework comes with an online monitoring tool in form of a module for data quality monitoring and immediate feedback to the shifter.
The \module{OnlineMonitor} is a relatively simple graphical user interface that displays and updates histograms and graphs produced by other modules during the run.
It should therefore be placed at the very end of the analysis chain in order to have access to all histograms previously registered by other modules.

\begin{figure}[tbp]
        \centering
        \includegraphics[width=0.9\textwidth]{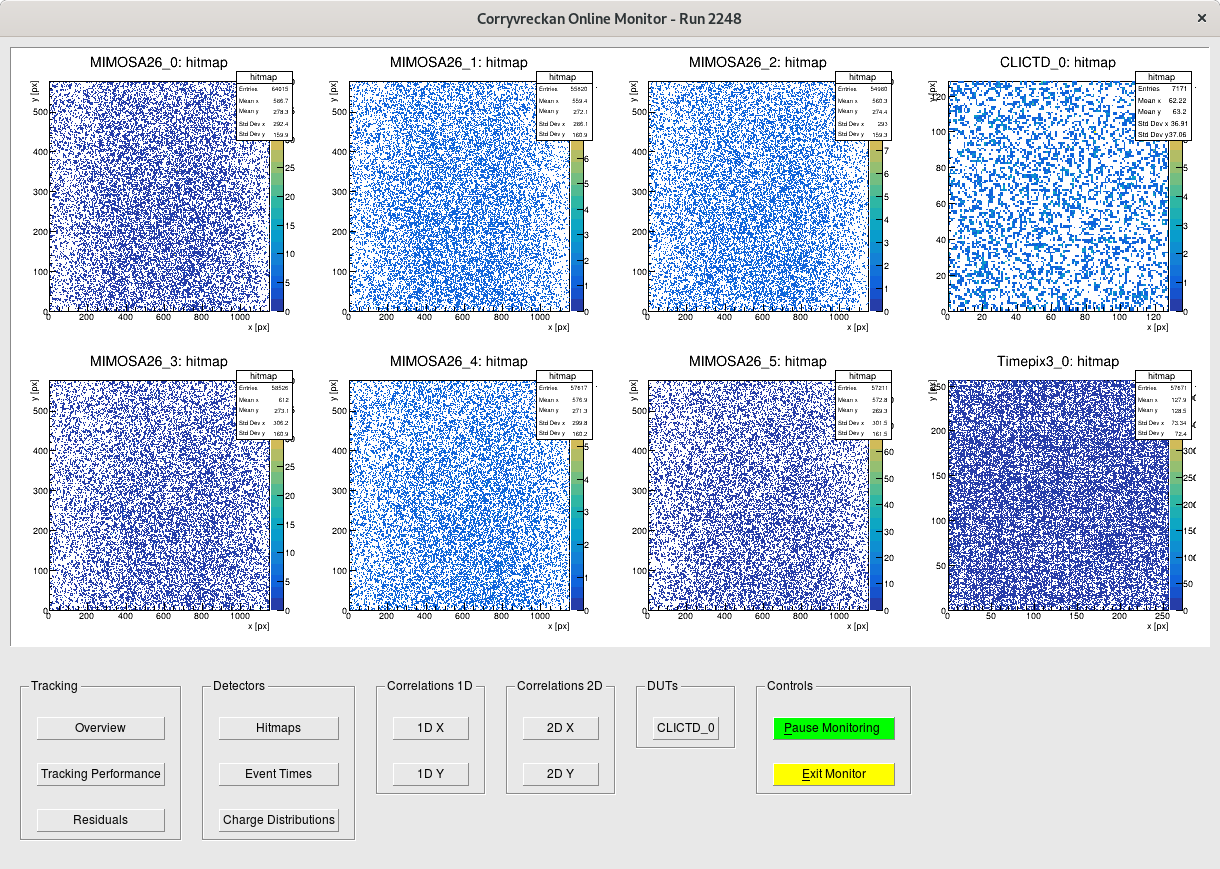}
        \caption{Screenshot of the OnlineMonitor module displaying reconstruction histograms during the \corry run.}
        \label{fig:onlinemon}
\end{figure}

A screenshot of the interface is displayed in Figure~\ref{fig:onlinemon}, showing histograms from the reconstruction of data recorded with the setup presented in Section~\ref{sec:reco_mixedmode}.
The histograms are distributed over several canvases according to their stage of production in the reconstruction chain.
It is possible to display histograms for all registered detectors through the \parameter{
Histograms only from all detectors marked as DUT can be added by placing \parameter{

The module has a default configuration that should match many reconstruction configurations, but each of the canvases and histograms can be freely configured as described in the documentation of the \module{OnlineMonitor} in Section~\ref{onlinemonitor}.

\section{Alignment Procedure}
\label{ch:howtoalign}
As decribed in Section~\ref{sec:detector_config}, an analysis with \corry requires a configuration file defining which detectors are present in the setup.
This file also contains the position and rotation of each detector plane.
The Z-positions of all planes can \textbf{and must} be measured by hand in the existing test beam setup and entered in this configuration file for the analysis.
The X- and Y-positions as well as the rotations cannot be measured precisely by hand.
However, these have a strong influence on the tracking since a misalignment of a fraction of a millimeter might already correspond to a shift by multiple pixel pitches.\\

Consequently, an alignment procedure is needed in which the detector planes are shifted and rotated iteratively relative to the detector with \parameter{role = reference} to increase the tracking quality.
More technically, the track residuals on all planes,~i.e. the distribution of the spatial distance between the interpolated track intercept and the associated cluster on the plane need to be centered around zero and 'as narrow as possible' -- the width of the distribution depends on the tracking resolution of the telescope and is influenced by many factors such as the beam energy, the material budget of the detector planes, the distance between the detector planes, etc.
It is important to correctly set the \parameter{spatial_resolution} specified in the detector configuration file described in Section~\ref{sec:detector_config} because is defining the uncertainty on the cluster positions and therefore influences the track $\chi^2$.\\

This chapter provides a description of how to use the alignment features of \corry.
It also includes step-by-step instructions on how to align the detector planes for new set of test beam data.
Example configuration files can be found in the \dir{testing/} directory of the repository.
These are based on a Timepix3~\cite{timepix3} telescope with an ATLASpix~\cite{atlaspix} DUT at the CERN SPS with a pion beam of \SI{120}{GeV}.\\

For the alignment of the \textbf{reference telescope} and \textbf{device-under-test (DUT)}, the following modules are available in \corry.
\begin{itemize}
\item \module{Prealignment} for both telescope and DUT prealignment (see Section~\ref{prealignment}).
\item \module{AlignmentTrackChi2} used for telescope alignment (see Section~\ref{alignmenttrackchi2}) and is relatively robust against an initial misalignment but usually needs several iterations.
\item \module{AlignmentMillepede}, an alternative telescope alignment algorithm (see Section~\ref{alignmentmillepede}) which requires fewer iterations to reach a precise alignment but needs a better prealignment.
\item \module{AlignmentDUTResidual} used for DUT alignment (see Section~\ref{alignmentdutresidual}).
\end{itemize}

The general procedure that needs to be followed for a successful alignment is outlined here and explained in detail below.
\begin{enumerate}
\item Prealignment of the telescope (ignoring the DUT).
\item Alignment of the telescope (ignoring the DUT).
\item Prealignment of the DUT (telescope geometry is frozen).
\item Alignment of the DUT (telescope alignment is frozen).
\end{enumerate}

\begin{warning}
When using the alignment modules, the new geometry is written out to a new geometry file which needs to be specified using the parameter \parameter{detectors_file_updated}.
For details, see Section~\ref{sec:framework_parameters}.
\end{warning}

\paragraph{Correlation vs. Residual}

A spatial \textbf{correlation} plot is filled with the spatial difference of any cluster on a given detector plane minus the position of any cluster on the reference plane. No tracking is required to fill these histograms.

A spatial \textbf{residual} plot shows the difference of the interpolated track intercept onto a given plane minus the position of its associated cluster.\\

Consequently, the goal of the alignment is to force the \textbf{residuals} to be centered around zero.
The \textbf{correlations} do not necessarily have to be centered at zero as a possible offset reflects the \emph{physical displacement} of a detector plane in X and Y with respect to the reference plane.
However, it can be useful to inspect the \textbf{correlation} plots especially in the beginning when the alignment is not yet good enough for a reasonable tracking.

\subsection{Aligning the Telescope}
\label{sec:align_tel}
Initially, the telescope needs to be aligned.
For this, the DUT is ignored.

\subsubsection*{Prealignment of the Telescope}
The \module{AlignmentTrackChi2} module requires a careful prealignment. Otherwise it does not converge and the alignment will fail.
The Z-positions of all planes need to be measured by hand \textbf{in the existing test beam setup} and then adjusted in the detectors file.
For X and Y, the alignment file from an already aligned run with the same telescope plane arrangement is a solid basis to start from.
If no previous alignment is available, all values for X and Y should be set to 0.

For the prealignment, two strategies can be applied:
\begin{itemize}
\item The \module{Prealignment} module can be used (see Section~\ref{prealignment}).
\item If the above does not bring the expected result, a manual prealignment can be performed as described below.
\end{itemize}

To have a first look at the initial alignment guess, one can run
\begin{verbatim}
$ /path/to/corryvreckan/bin/corry                     \
    -c analyse_telescope.conf                         \
   [-o detectors_file=<detectorsFile>                 \
    -o histogram_file=<histogramFile>                 \
    -o EventLoaderTimepix3.input_directory=<inputDir>]
\end{verbatim}

The \parameter{spatial_cut_abs/rel} in \module{Tracking4D} should be set to multiple ($\sim4$) pixel pitch.

One can inspect the spatial correlations in X and Ythe track $\chi^2$, and the residuals with the online monitoring or by opening the generated ROOT file after finishing the script.
These can be found in the modules \module{Correlations} (see Section~\ref{correlations}) and \module{Tracking4D} (see Section~\ref{tracking4d}).\\

To save time, one can limit the number of processed tracks. For instance, set \parameter{number_of_events = 10000} or \parameter{number_of_tracks = 10000} (see Section~\ref{sec:framework_parameters}).\\

If no peak at all is apparent in the correlations or residuals, the hitmaps can be checked to see if valid data is actually available for all planes.\\

Now, the \texttt{[Prealignment]} module can be used.
To prealign only the telescope, the DUT can be excluded by using \parameter{type = <detector_type_of_telescope>} (e.g.~\parameter{CLICPIX2}). For details, see Section~\ref{sec:module_manager}.

To use the module, \file{align_tel.conf} needs to be edited such that \texttt{[Prealignment]} is enabled and \texttt{[Alignment]} is disabled:
\begin{minted}[frame=single,framesep=3pt,breaklines=true,tabsize=2,linenos]{ini}
...
[Prealignment]
type = <detector_type_of_telescope> # <-- optional!
[Ignore]
#[AlignmentTrackChi2]
log_level=INFO
iterations = 4
align_orientation=true
align_position=true
\end{minted}

Then one can run
\begin{verbatim}
$ /path/to/corryvreckan/bin/corry                      \
    -c align_telescope.conf                            \
   [-o detectors_file=<detectorsFile>                  \
    -o detectors_file_updated=<detectorsFileUpdated>   \
    -o histogram_file=<histogramFile>                  \
    -o EventLoaderTimepix3.input_directory=<inputDir>]
\end{verbatim}

The actual prealignment is only performed after the events have been analyzed and written to the detectors file in the finalizing step.
This means to check whether the alignment has improved, one needs to re-run the analysis or the next iteration of the alignment as the previously generated ROOT file corresponds to the initial alignment.
This is the case for every iteration of the prealignment or alignment.

Generally, it suffices to run the \texttt{[Prealignment]} module once and then proceed with the next step.

\subsubsection*{Manual Prealignment of the Telescope}
If the prealignment using the module \texttt{[Prealignment]} does not bring the expected results, one can also perform the same steps manually by investigating the residuals of the DUT with respect to tracks.
For the residuals, the shift of the peak from 0 can be estimated with a precision of $\mathcal{O}(\SI{100}{\micro m})$ by zooming in using the \texttt{TBrowser}.
For instance, if the peak is shifted by +\SI{+300}{\micro m}, the detectors file needs to be edited and \SI{+300}{\micro m} should be added to the respective position, if \SI{-300}{\micro m}, subtracted.

After modifying the positions of individual planes in the configuration file, \corry can be re-run to check the correlation and residual plots for the updated geometry.
These steps need to be iterated a few times until the peaks of the \textbf{residuals} are centered around 0.

Rotational misalignments can be inferred from the slope of the 2D spatial correlation plots, the actual rotation angle has to be calculated using the respective pixel pitches of the devices.

\begin{warning}
It is important \textbf{not} to force the peak of the spatial \textbf{correlations} to be at exactly 0 because the position of the peak corresponds to the \textit{physical displacement} of a detector plane in X and Y with respect to the reference plane.
The spatial \textbf{correlations} should \textbf{only be used} if the spatial \textbf{residual} plots are not filled reasonable due to bad tracking.
Hence, the spatial correlations can be shifted towards zero in a first iteration.
\end{warning}

\subsubsection*{Alignment of the Telescope}
After the prealignment, the actual \textbf{precise} alignment can be performed using the \texttt{[AlignmentTrackChi2]} module (see Section~\ref{alignmenttrackchi2}).
To this end, \file{align_tel.conf} needs to be modified such that the prealignment is disabled and the alignment is enabled:
\begin{minted}[frame=single,framesep=3pt,breaklines=true,tabsize=2,linenos]{ini}
...
#[Prealignment]
#[Ignore]
[AlignmentTrackChi2]
log_level=INFO
iterations = 4
align_orientation=true
align_position=true
\end{minted}

The algorithm performs an optimisation of the track $\chi^2$.
Typically, the alignment needs to be iterated a handful of times until the residuals (which again can be inspected in the ROOT file after re-running the analysis) are nicely centered around 0 and 'as narrow as possible' -- the RMS of the residuals corresponds to the spatial resolution of each plane (convolved with the resolution of the telescope) and should thus be $\lesssim$ pixel pitch$/\sqrt{12}$.
Starting with a \parameter{spatial_cut_abs/rel} in \texttt{[Tracking4D]} (see Section~\ref{tracking4d}) of multiple ($\sim4$) pixel pitches, it should be decreased incrementally down to the pixel pitch (e.g. run \SI{200}{\micro\m} twice, then run \SI{150}{\micro\m} twice, then \SI{100}{\micro\m} twice, and then \SI{50}{\micro\m} twice).
This allows to perform the alignment with a tight selection of very high quality tracks only.
Also the \parameter{max_track_chi2ndof} should be decreased for the same reason.
For the further analysis, the cuts can be released again.

It may happen that the procedure runs into a 'false minimum', i.e. it converges in a wrong alignment in which the residuals are clearly not centered around 0.
In this case, it is required to go one step back and improve the prealignment.

Once the alignment is done, one should obtain narrow residuals centered around 0 and a good distribution of the track $\chi^2$ as shown in Figures~\ref{fig:exampleAlignment}.
If the alignment keeps to fail, it is possible to allow only for rotational or translational alignment while freezing the other for one or a few iterations.

\begin{minted}[frame=single,framesep=3pt,breaklines=true,tabsize=2,linenos]{ini}
...
#[Prealignment]
#[Ignore]
[AlignmentTrackChi2]
log_level=INFO
iterations = 4
align_orientation=false #<-- disable rotational alignment
align_position=true
\end{minted}

\begin{figure}
    \centering
    \begin{subfigure}[t]{0.66\textwidth}
        \includegraphics[width=\textwidth]{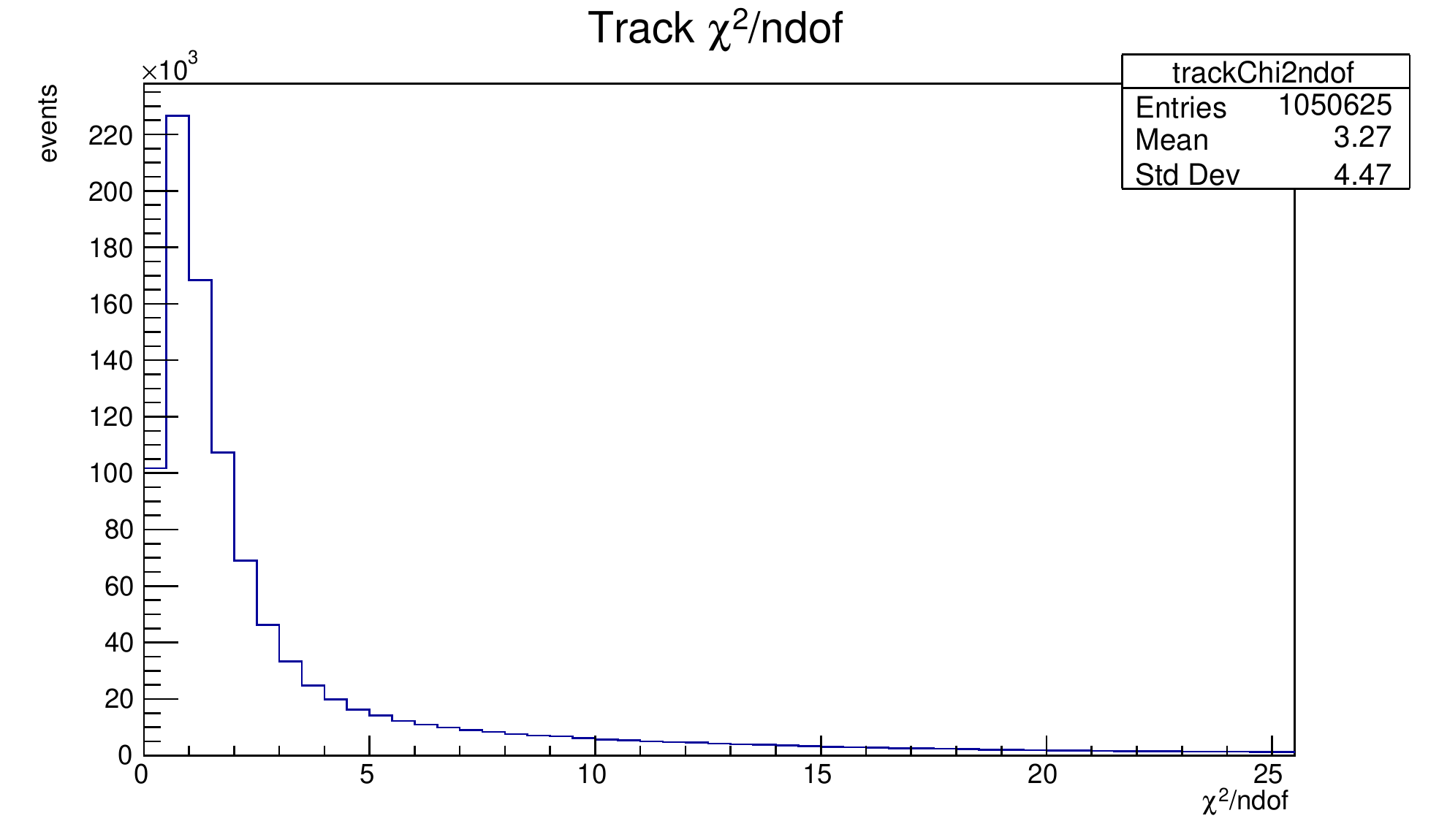}
        \caption{Good example of a track $\chi^2$/ndf distribution.}
        \label{fig:trackChi2}
    \end{subfigure}
    \begin{subfigure}[t]{0.66\textwidth}
        \includegraphics[width=\textwidth]{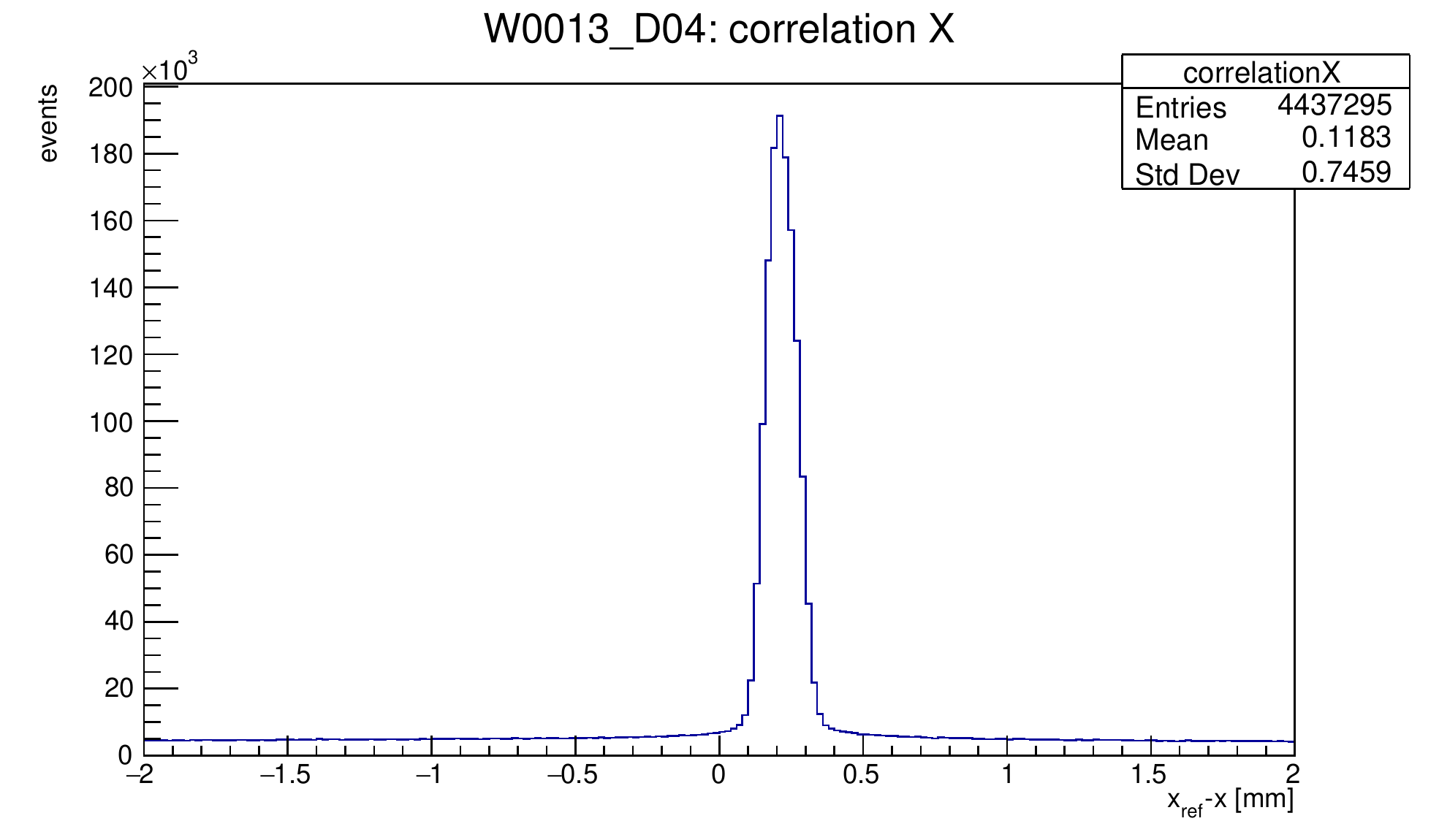}
        \caption{Good example of a spatial correlation plot between two telescope planes. The offset from zero corresponds to the \emph{physical displacement} of the plane  with respect to the reference plane.}
        \label{fig:correlationX}
    \end{subfigure}
    \begin{subfigure}[t]{0.66\textwidth}
        \includegraphics[width=\textwidth]{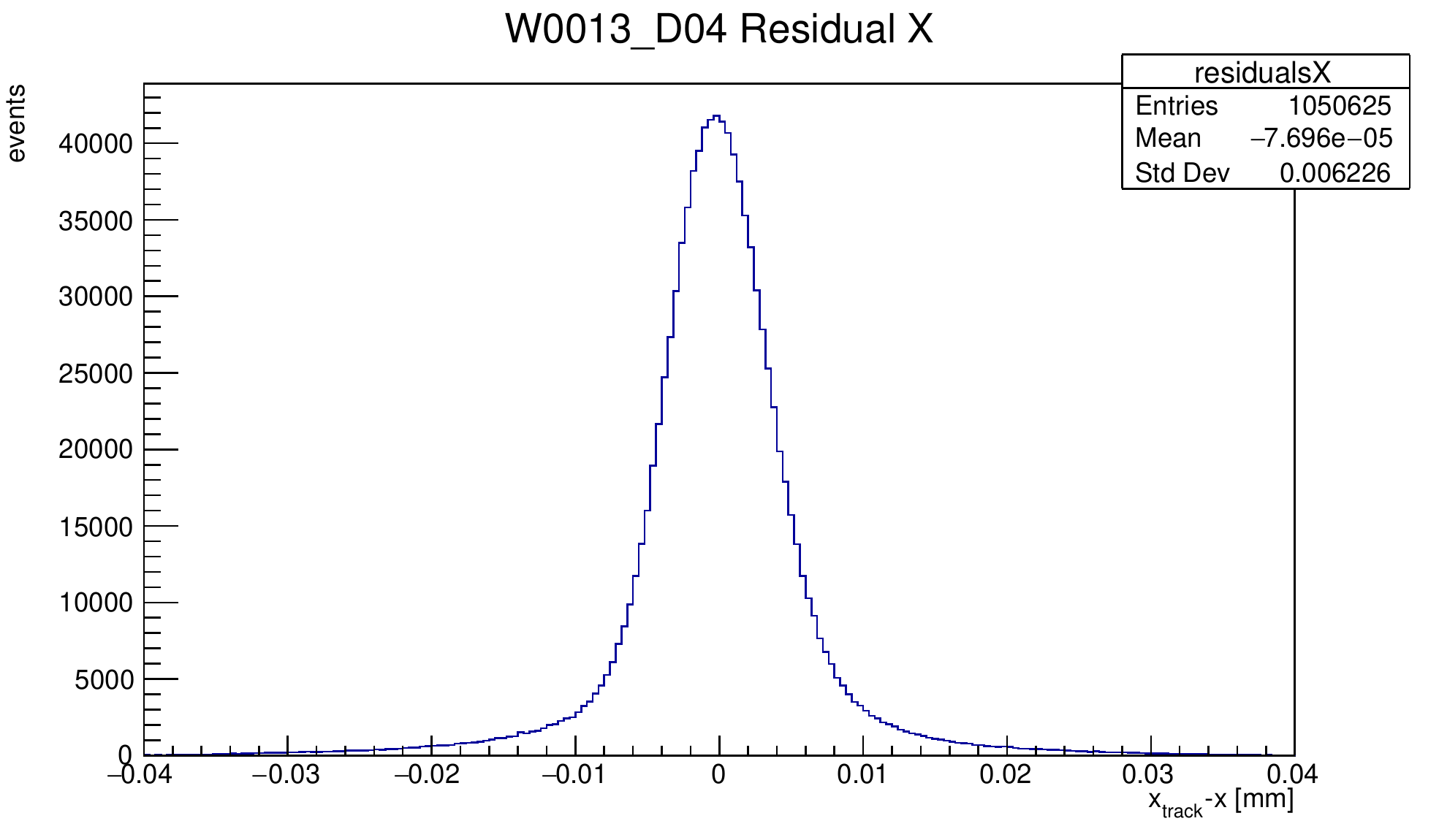}
        \caption{Good example of a spatial residual distribution. It is centered around zero.}
        \label{fig:residualX}
    \end{subfigure}
    \caption{Examples of distributions how they should look after a successful alignment of the Timepix3 telescope at the CERN SPS with \SI{120}{\GeV} pions.}
    \label{fig:exampleAlignment}
\end{figure}

Instead of using \texttt{[AlignmentTrackChi2]}, one can also use the module \texttt{[AlignmentMillepede]} (see Section~\ref{alignmentmillepede}).
It allows a simultaneous fit of both the tracks and the alignment constants.
The modules stops if the convergence, i.e.~the absolute sum of all corrections over the total number of parameters, is smaller than the configured value, and the aligment is complete.
It should be noted that this module requires a rather good prealignment already.

\subsection{Aligning the DUT}
\label{sec:align_dut}
Once the telescope is aligned, its geometry is not changed anymore. From now on, it is used to build tracks which are then matched to clusters on the DUT.

\subsubsection*{Prealignment of the DUT}
The prealignment of the DUT follows the same strategy as for the telescope. To look at the current alignment, the script
\begin{verbatim}
$ /path/to/corryvreckan/bin/corry                          \
    -c analyse_atlaspix.conf                               \
   [-o detectors_file=<detectorsFile>                      \
    -o histogram_file=<histogramFile>                      \
    -o EventLoaderTimepix3.input_directory=<inputDir_TPX>  \
    -o EventLoaderATLASpix.input_directory=<inputDir_APX>]
\end{verbatim}
needs to be run.
If no better guess is available, the initial alignment of the DUT should be set to $x=y=0$.

Then, by repeatedly running \corry and modifying the position of the DUT in the detectors file one should be able to bring the peaks of the spatial residuals in X and Y close to zero.
If no peak at all can be seen in the residual plots, the spatial correlations plots can be inspected. In addition, potentially parameters related to the corresponding event loader need to be corrected in the configuration file.

\begin{warning}
If using the \texttt{[Prealignment]} module, it is possible to prealign all planes at once as described above in Section~\ref{sec:align_tel}.
If only the DUT shall be prealigned here, the parameter \parameter{name = <name_of_dut>} or \parameter{type = <detector_type_of_dut>} need to be used.
Otherwise, the telescope planes are also shifted again destroying the telescope alignment.
\end{warning}

\begin{minted}[frame=single,framesep=3pt,breaklines=true,tabsize=2,linenos]{ini}
...
[Prealignment]
name = <name_of_dut> # <-- otherwise the telescope planes will be moved!

[Ignore]
#[AlignmentDUTResiduals]
log_level=INFO
iterations = 4
align_orientation=true
align_position=true
\end{minted}

\subsubsection*{Alignment of the DUT}
The alignment strategy for the DUT is similar as for the telescope and requires multiple iterations.
In \file{align_dut.conf}, the prealignment needs to be disabled and the alignment enabled.
Now, the algorithm optimizes the residuals of the tracks through the DUT.

\begin{minted}[frame=single,framesep=3pt,breaklines=true,tabsize=2,linenos]{ini}
...
#[Prealignment]
#[Ignore]
[AlignmentDUTResiduals]
log_level=INFO
iterations = 4
align_orientation=true
align_position=true
\end{minted}

Then run
\begin{verbatim}
$ /path/to/corryvreckan/bin/corry                          \
    -c align_dut.conf                                      \
   [-o detectors_file=<detectorsFile>                      \
    -o detectors_file_updated=<detectorsFileUpdated>       \
    -o histogram_file=<histogramFile>                      \
    -o EventLoaderTimepix3.input_directory=<inputDir_TPX>  \
    -o EventLoaderATLASpix.input_directory=<inputDir_APX>]
\end{verbatim}

Like for the telescope alignment, the RMS of the residuals can be interpreted as the spatial resolution of the DUT (convolved with the track resolution of the telescope at the position if the DUT) and should thus be $\lesssim$~pixel pitch$/\sqrt{12}$.
Again, starting with a \parameter{spatial_cut_abs/rel} in \texttt{[DUTAssociation]} (see Section~\ref{dutassociation}) of multiple ($\sim4$) pixel pitches, it should be decreased incrementally down to the pixel pitch. Note that an asymmetric pixel geometry requires the \parameter{spatial_cut_abs/rel} to be chosen accordingly.

If the alignment keeps to fail, it is possible to allow only for rotational or translational alignment while freezing the other for one or a few iterations.

\begin{minted}[frame=single,framesep=3pt,breaklines=true,tabsize=2,linenos]{ini}
...
#[Prealignment]
#[Ignore]
[AlignmentDUTResiduals]
log_level=INFO
iterations = 4
align_orientation=false #<-- disable rotational alignment
align_position=true
\end{minted}

\section{Modules}
\label{ch:modules}

This chapter describes the currently available \corry modules in detail.
It comprises a description of the implemented modules as well as possible configuration parameters along with their defaults.
Furthermore, an overview of output plots is provided.
For inquiries about certain modules or their documentation, the \corry issue tracker which can be found in the repository~\cite{corry-issue-tracker} should be used as described in Section~\ref{sub:contributing}.
The modules are listed in alphabetical order.

\lstset{language=Ini}
\includemodulesmd
\lstset{language=}

\section{Extending the \corry Framework}

This chapter provides some initial information for developers planning on extending the \corry framework.
\corry is a community project that benefits from active participation in the development and code contributions from users.
Users are encouraged to discuss their needs via the issue tracker of the repository~\cite{corry-issue-tracker} to receive ideas and guidance on how to implement a specific feature.
Getting in touch with other developers early in the development cycle avoids spending time on features which already exist or are currently under development by other users.

The repository contains a few tools to facilitate contributions and to ensure code quality as detailed in Chapter~\ref{ch:testing}.

\subsection{Writing Additional Modules}

Given the modular structure of the framework, its functionality can be easily extended by adding a new module.
To facilitate the creation of new modules including their CMake files and initial documentation, the script \file{addModule.sh} is provided in the \dir{etc/} directory of the repository.
It will ask for a name and type of the module as described in Section~\ref{sec:module_manager} and create all code necessary to compile a first (and empty) version of the files.

The content of each of the files is described in detail in the following paragraphs.

\subsubsection{Files of a Module}
\label{sec:module_files}
Every module directory should at minimum contain the following documents (with \texttt{<ModuleName>} replaced by the name of the module):
\begin{itemize}
\item \textbf{\file{CMakeLists.txt}}: The build script to load the dependencies and define the source files of the library.
\item \textbf{\file{README.md}}: Full documentation of the module.
\item \textbf{\file{<ModuleName>.h}}: The header file of the module.
\item \textbf{\file{<ModuleName>.cpp}}: The implementation file of the module.
\end{itemize}
These files are discussed in more detail below.
By default, all modules added to the \dir{src/modules/} directory will be built automatically by CMake.
If a module depends on additional packages which not every user may have installed, one can consider adding the following line to the top of the module's \file{CMakeLists.txt}:
\begin{minted}[frame=single,framesep=3pt,breaklines=true,tabsize=2,linenos]{cmake}
CORRYVRECKAN_ENABLE_DEFAULT(OFF)
\end{minted}

\paragraph{CMakeLists.txt}
Contains the build description of the module with the following components:
\begin{enumerate}
\item On the first line either \parameter{CORRYVRECKAN_DETECTOR_MODULE(MODULE_NAME)}, \parameter{CORRYVRECKAN_DUT_MODULE(MODULE_NAME)} or \parameter{CORRYVRECKAN_GLOBAL_MODULE(MODULE_NAME)} depending on the type of module defined.
The internal name of the module is automatically saved in the variable \parameter{${MODULE_NAME}} which should be used as an argument to other functions.
Another name can be used by overwriting the variable content, but in the examples below, \parameter{${MODULE_NAME}} is used exclusively and is the preferred method of implementation.
For DUT and Detector modules, the type of detector this module is capable of handling can be specified by adding so-called type restrictions, e.g.\
\begin{minted}[frame=single,framesep=3pt,breaklines=true,tabsize=2,linenos]{cmake}
CORRYVRECKAN_DETECTOR_TYPE(${MODULE_NAME} "Timepix3" "CLICpix2")
\end{minted}
The module will then only be instantiated for detectors of one of the given types. This is particularly useful for event loader modules which read a very specific file format.

\item The following lines should contain the logic to load possible dependencies of the module (below is an example to load Geant4).
Only ROOT is automatically included and linked to the module.
\item A line with \texttt{\textbf{CORRYVRECKAN\_MODULE\_SOURCES(\$\{MODULE\_NAME\} \textit{sources})}} defines the module source files. Here, \texttt{sources} should be replaced by a list of all source files relevant to this module.
\item Possible lines to include additional directories and to link libraries for dependencies loaded earlier.
\item A line containing \parameter{CORRYVRECKAN_MODULE_INSTALL(${MODULE_NAME})} to set up the required target for the module to be installed to.
\end{enumerate}

A simple \file{CMakeLists.txt} for a module named \parameter{Test} which should run only on DUT detectors of type \emph{Timepix3} is provided below as an example.
\vspace{5pt}

\begin{minted}[frame=single,framesep=3pt,breaklines=true,tabsize=2,linenos]{cmake}
# Define module and save name to MODULE_NAME
CORRYVRECKAN_DUT_MODULE(MODULE_NAME)
CORRYVRECKAN_DETECTOR_TYPE(${MODULE_NAME} "Timepix3")

# Add the sources for this module
CORRYVRECKAN_MODULE_SOURCES(${MODULE_NAME}
    Test.cpp
)

# Provide standard install target
CORRYVRECKAN_MODULE_INSTALL(${MODULE_NAME})
\end{minted}

\paragraph{README.md}
The \file{README.md} serves as the documentation for the module and should be written in Markdown format~\cite{markdown}.
It is automatically converted to \LaTeX~using Pandoc~\cite{pandoc} and included in the user manual in Chapter~\ref{ch:modules}.
By documenting the module functionality in Markdown, the information is also viewable with a web browser in the repository within the module sub-folder.

The \file{README.md} should follow the structure indicated in the \file{README.md} file of the \parameter{Dummy} module in \dir{src/modules/Dummy}, and should contain at least the following sections:
\begin{itemize}
\item The H1-size header with the name of the module and at least the following required elements: the \textbf{Maintainer}, the \textbf{Module Type} and the \textbf{Status} of the module.
The module type should be either \textbf{GLOBAL}, \textbf{DETECTOR}, \textbf{DUT}.
If the module is working and well-tested, the status of the module should be \textbf{Functional}.
By default, new modules are given the status \textbf{Immature}.
The maintainer should mention the full name of the module maintainer, with their email address in parentheses.
A minimal header is therefore:
\begin{verbatim}
# ModuleName
Maintainer: Example Author (<example@example.org>)
Module Type: GLOBAL
Status: Functional
\end{verbatim}
In addition, the \textbf{Detector Type} should be mentioned for modules of types \textbf{DETECTOR} and \textbf{DUT}.
\item An H3-size section named \textbf{Description}, containing a short description of the module.
\item An H3-size section named \textbf{Parameters}, with all available configuration parameters of the module.
The parameters should be briefly explained in an itemised list with the name of the parameter set as an inline code block.
\item An H3-size section named \textbf{Plots Created}, listing all plots created by this module.
\item An H3-size section with the title \textbf{Usage} which should contain at least one simple example of a valid configuration for the module.
\end{itemize}

\paragraph{ModuleName.h and ModuleName.cpp}
All modules should consist of both a header file and a source file.
In the header file, the module is defined together with all of its methods.
Doxygen documentation should be added to explain what each method does.
The source file should provide the implementation of every method.
Methods should only be declared in the header and defined in the source file in order to keep the interface clean.

\subsubsection{Module structure}
\label{sec:module_structure}
All modules must inherit from the \parameter{Module} base class, which can be found in \dir{src/core/module/Module.hpp}.
The module base class provides two base constructors, a few convenient methods and several methods which the user is required to override.
Each module should provide a constructor using the fixed set of arguments defined by the framework; this particular constructor is always called during by the module instantiation logic.
These arguments for the constructor differ for global and detector/DUT modules.

For global modules, the constructor for a \module{TestModule} should be:
\begin{minted}[frame=single,framesep=3pt,breaklines=true,tabsize=2,linenos]{c++}
TestModule(Configuration& config, std::vector<std::shared_ptr<Detector>> detectors): Module(std::move(config), detectors) {}
\end{minted}

For detector and DUT modules, the first argument are the same, but the last argument is a \texttt{std::shared\_ptr} to the linked detector.
It should always forward this detector to the base class together with the configuration object.
Thus, the constructor of a detector module is:
\begin{minted}[frame=single,framesep=3pt,breaklines=true,tabsize=2,linenos]{c++}
TestModule(Configuration& config, std::shared_ptr<Detector> detector): Module(std::move(config), std::move(detector)) {}
\end{minted}

In addition to the constructor, each module can override the following methods:
\begin{itemize}
\item \parameter{initialise()}: Called after loading and constructing all modules and before starting the analysis loop.
This method can for example be used to initialize histograms.
\item \parameter{run(std::shared_ptr<Clipboard> clipboard)}: Called for every time frame or triggered event to be analyzed. The argument represents a pointer to the clipboard where the event data is stored.
A status code is returned to signal the framework whether to continue processing data or to end the run.
\item \parameter{finalise()}: Called after processing all events in the run and before destructing the module.
Typically used to summarize statistics like the number of tracks used in the analysis or analysis results like the chip efficiency.
Any exceptions should be thrown from here instead of the destructor.
\end{itemize}

\section{Development Tools \& Continuous Integration}
\label{ch:testing}

The following chapter will introduce a few tools included in the framework to ease development and help to maintain a high code quality. This comprises tools for the developer to be used while coding, as well as a continuous integration (CI) and automated test cases of various framework and module functionalities.

\subsection{Additional Targets}
\label{sec:targets}

A set of testing targets in addition to the standard compilation targets are automatically created by CMake to enable additional code quality checks and testing.
Some of these targets are used by the project's CI, others are intended for manual checks.
Currently, the following targets are provided:

\begin{description}
  \item[\command{make format}] invokes the \command{clang-format} tool to apply the project's coding style convention to all files of the code base. The format is defined in the \file{.clang-format} file in the root directory of the repository and mostly follows the suggestions defined by the standard LLVM style with minor modifications. Most notably are the consistent usage of four whitespace characters as indentation and the column limit of 125 characters.
  This can be further simplified by installing the \emph{git hook} provided in the directory \dir{/etc/git-hooks/}. A hook is a script called by \command{git} before a certain action. In this case, it is a pre-commit hook which automatically runs \command{clang-format} in the background and offers to update the formatting of the code to be committed. It can be installed by calling
  \begin{minted}[frame=single,framesep=3pt,breaklines=true,tabsize=2,linenos]{bash}
  ./etc/git-hooks/install-hooks.sh
  \end{minted}
  once.
  \item[\command{make check-format}] also invokes the \command{clang-format} tool but does not apply the required changes to the code. Instead, it returns an exit code 0 (pass) if no changes are necessary and exit code 1 (fail) if changes are to be applied. This is used by the CI.
  \item[\command{make lint}] invokes the \command{clang-tidy} tool to provide additional linting of the source code. The tool tries to detect possible errors (and thus potential bugs), dangerous constructs (such as uninitialized variables) as well as stylistic errors. In addition, it ensures proper usage of modern \CPP standards. The configuration used for the \command{clang-tidy} command can be found in the \file{.clang-tidy} file in the root directory of the repository.
  \item[\command{make check-lint}] also invokes the \command{clang-tidy} tool but does not report the issues found while parsing the code. Instead, it returns an exit code 0 (pass) if no errors have been produced and exit code 1 (fail) if issues are present. This is used by the CI.
  \item[\command{make cppcheck}] runs the \command{cppcheck} command for additional static code analysis. The output is stored in the file \file{cppcheck_results.xml} in XML2.0 format. It should be noted that some of the issues reported by the tool are to be considered false positives.
  \item[\command{make cppcheck-html}] compiles a HTML report from the defects list gathered by \command{make cppcheck}. This target is only available if the \command{cppcheck-htmlreport} executable is found in the \dir{PATH}.
  \item[\command{make package}] creates a binary release tarball as described in Section~\ref{sec:packaging}.
\end{description}

\subsection{Packaging}
\label{sec:packaging}
\corry comes with a basic configuration to generate tarballs from the compiled binaries using the CPack command. In order to generate a working tarball from the current \corry build, the \parameter{RPATH} of the executable should not be set, otherwise the \command{corry} binary will not be able to locate the dynamic libraries. If not set, the global \parameter{LD_LIBRARY_PATH} is used to search for the required libraries:

\begin{verbatim}
$ mkdir build
$ cd build
$ cmake -DCMAKE_SKIP_RPATH=ON ..
$ make package
\end{verbatim}

The content of the produced tarball can be extracted to any location of the file system, but requires the ROOT6 and Geant4 libraries as well as possibly additional libraries linked by individual at runtime.

For this purpose, a \file{setup.sh} shell script is automatically generated and added to the tarball.
By default, it contains the ROOT6 path used for the compilation of the binaries.
Additional dependencies, either library paths or shell scripts to be sourced, can be added via CMake for individual modules using the CMake functions described below.
The paths stored correspond to the dependencies used at compile time, it might be necessary to change them manually when deploying on a different computer.

\paragraph{\texttt{\textbf{ADD\_RUNTIME\_DEP(name)}}}

This CMake command can be used to add a shell script to be sourced to the setup file.
The mandatory argument \parameter{name} can either be an absolute path to the corresponding file, or only the file name when located in a search path known to CMake, for example:

\begin{minted}[frame=single,framesep=3pt,breaklines=true,tabsize=2,linenos]{cmake}
# Add "thisroot.sh" of the ROOT framework as runtime dependency for setup.sh file:
ADD_RUNTIME_DEP(thisroot.sh)
\end{minted}

The command uses the \command{GET_FILENAME_COMPONENT} command of CMake with the \parameter{PROGRAM} option.
Duplicates are removed from the list automatically.
Each file found will be written to the setup file as

\begin{verbatim}
source <absolute path to the file>
\end{verbatim}

\paragraph{\texttt{\textbf{ADD\_RUNTIME\_LIB(names)}}}

This CMake command can be used to add additional libraries to the global search path.
The mandatory argument \parameter{names} should be the absolute path of a library or a list of paths, such as:

\begin{minted}[frame=single,framesep=3pt,breaklines=true,tabsize=2,linenos]{cmake}
# This module requires the LCIO library:
FIND_PACKAGE(LCIO REQUIRED)
# The FIND routine provides all libraries in the LCIO_LIBRARIES variable:
ADD_RUNTIME_LIB(${LCIO_LIBRARIES})
\end{minted}

The command uses the \command{GET_FILENAME_COMPONENT} command of CMake with the \parameter{DIRECTORY} option to determine the directory of the corresponding shared library.
Duplicates are removed from the list automatically.
Each directory found will be added to the global library search path by adding the following line to the setup file:

\begin{verbatim}
export LD_LIBRARY_PATH="<library directory>:$LD_LIBRARY_PATH"
\end{verbatim}

\subsection{Continuous Integration}
\label{sec:ci}

Quality and compatibility of the \corry framework is ensured by an elaborate continuous integration (CI) which builds and tests the software on all supported platforms.
The \corry CI uses the GitLab Continuous Integration features and consists of six distinct stages.
It is configured via the \file{.gitlab-ci.yml} file in the repository's root directory, while additional setup scripts for the GitLab CI Runner machines and the Docker instances can be found in the \dir{.gitlab-ci.d} directory.

The \textbf{compilation} stage builds the framework from the source on different platforms.
Currently, builds are performed on Scientific Linux 6, CentOS7, and Mac OS X.
On Linux type platforms, the framework is compiled with recent versions of GCC and Clang, while the latest AppleClang is used on Mac OS X.
The build is always performed with the default compiler flags enabled for the project:
\begin{verbatim}
    -pedantic -Wall -Wextra -Wcast-align -Wcast-qual -Wconversion
    -Wuseless-cast -Wctor-dtor-privacy -Wzero-as-null-pointer-constant
    -Wdisabled-optimization -Wformat=2 -Winit-self -Wlogical-op
    -Wmissing-declarations -Wmissing-include-dirs -Wnoexcept
    -Wold-style-cast -Woverloaded-virtual -Wredundant-decls
    -Wsign-conversion -Wsign-promo -Wstrict-null-sentinel
    -Wstrict-overflow=5 -Wswitch-default -Wundef -Werror -Wshadow
    -Wformat-security -Wdeprecated -fdiagnostics-color=auto
    -Wheader-hygiene
\end{verbatim}

The \textbf{testing} stage executes the framework tests described in Section~\ref{sec:tests}.
All tests are expected to pass, and no code that fails to satisfy all tests will be merged into the repository.

The \textbf{formatting} stage ensures proper formatting of the source code using the \command{clang-format} and following the coding conventions defined in the \file{.clang-format} file in the repository.
In addition, the \command{clang-tidy} tool is used for ``linting'' of the source code.
This means, the source code undergoes a static code analysis in order to identify possible sources of bugs by flagging suspicious and non-portable constructs used.
Tests are marked as failed if either of the CMake targets \command{make check-format} or \command{make check-lint} fail.
No code that fails to satisfy the coding conventions and formatting tests will be merged into the repository.

The \textbf{documentation} stage prepares this user manual as well as the Doxygen source code documentation for publication.
This also allows to identify e.g.\ failing compilation of the \LaTeX~documents or additional files which accidentally have not been committed to the repository.

The \textbf{packaging} stage wraps the compiled binaries up into distributable tarballs for several platforms.
This includes adding all libraries and executables to the tarball as well as preparing the \file{setup.sh} script to prepare run-time dependencies using the information provided to the build system.
This procedure is described in more detail in Section~\ref{sec:packaging}.

Finally, the \textbf{deployment} stage is only executed for new tags in the repository.
Whenever a tag is pushed, this stages receives the build artifacts of previous stages and publishes them to the \corry project website through the EOS file system~\cite{eos}. More detailed information on deployments is provided in the following.

\subsection{Automatic Deployment}

The CI is configured to automatically deploy new versions of \corry and its user manual and code reference to different places to make them available to users.
This section briefly describes the different deployment end-points currently configured and in use.
The individual targets are triggered either by automatic nightly builds or by publishing new tags.
In order to prevent accidental publications, the creation of tags is protected.
Only users with \emph{Maintainer} privileges can push new tags to the repository.
For new tagged versions, all deployment targets are executed.

\subsubsection{Software deployment to CVMFS}
\label{sec:cvmfs}

The software is automatically deployed to CERN's VM file system (CVMFS)~\cite{cvmfs} for every new tag.
In addition, the \parameter{master} branch is built and deployed every night.
New versions are published to the folder \dir{/cvmfs/clicdp.cern.ch/software/corryvreckan/} where a new folder is created for every new tag, while updates via the \parameter{master} branch are always stored in the \dir{latest} folder.

The deployed version currently comprises of all modules as well as the detector models shipped with the framework.
An additional \file{setup.sh} is placed in the root folder of the respective release, which allows all the runtime dependencies necessary for executing this version to be set up.
Versions for both SLC\,6 and CentOS\,7 are provided.

The deployment CI job runs on a dedicated computer with a GitLab SSH runner.
Job artifacts from the packaging stage of the CI are downloaded via their ID using the script found in \dir{.gitlab-ci.d/download_artifacts.py}, and are made available to the \emph{cvclicdp} user who has access to the CVMFS interface.
The job checks for concurrent deployments to CVMFS and then unpacks the tarball releases and publishes them to the CLICdp experiment CVMFS space, the corresponding script for the deployment can be found in \dir{.gitlab-ci.d/gitlab_deployment.sh}.
This job requires a private API token to be set as secret project variable through the GitLab interface, currently this token belongs to the service account user \emph{corry}.

\subsubsection{Documentation deployment to EOS}

The project documentation is deployed to the project's EOS space at \dir{/eos/project/c/corryvreckan/www/} for publication on the project website.
This comprises both the PDF and HTML versions of the user manual (subdirectory \dir{usermanual}) as well as the Doxygen code reference (subdirectory \dir{reference/}).
The documentation is only published only for new tagged versions of the framework.

The CI jobs uses the \parameter{ci-web-deployer} Docker image from the CERN GitLab CI tools to access EOS, which requires a specific file structure of the artifact.
All files in the artifact's \dir{public/} folder will be published to the \dir{www/} folder of the given project.
This job requires the secret project variables \parameter{EOS_ACCOUNT_USERNAME} and \parameter{EOS_ACCOUNT_PASSWORD} to be set via the GitLab web interface.
Currently, this uses the credentials of the service account user \emph{corry}.

\subsubsection{Release tarball deployment to EOS}

Binary release tarballs are deployed to EOS to serve as downloads from the website to the directory \dir{/eos/project/c/corryvreckan/www/releases}.
New tarballs are produced for every tag as well as for nightly builds of the \parameter{master} branch, which are deployed with the name \file{corryvreckan-latest-<system-tag>-opt.tar.gz}.

The files are taken from the packaging jobs and published via the \parameter{ci-web-deployer} Docker image from the CERN GitLab CI tools.
This job requires the secret project variables \parameter{EOS_ACCOUNT_USERNAME} and \parameter{EOS_ACCOUNT_PASSWORD} to be set via the GitLab web interface.
Currently, this uses the credentials of the service account user \emph{corry}.

\subsubsection{Building Docker images}
\label{sec:build-docker}

New \corry Docker images are automatically created and deployed by the CI for every new tag and as a nightly build from the \parameter{master} branch.
New versions are published to project Docker container registry~\cite{corry-container-registry}.
Tagged versions can be found via their respective tag name, while updates via the nightly build are always stored with the \parameter{latest} tag attached.

The final Docker image is formed from three consecutive images with different layers of software added.
The `base` image contains all build dependencies such as compilers, CMake, and git.
It derives from a CentOS7 Docker image and can be build using the \file{etc/docker/Dockerfile.base} file via the following commands:

\begin{verbatim}
# Log into the CERN GitLab Docker registry:
$ docker login gitlab-registry.cern.ch
# Compile the new image version:
$ docker build --file etc/docker/Dockerfile.base          \
               --tag gitlab-registry.cern.ch/corryvreckan/\
                     corryvreckan/corryvreckan-base       \
               .
# Upload the image to the registry:
$ docker push gitlab-registry.cern.ch/corryvreckan/\
              corryvreckan/corryvreckan-base
\end{verbatim}

The main dependency of the framework us ROOT6, which is added to the base image via the \parameter{deps} Docker image created from the file \file{etc/docker/Dockerfile.deps} via:
\begin{verbatim}
$ docker build --file etc/docker/Dockerfile.deps          \
               --tag gitlab-registry.cern.ch/corryvreckan/\
               corryvreckan/corryvreckan-deps             \
              .
$ docker push gitlab-registry.cern.ch/corryvreckan/\
              corryvreckan/corryvreckan-deps
\end{verbatim}
These images are created manually and only updated when necessary, i.e.\ if major new version of the underlying dependencies are available.

Finally, the latest revision of \corry is built using the file \file{etc/docker/Dockerfile}.
This job is performed automatically by the continuous integration and the created containers are directly uploaded to the project's Docker registry.
\begin{verbatim}
$ docker build --file etc/docker/Dockerfile                            \
               --tag gitlab-registry.cern.ch/corryvreckan/corryvreckan \
              .
\end{verbatim}

A short summary of potential use cases for Docker images is provided in Section~\ref{sec:docker}.

\subsection{Data-driven Functionality Tests}
\label{sec:tests}

The build system of the framework provides a set of automated tests which are executed by the CI to ensure proper functioning of the framework and its modules.
The tests can also be manually invoked from the build directory of \corry with
\begin{verbatim}
$ ctest
\end{verbatim}

Individual tests be executed or ignored using the \command{-E} (exclude) and \command{-R} (run) switches of the \command{ctest} program:
\begin{verbatim}
$ ctest -R test_timepix3tel
\end{verbatim}

The configurations of the tests can be found in the \dir{testing/} directory of the repository and are automatically discovered by CMake.
CMake automatically searches for \corry configuration files in the respective directory and passes them to the \corry executable~(cf.\ Section~\ref{sec:executable}).

Adding a new test requires placing the configuration file in the directory, specifying the pass or fail conditions based on the tags described in the following paragraph, and providing reference data as described below.

\paragraph{Pass and Fail Conditions}

The output of any test is compared to a search string in order to determine whether it passed or failed.
These expressions are simply placed in the configuration file of the corresponding tests, a tag at the beginning of the line indicates whether it should be used for passing or failing the test.
Each test can only contain \emph{one passing} and \emph{one failing} expression.
If different functionality and thus outputs need to be tested, a second test should be added to cover the corresponding expression.

\begin{description}
  \item[Passing a test] The expression marked with the tag \parameter{#PASS} has to be found in the output in order for the test to pass. If the expression is not found, the test fails.
  \item[Failing a test] If the expression tagged with \parameter{#FAIL} is found in the output, the test fails. If the expression is not found, the test passes.
  \item[Depending on another test] The tag \parameter{#DEPENDS} can be used to indicate dependencies between tests, e.g.\ if a test requires a data file produced by another test.
  \item[Defining a timeout] For performance tests the runtime of the application is monitored, and the test fails if it exceeds the number of seconds defined using the \parameter{#TIMEOUT} tag.
  \item[Adding additional CLI options] Additional module command line options can be specified for the \parameter{corry} executable using the \parameter{#OPTION} tag, following the format found in Section~\ref{sec:executable}. Multiple options can be supplied by repeating the \parameter{#OPTION} tag in the configuration file, only one option per tag is allowed.
  \item[Providing datasets] The \parameter{#DATASET} tag allows to specify a configured data set which has to be available in order for the test to be executed. Datasets and their configuration is described below. Only one data set per tag is allowed, multiple tags can be used.
\end{description}

\paragraph{Providing Reference Datasets}

Reference datasets for testing are centrally stored on EOS at \dir{/eos/project/c/corryvreckan/www/data/} and are accessible over the internet. The \file{download_data.py} tool provided in the \dir{testing/} directory of the framework is capable of downloading individual data files, checking their integrity via an SHA256 hash and decompressing the tar archives.

Each test can specify one or several data sets it requires to be present in order to successfully run via the \parameter{#DATASET} tag in the configuration file. The datasets have to be made known to the download script together with their calculated SHA256 hashes when adding a new test to the repository. The file name and hash have to be added to the Python \parameter{DATASETS} array, and the file of the dataset has to be uploaded manually to EOS by one of the framework maintainers. The SHA256 hash can be generated by executing:
\begin{verbatim}
openssl sha256 <dataset>
\end{verbatim}

Paths in the test configuration files should be provided relative to the \dir{testing/} directory, all downloaded data will be stored in individual subdirectories per dataset following the naming scheme \dir{testing/data/<dataset>}.

\section{Additional Tools \& Resources}
\label{ch:additional_tools_resources}

The following section briefly describes tools provided with the Corryvreckan framework.

\hypertarget{jobsub}{%
\subsection{Jobsub}\label{jobsub}}

\hypertarget{overview}{%
\subsubsection{Overview}\label{overview}}

\parameter{jobsub} is a tool for the convenient
run-specific modification of Corryvreckan configuration files and their
execution through the \parameter{corry} executable. 
It can be used for automated processing of several data files and/or scans of reconstruction parameters.
It is derived from the original \parameter{jobsub} written for
EUTelescope~\cite{eutel-website} by Hanno Perrey, Lund University.

It should be noted that when using \parameter{jobsub} on a local machine, the jobs are processed one by one. When running in \texttt{batch} mode, all jobs are submitted to HTCondor and processed in parallel.

\hypertarget{usage}{%
\subsubsection{Usage}\label{usage}}

The following help text is printed when invoking
\parameter{jobsub} with the \parameter{-h}
argument:

\begin{verbatim}
usage: jobsub.py [-h] [--option NAME=VALUE] [-c FILE] [-csv FILE]
                 [--log-file FILE] [-l LEVEL] [-s] [--dry-run]
         [--batch FILE] [--subdir]
                 jobtask [runs [runs ...]]

A tool for the convenient run-specific modification of Corryvreckan 
steering files and their execution through the corry executable

positional arguments:
  runs                  The runs to be analyzed; can be a list of
                        single runs and/or a range, e.g. 1056-1060.

optional arguments:
  -h, --help            show this help message and exit
  --version             show program's version number and exit
  -c FILE, --conf-file FILE, --config FILE
                        Configuration file with all Corryvreckan 
                        algorithms defined
  --option NAME=VALUE, -o NAME=VALUE
                        Specify further options such as 
                        'beamenergy=5.3'. This switch be specified 
                        several times for multiple options
                        or can parse a comma-separated list of options. 
                        This switch overrides any configuration file 
                        options and also overwrites hard-coded settings 
                        on the Corryvreckan configration file.
  -htc FILE, --htcondor-file FILE, --batch FILE
                        Specify condor_submit parameter file for HTCondor
                        submission. Run HTCondor submission via 
                        condor_submit instead of calling Corryvreckan 
                        directly
  -csv FILE, --csv-file FILE
                        Load additional run-specific variables from
                        table (text file in csv format)
  --log-file FILE       Save submission log to specified file
  -v LEVEL, --verbosity LEVEL
                        Sets the verbosity of log messages during job
                        submission where LEVEL is either debug, info, 
                        warning or error
  -s, --silent          Suppress non-error (stdout) Corryvreckan 
                        output to console
  --dry-run             Write configuration files but skip actual 
                        Corryvreckan execution
  --subdir              Execute every job in its own subdirectory 
                        instead of all in the base path
  --plain               Output written to stdout/stderr and log
                        file in prefix-less format i.e. without 
                        time stamping
  --zfill N             Fill run number with zeros up to the defined 
                        number of digits
\end{verbatim}

If running on lxplus, the environment variables need to be set using
\parameter{source etc/setup_lxplus.sh}. When using a
submission file, \parameter{getenv = True} should be used
(see example.sub).

\hypertarget{preparation-of-configuration-file-templates}{%
\subsubsection{Preparation of Configuration File
Templates}\label{preparation-of-configuration-file-templates}}

Configuration file templates are valid Corryvreckan configuration files
in TOML format, where single values are replaced by variables in the
form \parameter{@SomeVariable@}. A more detailed
description of the configuration file format can be found in
Chapter~\ref{ch:configuration_files}. The section of a configuration file template with
variable geometry file could for instance look like

\begin{minted}[frame=single,framesep=3pt,breaklines=true,tabsize=2,linenos]{ini}
[Corryvreckan]
detectors_file = "@telescopeGeometry@"
histogram_file = "histograms_@RunNumber@.root"

number_of_events = 5000000

log_level = WARNING
\end{minted}

When \parameter{jobsub} is executed, these placeholders
are replaced with user-defined values that can be specified through
command-line arguments or a table with a row for each run number
processed, and a final configuration file is produced for each run
separately, e.g.

\begin{minted}[frame=single,framesep=3pt,breaklines=true,tabsize=2,linenos]{ini}
[Corryvreckan]
detectors_file = "my_telescope_Nov2017_1.conf"
histogram_file = "histograms_run999.root"

number_of_events = 5000000

log_level = WARNING
\end{minted}

There is only one predefined placeholder,
\parameter{@RunNumber@}, which will be substituted with
the current run number. Run numbers are not padded with leading zeros
unless the \parameter{--zfill} option is provided.

\begin{minted}[frame=single,framesep=3pt,breaklines=true,tabsize=2,linenos]{ini}
spatialCut = @spatialCutX@, @spatialCutY@
\end{minted}

\hypertarget{using-configuration-variables}{%
\subsubsection{Using Configuration
Variables}\label{using-configuration-variables}}

As described in the previous paragraph, variables in the configuration
file template are replaced with values at run time. Two sources of
values are currently supported, and are described in the following.

\hypertarget{command-line}{%
\paragraph{Command Line}\label{command-line}}

Variable substitutions can be specified using the
\parameter{--option} or \parameter{-o}
command line switches, e.g.

\begin{minted}[frame=single,framesep=3pt,breaklines=true,tabsize=2,linenos]{ini}
jobsub.py --option beamenergy=5.3 -c alignment.conf 1234
\end{minted}

This switch can be specified several times for multiple options or can
parse a comma-separated list of options. This switch overrides any
configuration file options.

\hypertarget{table-comma-separated-text-file}{%
\paragraph{Table (comma-separated text
file)}\label{table-comma-separated-text-file}}

Tables in the form of CSV files can be used to replace placeholders with
the \passthrough{\lstinline!-csv!} option. For the correct format, the
following tools can be used:
\begin{itemize}
\tightlist
\item
  export from LibreOffice with default settings (UTF-8, comma-separated, text-field delimiter:~")
\item
  emacs org-mode table (see http://orgmode.org/manual/Tables.html)
\item
  use Atom's \emph{tablr} extension
\item
  or the CSV file can be edited in a text editor of choice.
\end{itemize}

The following rules apply:
\begin{itemize}
\tightlist
\item
  commented lines (starting with \#) are ignored
\item
  first row (after comments) has to provide column headers which
  identify the variables in the steering template to replace
  (case-insensitive)
\item
  requires one column labeled ``RunNumber''
\item
  only considers placeholders left in the steering template after
  processing command-line arguments and configuration file options
\end{itemize}

Strings can be passed by the user of double-quotes \parameter{" "} which also avoid the separation by commas.
A double-quote can be used as part of a string when using the escape
character backslash \parameter{\} in front of the double-quote.

It is also possible to specify multiple different settings for the same
run number by making use of the following syntax to specify a set or
range of parameters. Curly brackets in double-quotes
\parameter{"{ }"} can be used to indicate a set
(indicated by a comma \parameter{,}) or range (indicated
by a dash \parameter{-}) of parameters which will be split
up and processed one after the other. If a set or a range is detected,
the parameter plus its value are attached to the name of the
configuration file. Ranges can only be used for integer values (without
units). However, a set or range can oly be used for one parameter,
i.e.~multi-dimensional parameters scans are not supported and have to be
separated into individual CSV files.

\begin{itemize}
\tightlist
\item
  \parameter{"{10,12-14}"} translates to
  \parameter{10}, \parameter{12}, \parameter{13}, \parameter{14} in consecutive jobs for the same run number
\item
  \parameter{"{10ns, 20ns}"} tranlates to \parameter{10ns}, \parameter{20ns} in consecutive jobs for the same run number
\item
  \parameter{"string,with,comma"} translates to \parameter{string,with,comma!} in one job
\item
  \parameter{"{string,with,comma}"!} which translates to \parameter{string}, \parameter{with}, \parameter{comma} in consecutive jobs for the same run number
\item
  \parameter{"\"string in quotes\""} translates to \parameter{"string in quotes"}
\end{itemize}

If a range or set of parameters is detector, the naming scheme of the auto-generated configuration files is extended from
\parameter{MyAnalysis_run@RunNUmber@.conf} to \parameter{MyAnalysis_run@RunNUmber@_OtherParameter@OtherParameter@.conf}

It must be insured by the user that the output ROOT file is not simply called \parameter{histograms_@RunNumber@.root} but rather
\parameter{histograms_@RunNumber@_OtherParameter@OtherParameter@.root} to prevent overwriting the output file.

\hypertarget{example}{%
\subparagraph{Example}\label{example}}

The CSV file could have the following form:

\begin{minted}[frame=single,framesep=3pt,breaklines=true,tabsize=2,linenos]{ini}
# AnalysisExample.csv
# This is an example.
RunNumber,  ExampleParameter,   AnotherParameter
100,        "{3-5}",            10ns
101,        3,                  "{10ns, 20ns}"
\end{minted}

Using this table, the placeholders
\parameter{@RunNUmber@}, \parameter{@ExampleParameter@}, and \parameter{@AnotherParameter@} in the template file \parameter{AnalysisExample.conf} would be replaced by the values corresponding to the current run number and the following configuration files would be generated:

\begin{itemize}
\tightlist
\item AnalysisExample\_run100\_exampleparameter3.conf
\item AnalysisExample\_run100\_exampleparameter4.conf
\item AnalysisExample\_run100\_exampleparameter5.conf
\item AnalysisExample\_run101\_anotherparameter10ns.conf
\item AnalysisExample\_run101\_anotherparameter20ns.conf
\end{itemize}

\hypertarget{example-usage-with-a-batch-file}{%
\subsubsection{Example Usage with a Batch
File:}\label{example-usage-with-a-batch-file}}

Example command line usage:

\begin{minted}[frame=single,framesep=3pt,breaklines=true,tabsize=2,linenos]{ini}
./jobsub.py -c /path/to/example.conf -v DEBUG --batch /path/to/example.sub \
	--subdir <run_number>
\end{minted}

An example batch file is provided in the repository as
\parameter{htcondor.sub}. Complicated and error-prone
\parameter{transfer_output_files} commands can be
avoided. It is much simpler to set an absolute path like

\begin{minted}[frame=single,framesep=3pt,breaklines=true,tabsize=2,linenos]{ini}
output_directory = "/eos/user/y/yourname/whateveryouwant/run@RunNumber@"
\end{minted}

directly in the Corryvreckan configuration file.

\label{sec:jobsub}

\clearpage

\appendix
\section{Acknowledgements}

\corry has been developed and is maintained by:

\begin{itemize}
    \item Morag Williams, University of Glasgow/CERN
    \item Jens Kroeger, University of Heidelberg/CERN
    \item Daniel Hynds, NIKHEF
    \item Simon Spannagel, CERN
\end{itemize}

The following authors, in alphabetical order, have contributed to \corry:

\begin{itemize}
    \item Matthew Daniel Buckland, University of Liverpool
    \item Estel Perez Codina, CERN
    \item Dominik Dannheim, CERN
    \item Katharina Dort, University of Gie\ss en/CERN
    \item Adrian Fiergolski, CERN
    \item Lennart Huth, DESY
    \item Andreas Matthias Nürnberg, KIT
    \item Florian Pitters, HEPHY Vienna
    \item Paul Schütze, DESY
    \item Tomas Vanat, CERN
\end{itemize}

\clearpage
\printbibliography[heading=bibintoc,title={References}]

\end{document}